\documentclass[lettersize,journal]{IEEEtran}

\ifCLASSOPTIONcompsoc
  \usepackage[nocompress]{cite}
\else
  \usepackage{cite}

\fi

\usepackage{cite}
\usepackage{amsmath,amsfonts}
\usepackage{array}
\usepackage{amssymb}
\usepackage{amsthm}
\usepackage{mathpazo}
\usepackage{textcomp}
\usepackage{stfloats}
\usepackage{url}
\usepackage{verbatim}
\usepackage{subfigure}
\usepackage{graphicx}
\usepackage{framed}
\usepackage{enumitem}
\usepackage{color}
\usepackage{multirow}
\usepackage{etoolbox}
\usepackage{booktabs}
\usepackage{changepage}
\usepackage{bm}

\usepackage{setspace} 

\usepackage{pifont}
\usepackage{utfsym}
\usepackage{bbding}

\usepackage[ruled,vlined,linesnumbered]{algorithm2e}

\usepackage[mathscr]{eucal}

\usepackage{pgfplots}
\usepackage{colortbl}
\usepackage{hyperref}
\usepackage[flushleft]{threeparttable}
\usepackage{makecell}
\usepackage{rotating}

\usepackage[mathscr]{eucal}

\newtheorem{definition}{Definition}
\theoremstyle{definition}
\newtheorem{nameddefinition}[definition]{Definition}

\definecolor{customteal}{RGB}{0, 169, 131}

\definecolor{darkgreen}{rgb}{0.0, 0.5, 0.0}

\newcommand{\Tx}[0]{\ensuremath{\mathtt{Tx}}}
\newcommand{\TxCC}[0]{\ensuremath{\mathtt{Tx\_CC}}}

\def\BibTeX{{\rm B\kern-.05em{\sc i\kern-.025em b}\kern-.08em
    T\kern-.1667em\lower.7ex\hbox{E}\kern-.125emX}}
\usepackage{balance}

\begin{document}

\title{\LARGE{\textbf{\textsf{Enhancing Blockchain Cross-Chain Interoperability:\\ A Comprehensive Survey}}}}

\author{
Zhihong Deng\textsuperscript{†},
Chunming Tang\textsuperscript{†*},
Taotao Li\textsuperscript{‡},
Parhat Abla\textsuperscript{§},
Qi Chen\textsuperscript{¶},
Wei Liang\textsuperscript{‖},
Debiao He\textsuperscript{$\flat$}
\thanks{\textsuperscript{†} School of Mathematics and Information Science, Guangzhou University, Guangzhou, China.}
\thanks{\textsuperscript{‡} School of Software Engineering, Sun Yat-Sen University, Zhuhai, China.}
\thanks{\textsuperscript{§} School of Software, South China Normal University, Foshan, China.}
\thanks{\textsuperscript{¶} Institute of Artificial Intelligence, Guangzhou University, Guangzhou, China.}
\thanks{\textsuperscript{‖} School of Computer Science and Engineering, Hunan University of Science and Technology, Xiangtan, China.}
\thanks{\textsuperscript{$\flat$} School of Cyber Science and Engineering, Wuhan University, Wuhan, China.}
\thanks{* Corresponding author: ctang@gzhu.edu.cn.}
}



\markboth{}%
{How to Use the IEEEtran \LaTeX \ Templates}

\maketitle

\begin{abstract}

 Blockchain technology, introduced in 2008, has revolutionized data storage and transfer across sectors such as finance, healthcare, intelligent transportation, and the metaverse. However, the proliferation of blockchain systems has led to discrepancies in architectures, consensus mechanisms, and data standards, creating "\emph{data and value silos}" that hinder the development of an integrated multi-chain ecosystem. Blockchain interoperability (a.k.a cross-chain interoperability) has thus emerged as a solution to enable seamless data and asset exchange across disparate blockchains. 
 In this survey, we systematically analyze over 150 high-impact sources from academic journals, digital libraries, and grey literature to provide an in-depth examination of blockchain interoperability. By exploring the existing methods, technologies, and architectures, we offer a classification of interoperability approaches including Atomic Swaps, Sidechains, Light Clients, and so on, which represent the most comprehensive overview to date. Furthermore, we investigate the convergence of academic research with industry practices, underscoring the importance of collaborative efforts in advancing blockchain innovation. Finally, we identify key strategic insights, challenges, and future research trajectories in this field. Our findings aim to support researchers, policymakers, and industry leaders in understanding and harnessing the transformative potential of blockchain interoperability to address current challenges and drive forward a cohesive multi-chain ecosystem.
\end{abstract}


\section{Introduction}

\begin{figure}[!ht] 
\small
\centering
\includegraphics[width=3.4in]{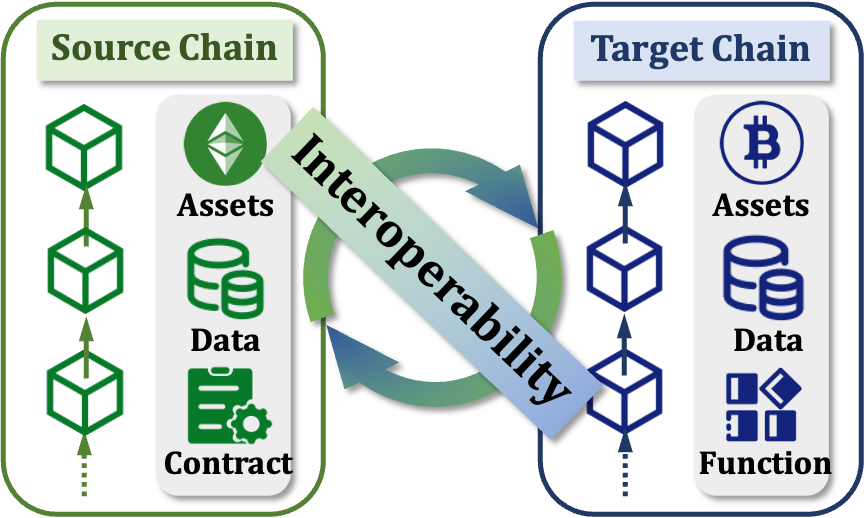} 
   \vspace{-2mm}
\caption{Blockchain interoperability.}
\label{attack}
   \vspace{-2mm}
\end{figure} 

Since its inception in 2008 \cite{nakamoto2008peer}, blockchain technology has rapidly evolved, finding applications across diverse fields such as finance \cite{wu2024blockchain}, intelligent transportation \cite{wang2023blockchain}, healthcare \cite{arbabi2022survey}, Artificial Intelligence (AI) \cite{zuo2023survey}, and the metaverse \cite{li2023metaopera}. This decentralized, transparent, and efficient distributed ledger technology has revolutionized modern data storage and transmission. However, as more blockchain systems emerge \cite{buterin2013ethereum,androulaki2018hyperledger,kwon2014tendermint,wang2019monoxide,pass2016hybrid}, significant disparities in underlying architectures, consensus mechanisms, smart contracts, and data formats have resulted in the formation of “\textbf{\emph{data and value silos}}” between systems. This fragmentation severely constrains the application potential of blockchain and impedes the development of a cohesive multi-chain ecosystem.

To address this issue, the concept of blockchain interoperability (or cross-chain interoperability, $\mathcal{CCI}$) has emerged, aimed at facilitating seamless data and asset transfers between disparate blockchain systems. Some scholars argue that interoperability is a crucial factor in unlocking the full potential of blockchain technology \cite{augusto2024sok}. As the scope of applications expands, the value and commitments of interoperability become increasingly apparent across various scenarios, including enhancing asset liquidity, reducing transaction costs, improving user experience, and breaking network silos, as outlined in Tab. 
\ref{commitmenttab}. 
However, achieving interoperability between blockchains introduces considerable complexity—not only does it require synchronizing multiple software components, but it also necessitates integrating diverse distributed systems while addressing unique challenges in security, liveness, data consistency, atomicity, and decentralization. This cross-chain collaboration primarily relies on interoperability mechanisms. Nevertheless, the implementation of interoperability faces numerous obstacles due to the diversity of blockchain systems in network structures, cryptographic primitives, and application contexts. Despite these challenges, academic contributions have yielded a variety of protocols \cite{poon2016bitcoin,klamti2022post,xie2022zkbridge}, novel architectures \cite{sidechainpeg2014,li2020scalable,luu2016secure}, and practical applications \cite{arbabi2022survey,wu2024blockchain,wang2023blockchain}, underscoring the importance of $\mathcal{CCI}$.

  \begin{table*}[!htbp]
 \caption{A Vision of Interoperability Commitments.}
 \centering
 \scriptsize
 \setlength{\tabcolsep}{16.2pt}
 \renewcommand{\arraystretch}{1.5}
 \label{commitmenttab}
 \begin{threeparttable}    
  \begin{tabular}{>{\columncolor{blue!6}}p{1.2cm}|p{6.5cm}|p{6.8cm}}
   \rowcolor{blue!6}
   \bottomrule
   
   \textsl{\textbf{Benefit}}  & \multicolumn{1}{c}{\textsl{\textbf{Description}}} & \multicolumn{1}{|c}{\textsl{\textbf{Example Application}}}  \\
   \hline
   
\qquad  \emph{\textbf{Enhancing}}\newline  \emph{\textbf{Asset}} \newline  \emph{\textbf{Liquidity}} & \qquad 
Liquidity fragmentation across blockchains hampers overall market efficiency. Cross-chain platforms address this issue by enabling seamless asset transfers between different chains, fostering a more unified and liquid cryptocurrency market. Enhanced liquidity allows users to select the most optimal platforms for trading and lending, thereby maximizing their investment returns. & \qquad Uniswap v3's recent integration of Layer 2 solutions \cite{adams2024layer,nassr2024concentration}, such as Optimism and Arbitrum, has enabled users to transfer assets from the Ethereum mainnet to these Layer-2 networks via cross-chain bridge. This allows them to benefit from lower transaction fees and faster transaction speeds. Such liquidity migration empowers users to move funds swiftly and efficiently across different chains, thereby optimizing their investment strategies.  \\ \hline

\qquad  \emph{\textbf{Reducing}}\newline  \emph{\textbf{Transaction}} \newline  \emph{\textbf{Costs}} & \qquad $\mathcal{CCI}$ enables the transfer of assets from blockchains with high transaction fees to those with lower fees, thereby reducing transaction costs for users. For digital finance users, this results in lower transaction friction and greater capital efficiency. & \qquad SushiSwap and Uniswap \cite{berg2022empirical} offer interoperability solutions that allow users to trade and provide liquidity across multiple blockchains. Users can choose to conduct transactions on chains with lower fees, reducing transaction costs and enhancing the efficiency of capital utilization.  \\ \hline

\qquad  \emph{\textbf{Improving}}\newline  \emph{\textbf{User}} \newline  \emph{\textbf{Experience}} & \qquad  The frequent need to operate across diverse blockchain networks necessitates the management of multiple wallets and the navigation of intricate user interfaces. By integrating $\mathcal{CCI}$, these challenges are significantly mitigated, providing users with a seamless and enhanced operational experience.
   & \qquad Protocols such as Axelar Network \cite{agrawal2023grant} facilitate cross-chain messaging, empowering developers to design applications capable of interacting with assets and data across disparate blockchain networks, thereby obviating the need for users to handle intricate infrastructure management.\\ \hline

\qquad  \emph{\textbf{Breaking}}\newline  \emph{\textbf{Network}} \newline  \emph{\textbf{Silos}} & \qquad  In traditional blockchain systems, information is often isolated within individual chains, preventing direct communication between them. $\mathcal{CCI}$ technology breaks down these barriers, enabling the free exchange and sharing of information across different chains, leading to more efficient data collaboration.   & \qquad  The Cosmos network \cite{kwon2019cosmos}, through its IBC protocol, allows different blockchains to transmit information without the need for third parties. For instance, the source chain can send its state or data to target chain via IBC \cite{wei2023formal}, enabling target chain to trigger its own smart contracts based on this information, thereby facilitating cross-chain data interaction. \\

   \toprule
  \end{tabular}

 \end{threeparttable}    
 \label{table fault detection}
\end{table*}

\begin{figure*}[!ht] 
\small
\centering
\includegraphics[width=6.8in]{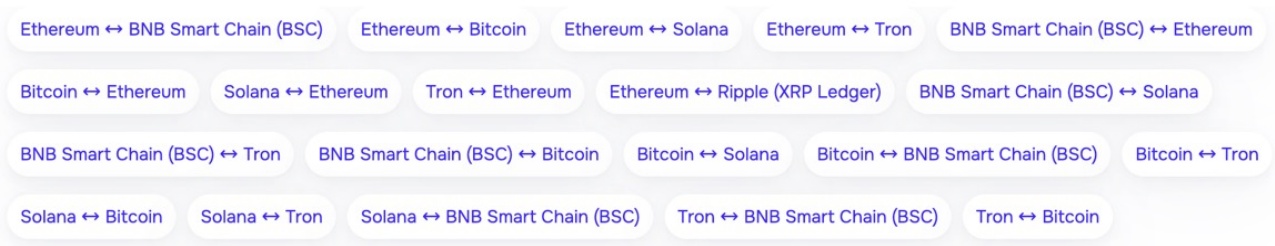} 
   \vspace{-2mm}
\caption{The most popular interoperability routes.}
\label{popular_fig}
   \vspace{-2mm}
\end{figure*}

\begin{figure}[!ht] 
\small
\centering
\includegraphics[width=3.4in]{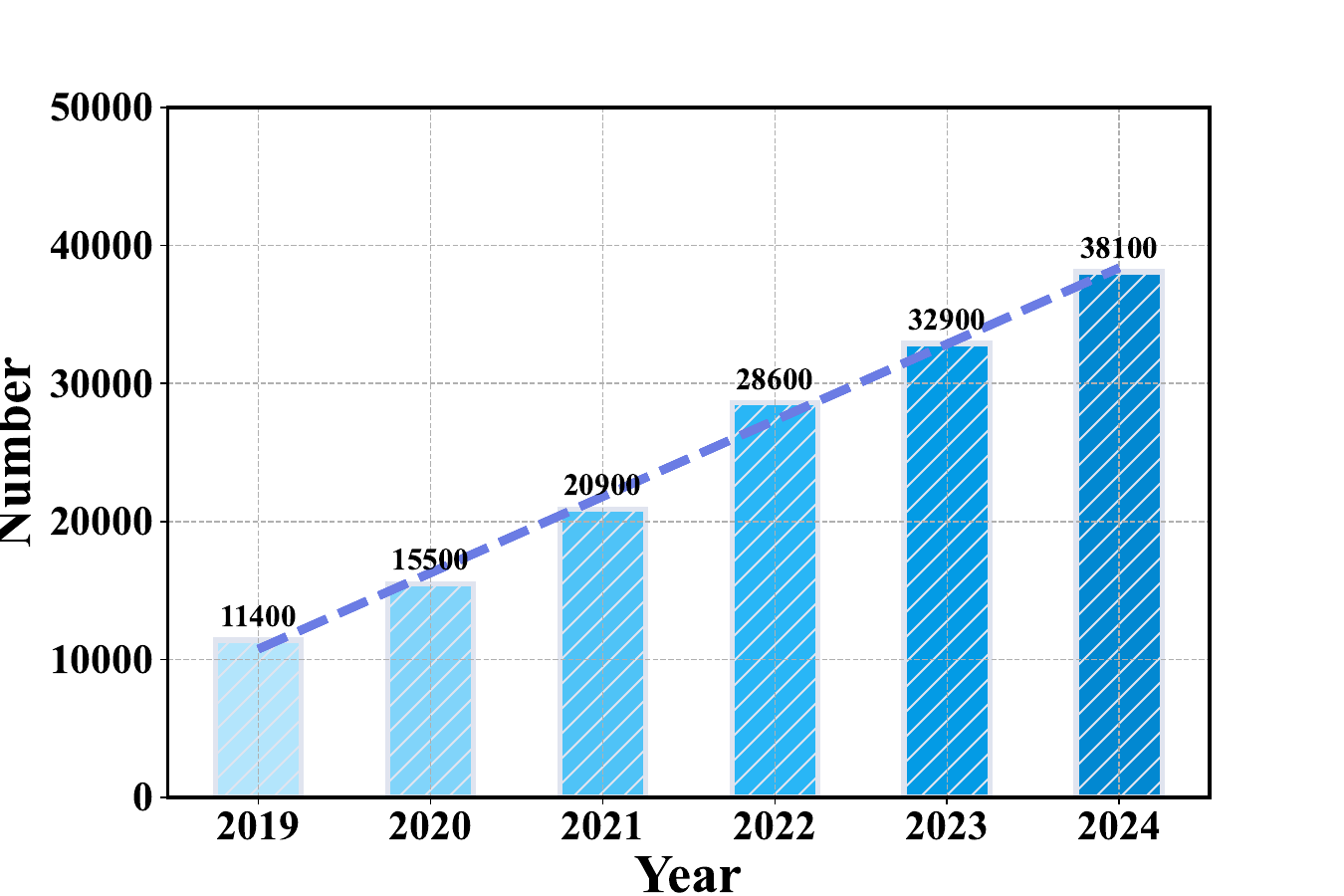} 
   \vspace{-2mm}
\caption{The number of interoperability studies on Google Scholar over the past six years.}
\label{yearnumberfig}
   \vspace{-2mm}
\end{figure}

\subsection{Motivation}

The significance of researching blockchain interoperability technologies cannot be overstated. Consider the following data: \ding{172} By the end of 2024, the total market capitalization of digital currencies is expected to reach \$2.32 trillion, encompassing over 9,982 digital currencies and at least 8,000 blockchains, with no fewer than 759 exchanges involved \cite{coinmarketcap}. Fig. \ref{popular_fig} illustrates the most prevalent digital currency interoperability routes. \ding{173} A Google Scholar search for the term “Blockchain \& Interoperability $\mid$ Cross Chain” reveals over 219,000 relevant studies. As shown in Fig. \ref{yearnumberfig}, the number of research publications has steadily increased from 2019 to November 2024, with projections indicating a new peak in 2025. \ding{174} Experts forecast that the global blockchain interoperability market will expand from \$375.46 million in 2024 to \$8.48 billion by the end of 2037, witnessing around anticipated annual growth rate of 27.1\% \cite{researchnester}. This growth is primarily driven by the urgent demand for asset conversion, widespread adoption of decentralized applications (dApps) across industries, and ongoing improvements in regulation and standardization.

However, current solutions in the field are diverse and relatively fragmented, lacking systematic and comprehensive integration. This paper aims to conduct a fine-grained analysis of the existing literature, revealing current research characteristics and limitations while exploring future development potential. Additionally, we seek to foster interdisciplinary dialogue and collaboration, assisting researchers across various domains in recognizing the critical importance of blockchain interoperability in their work, thereby promoting more comprehensive and systematic research endeavors.

\subsection{Research Paradigm}

\begin{figure}[!ht] 
\small
\centering
\includegraphics[width=3.5in]{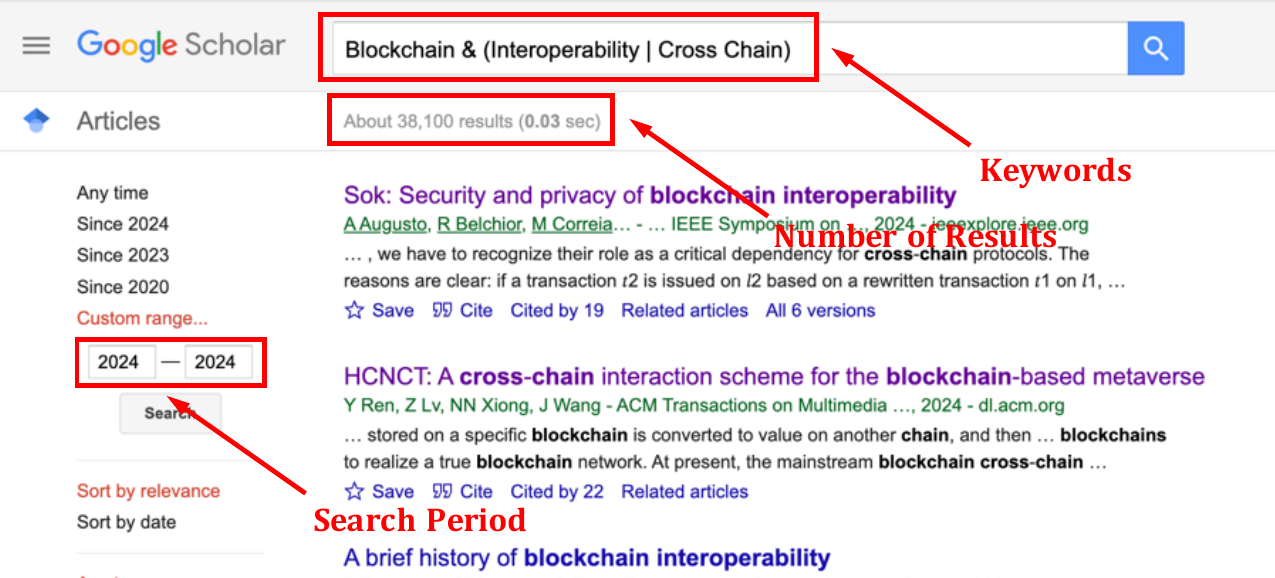} 
   \vspace{-2mm}
\caption{Search Procedure in Google Scholar.}
\label{googlefig}
   \vspace{-2mm}
\end{figure}

This survey employs a systematic literature review methodology to conduct an in-depth analysis of research on blockchain interoperability. The literature collection is centered on Google Scholar, which aggregates content from prominent digital libraries and conference proceedings (e.g., IEEE Xplore, ACM Digital Library, Springer Nature Lecture Notes), alongside gray literature sources (e.g., arXiv, Cryptology ePrint Archive) and other auxiliary resources (e.g., books and repositories like GitHub). By utilizing the keyword
\begin{equation*}
 \texttt{Blockchain} \& \texttt{(Interoperability} \mid \texttt{Cross-Chain)}
\end{equation*}
in our search (as illustrated in Fig. \ref{googlefig}), we restricted the time frame to publications from 2016 to 2024, prioritizing highly cited works relevant to the topic. Following an initial screening and quality assessment, over 150 pertinent studies were selected for in-depth analysis. We also acknowledge the limitations of our research methodology, including potential publication bias and the risk of omitting significant studies.

\begin{figure}[!ht] 
\small
\centering
\includegraphics[width=3.4in]{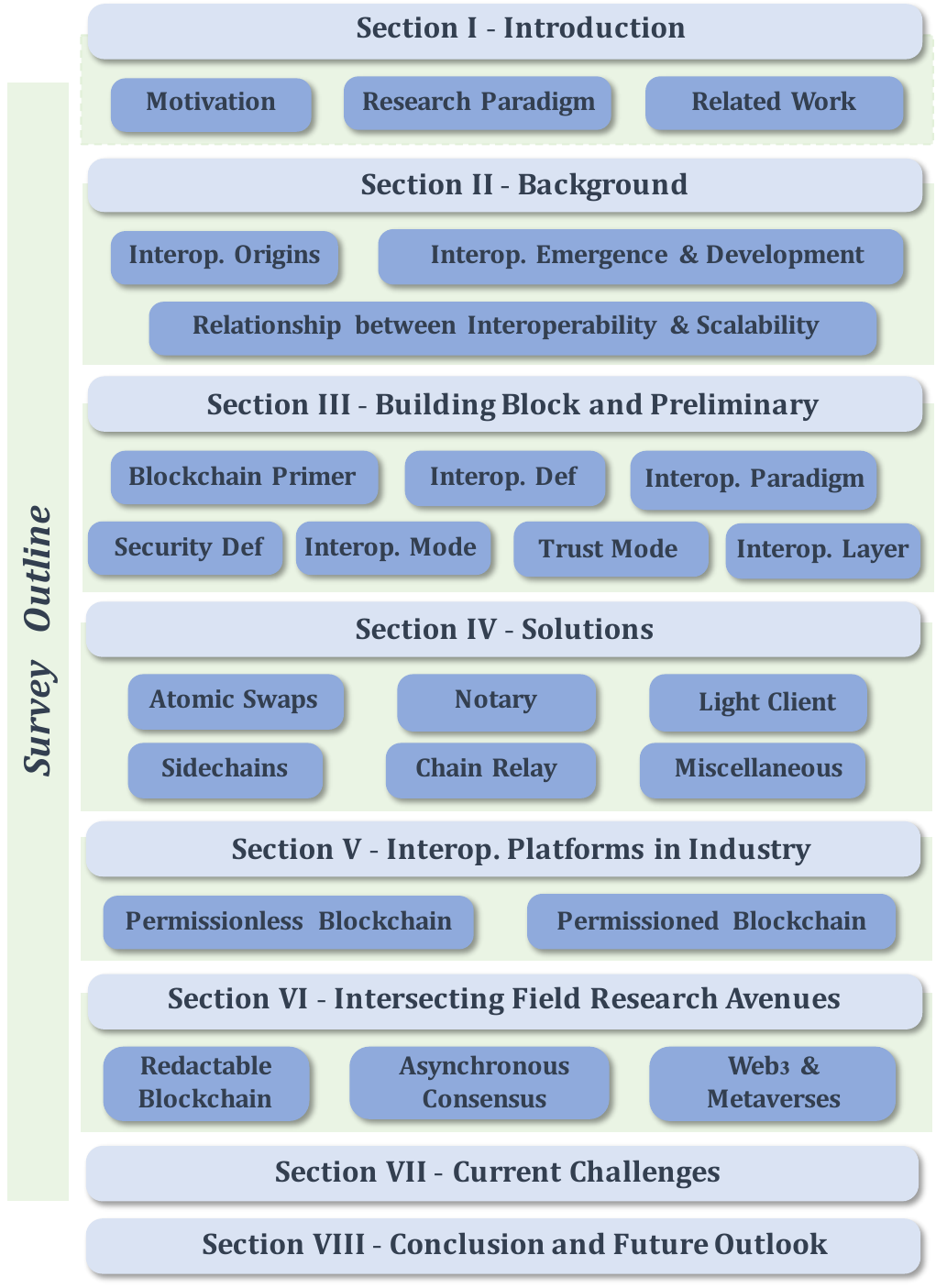} 
   \vspace{-2mm}
\caption{Survey Outline.}
\label{surveyoutlinetab}
   \vspace{-2mm}
\end{figure}

\subsection{Related Works and Contributions}

In recent years, numerous reviews addressing interoperability and cross-chain technologies have emerged. Augusto et al. \cite{augusto2024sok} conducted the most comprehensive investigation to date on the security and privacy of blockchain systems. Belchior et al. \cite{belchior2021survey} classified interoperability into Public Connectors, Blockchain of Blockchains, and Hybrid Connectors. Wang et al. \cite{wang2023exploring} categorized it into chain-based, bridge-based, and dApp-based interoperability. Ren et al. \cite{ren2023interoperability} proposed a performance evaluation mechanism for interoperability approaches. Zamyatin et al. \cite{zamyatin2021sok} systematically articulated cross-chain communication (CCC) protocol for the first time, formalizing the argument that the implementation of CCC is unachievable without a trusted third party. Additional relevant reviews include \cite{koens2019assessing,bhatia2020interoperability,ou2022overview,kotey2023blockchain,zhou2023exploring,li2024blockchain}.

We summarize these surveys in Tab. \ref{surveytatoltab}, evaluating each study from methodological, technical, and discussion perspectives. Furthermore, this paper expands upon the aforementioned research by providing a more comprehensive and systematic examination of the field, with a particular focus on technology types, comparative analyses, and interdisciplinary research. 

This survey provides the following \textbf{contributions}, which are outlined below:

\begin{itemize}[itemsep=0.5pt]

\item  \emph{Systematic Knowledge Construction.} This survey integrates and organizes existing literature, offering a comprehensive description that encompasses building blocks, methodologies, and architectural frameworks. We compare the characteristics and relationships of various technologies for scalability with interoperability and categorize interoperability techniques into native, local, and external validation mechanisms. Furthermore, we provide an in-depth analysis of no fewer than ten technology categories (including HTLC, Adaptor Signatures, Notary, Light Client, Sidechains, Chain Relay, and so on), representing the most extensive classification to date. This systematic construction of knowledge not only mitigates fragmentation within the field but also establishes a solid foundation for future research.
\item \emph{Academic and Industrial Collaboration.} This survey emphasizes the close relationship between academic research and industrial practice. By analyzing multiple classic interoperability platforms, we encourage collaborative efforts between academia and industry to advance the practical application and innovation of blockchain technology.
\item \emph{Strategic Insights.} This survey offers profound insights into the future development of blockchain interoperability, identifying key research directions and potential technological trends. We pay particular attention to areas such as editable blockchains, asynchronous consensus, and the metaverse, as well as challenges related to regulation and knowledge frameworks. These strategic insights aim to guide decision-makers, researchers, and industry leaders in making informed choices within the rapidly evolving technological landscape.

\end{itemize}

The organizational framework of this survey is depicted in Fig. \ref{surveyoutlinetab}.

\begin{table*}[!ht]
 \caption{Comparison of Existing Surveys and Our Work.}
 \centering
 \scriptsize
 \setlength{\tabcolsep}{11.2pt}
 \renewcommand{\arraystretch}{1.5}
 \label{surveytatoltab}
 \begin{threeparttable}    
  \begin{tabular}{>{\columncolor{blue!6}}lcccc|ccccc|ccc}
   \rowcolor{blue!6}
   \bottomrule

\rotatebox{0}{\textsl{\textbf{Reference}}}  &\multicolumn{4}{c}{\textsl{\textbf{Methodology}}}& \multicolumn{5}{c}{\textsl{\textbf{Technique Analysis}}}  &  \multicolumn{3}{c}{\textsl{\textbf{Discussion}}}  \\

 & \textbf{SC} & \textbf{SD} & \textbf{GP} & \textbf{AM} & \textbf{TC}   &  \textbf{TT}  & \textbf{CB}  &  \textbf{IC} & \textbf{IS}  & \textbf{IF} & \textbf{IO} & \textbf{FC}  \\
   \hline

  Koens et al. (2019)  \cite{koens2019assessing} &  \cellcolor{gray!35}\ding{109} & \cellcolor{gray!35}\ding{109} & \cellcolor{gray!35}\ding{109} &  \cellcolor{green!30}\ding{119} &  \cellcolor{green!30}\ding{119} &  \cellcolor{green!30}\ding{119} &  \cellcolor{green!30}\ding{119} &   \cellcolor{green!30}\ding{119} & \cellcolor{gray!35}\ding{109} & \cellcolor{gray!35}\ding{109} & \cellcolor{green!30}\ding{119} & \cellcolor{gray!35}\ding{109} \\

  Bhatia et al. (2020) \cite{bhatia2020interoperability} & \cellcolor{gray!35}\ding{109} & \cellcolor{gray!35}\ding{109} & \cellcolor{gray!35}\ding{109} &  \cellcolor{green!30}\ding{119} & \cellcolor{green!30}\ding{119} & \cellcolor{green!30}\ding{119} & \cellcolor{green!30}\ding{119} & \cellcolor{green!30}\ding{119} & \cellcolor{gray!35}\ding{109} & \cellcolor{gray!35}\ding{109} & \cellcolor{gray!35}\ding{109} & \cellcolor{gray!35}\ding{109} \\

   Belchior et al. (2021) \cite{belchior2021survey} &  \cellcolor{blue!40}\ding{108} &   \cellcolor{green!30}\ding{119} & \cellcolor{green!30}\ding{119} &  \cellcolor{blue!40}\ding{108} &  \cellcolor{green!30}\ding{119} &  \cellcolor{green!30}\ding{119} & \cellcolor{blue!40}\ding{108}  &  \cellcolor{green!30}\ding{119} &  \cellcolor{green!30}\ding{119}  & \cellcolor{gray!35}\ding{109}  &  \cellcolor{blue!40}\ding{108} & \cellcolor{blue!40}\ding{108}   \\
   
   Zamyatin et al. (2021) \cite{zamyatin2021sok}  & \cellcolor{green!30}\ding{119} &   \cellcolor{blue!40}\ding{108} &  \cellcolor{blue!40}\ding{108} &  \cellcolor{blue!40}\ding{108} &  \cellcolor{green!30}\ding{119} & \cellcolor{green!30}\ding{119} & \cellcolor{green!30}\ding{119} & \cellcolor{green!30}\ding{119} & \cellcolor{gray!35}\ding{109} & \cellcolor{gray!35}\ding{109} & \cellcolor{blue!40}\ding{108} & \cellcolor{blue!40}\ding{108} \\

   Ou et al. (2022) \cite{ou2022overview} &  \cellcolor{gray!35}\ding{109} &  \cellcolor{gray!35}\ding{109} &  \cellcolor{gray!35}\ding{109} &  \cellcolor{green!30}\ding{119} &  \cellcolor{green!30}\ding{119} &  \cellcolor{green!30}\ding{119} &  \cellcolor{blue!40}\ding{108} &  \cellcolor{blue!40}\ding{108} &  \cellcolor{gray!35}\ding{109} & \cellcolor{gray!35}\ding{109} &  \cellcolor{blue!40}\ding{108} & \cellcolor{blue!40}\ding{108} \\

   Ren et al. (2023) \cite{ren2023interoperability} & \cellcolor{green!30}\ding{119} & \cellcolor{green!30}\ding{119} &  \cellcolor{gray!35}\ding{109}  & \cellcolor{blue!40}\ding{108} & \cellcolor{green!30}\ding{119} & \cellcolor{green!30}\ding{119} & \cellcolor{green!30}\ding{119} & \cellcolor{blue!40}\ding{108} &  \cellcolor{gray!35}\ding{109} & \cellcolor{gray!35}\ding{109} & \cellcolor{green!30}\ding{119} &  \cellcolor{blue!40}\ding{108} \\
   
    Kotey et al. (2023) \cite{kotey2023blockchain}  &  \cellcolor{blue!40}\ding{108} &   \cellcolor{gray!35}\ding{109} &  \cellcolor{gray!35}\ding{109} &  \cellcolor{green!30}\ding{119} & \cellcolor{green!30}\ding{119} & \cellcolor{green!30}\ding{119} & \cellcolor{green!30}\ding{119} & \cellcolor{blue!40}\ding{108} & \cellcolor{gray!35}\ding{109} & \cellcolor{green!30}\ding{119}  & \cellcolor{green!30}\ding{119} &  \cellcolor{gray!35}\ding{109}   \\

 Zhou et al. (2023) \cite{zhou2023exploring} & \cellcolor{gray!35}\ding{109} &   \cellcolor{gray!35}\ding{109} & \cellcolor{gray!35}\ding{109} & \cellcolor{green!30}\ding{119} & \cellcolor{green!30}\ding{119} & \cellcolor{green!30}\ding{119} & \cellcolor{green!30}\ding{119} &   \cellcolor{green!30}\ding{119} &  \cellcolor{gray!35}\ding{109} & \cellcolor{gray!35}\ding{109} & \cellcolor{gray!35}\ding{109} &  \cellcolor{green!30}\ding{119} \\

   Wang et al. (2023) \cite{wang2023exploring}  & \cellcolor{blue!40}\ding{108}  &  \cellcolor{green!30}\ding{119}  & \cellcolor{blue!40}\ding{108}  & \cellcolor{blue!40}\ding{108} &  \cellcolor{green!30}\ding{119} & \cellcolor{green!30}\ding{119}  &  \cellcolor{green!30}\ding{119}  &  \cellcolor{green!30}\ding{119}  & \cellcolor{gray!35}\ding{109} &  \cellcolor{gray!35}\ding{109}  & \cellcolor{blue!40}\ding{108} & \cellcolor{blue!40}\ding{108}  \\

   Li et al. (2024) \cite{li2024blockchain}  &  \cellcolor{gray!35}\ding{109} & \cellcolor{gray!35}\ding{109} & \cellcolor{gray!35}\ding{109} &   \cellcolor{green!30}\ding{119} &  \cellcolor{green!30}\ding{119} &  \cellcolor{green!30}\ding{119} &  \cellcolor{green!30}\ding{119} & \cellcolor{blue!40}\ding{108} &  \cellcolor{green!30}\ding{119} &   \cellcolor{gray!35}\ding{109} &   \cellcolor{blue!40}\ding{108} &   \cellcolor{blue!40}\ding{108}\\

   \textbf{Our Survey}  & \cellcolor{blue!40}\ding{108} & \cellcolor{blue!40}\ding{108} & \cellcolor{blue!40}\ding{108} & \cellcolor{blue!40}\ding{108} & \cellcolor{blue!40}\ding{108} & \cellcolor{blue!40}\ding{108} & \cellcolor{blue!40}\ding{108} & \cellcolor{blue!40}\ding{108} & \cellcolor{blue!40}\ding{108}  & \cellcolor{blue!40}\ding{108} & \cellcolor{blue!40}\ding{108} & \cellcolor{blue!40}\ding{108} \\


   \toprule
  \end{tabular}
  
  \begin{tablenotes}
   \footnotesize
   \item[$\bigstar$] $Symbol$. “covered” (\ding{108} with \textcolor{blue!60}{blue background}); “partially covered” (\ding{119} with \textcolor{darkgreen}{green background}); “not covered” (\ding{109} with \textcolor{gray!80}{gray background}).
   \item[$\bigstar$] $Abbreviation$. SC: Survey Comparison; SD: Security Definition; GP: Generic Paradigm; AM: Abstraction with Modeling; TC: Technology Coverage; TT: Technology Types; CB: Comparison between technologies; IC: Industry Case Analysis; IS: Interoperability vs. Scalability Comparison; IF: Intersecting Field Research Avenues; IO: Open Issues; FC: Future Challenges.

\end{tablenotes} 
\end{threeparttable}    
\label{totalcomparison}
\end{table*}

\section{Background}

\subsection{Origins of Interoperability}

Interoperability originated as early as the 1980s in the field of computer science \cite{lavean1980interoperability,wegner1996interoperability}, primarily to address compatibility issues between systems, enabling different systems and applications to collaborate despite differences in hardware, operating systems, and programming languages. For example, classic standards like the TCP/IP protocol \cite{postel1981rfc0793} and the OSI model \cite{zimmermann1980osi} became the foundation for cross-system network communication. By the early 21st century, the widespread adoption of web services and cloud computing further advanced interoperability technologies. Web services provided standardized interfaces based on HTTP and XML \cite{coyle2002xml}, allowing applications to easily integrate over the network, while cloud computing, through virtualization and service-oriented technologies, enabled unified management and interaction between different applications in the internet environment \cite{gong2010characteristics}. The rise of Web 2.0 \cite{o2009web} further promoted API-based communication between systems, allowing different applications to interact and share data in an open environment, laying the theoretical groundwork for future explorations in blockchain interoperability \cite{park2023interoperability} and Web 3.0 \cite{zhu2024survey}. Unsurprisingly, the principles established in the architecture of the Internet have guided the development of interoperability protocols and standards \cite{lee2021survey}, with direct application to the blockchain domain. Given the trajectory of the Internet and computer networks throughout their historical development, the shift toward $\mathcal{CCI}$ is unsurprising. As a result, a global landscape of multi-chain blockchain networks \cite{kan2018multiple,liu2024distributed} connected by cross-chain solutions has gradually emerged \cite{wang2023exploring,ren2023interoperability}.

\subsection{Emergence \& Development of Interoperability}
 \begin{figure*}
\centering
\includegraphics[width=7.2in]{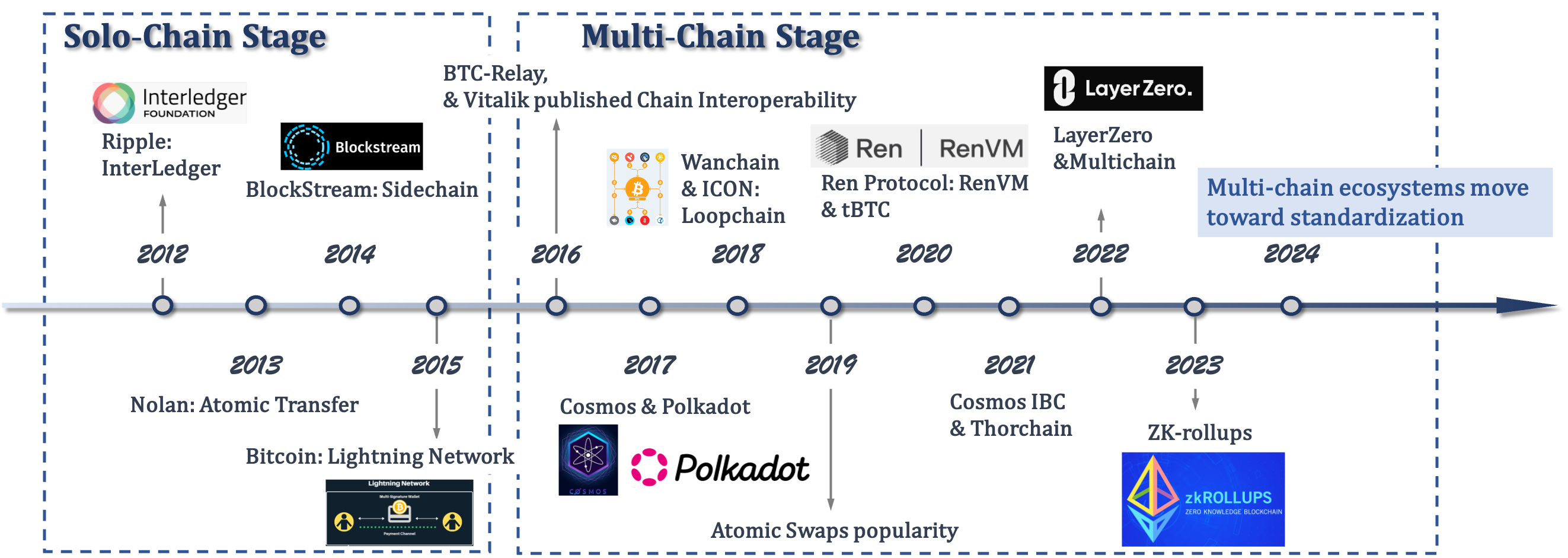} 
\caption{Timeline: the development of interoperability.}
\label{development fig}
\vspace{-2mm}
\end{figure*}

The exploration of interoperability technologies has been ongoing since the inception of Bitcoin \cite{nakamoto2008peer}. The development of $\mathcal{CCI}$ can be divided into two distinct phases: Solo-Chain Stage and Multi-Chain Stage. Fig. \ref{development fig} illustrates key events from the early stages of development to the present.

\subsubsection{\textbf{Solo-Chain Stage (\texttt{2012-2015})}}
As early as \textbf{\texttt{2012}}, Ripple Labs proposed the InterLedger Protocol (ILP) \cite{interledger,hope2016interledger}, which was formally implemented on the Ripple blockchain in 2015 \cite{armknecht2015ripple}. The protocol introduced a third-party entity, termed Connector, to manage custody and transaction verification between cross-chain participants.

In \textbf{\texttt{2013}}, Holan et al. \cite{nolan2013alt} proposed the concept of atomic transfers based on the Bitcoin network and Alt chains. This approach utilized hash-locking technology, where a script was triggered upon the revelation of a hash pre-image, enabling atomic cross-chain operations across Bitcoin and other blockchain networks.

In \textbf{\texttt{2014}}, BlockStream introduced the concept of pegged sidechains \cite{back2014enabling}, utilizing a two-way peg mechanism to transfer crypto assets between the sidechain and the main chain. This innovation enabled developers to create new blockchain systems on Bitcoin while maintaining interoperability with the Bitcoin network (By 2016, BlockStream further developed federated sidechains by introducing multi-signature technology, which reduced latency and enhanced interoperability between the sidechain and the main chain).

In \textbf{\texttt{2015}}, Poon et al. introduced the concept of off-chain transaction technology in their Lightning Network whitepaper \cite{poon2016bitcoin}, which facilitated the transfer of value off-chain via micropayment channels \cite{decker2015fast}. This innovation significantly improved transaction efficiency within the Bitcoin ecosystem by providing a mechanism for intra-chain atomic cross-chain operations.

\subsubsection{\textbf{Multi-Chain Stage (\texttt{2016-Present})}}

In \textbf{\texttt{2016}}, ConsenSys, an Ethereum blockchain software company, developed BTC Relay \cite{btcrelay}, allowing users to interact directly with the Bitcoin network via Ethereum, thus enabling cross-chain operations between ETH and BTC. BTC Relay's implementation leveraged BTC block header information and Ethereum smart contract functionality to securely verify Bitcoin transactions without the need for third-party intermediaries. In the same year, Vitalik Buterin \cite{buterin2016chain} provided an in-depth analysis of blockchain interoperability challenges.

In \textbf{\texttt{2017}}, the Cosmos project published its whitepaper \cite{kwon2019cosmos}, outlining a vision for $\mathcal{CCI}$ using a Hub-and-Zone architecture. Shortly after, Tendermint \cite{kwon2014tendermint} secured its first round of funding and began developing the Cosmos network. That same year, Polkadot introduced its whitepaper \cite{wood2016polkadot}, presenting its parachain and relay chain architecture, along with key concepts such as shared security and Cross-Chain Message Passing (XCMP).

In \textbf{\texttt{2018}}, Wanchain launched its mainnet, aiming to create a distributed "bank" that facilitates cross-chain asset and data transfers. By leveraging cross-chain smart contracts and privacy-preserving technologies, Wanchain achieved interoperability for various assets. ICON, also in 2018, released its Loopchain protocol \cite{qasse2019inter}, which connected independent blockchains across multiple industries, enhancing cross-industry interoperability, particularly in finance, healthcare, and government applications.

Atomic swap technology \cite{herlihy2018atomic} gained significant traction in \textbf{\texttt{2019}}, becoming widely adopted within the cryptocurrency community. This advancement allowed direct, trustless exchanges of cryptocurrencies across different blockchains, fostering the growth of decentralized exchanges (DEXs) and cross-chain trading.

In \textbf{\texttt{2020}}, Ren Protocol launched RenVM \cite{burgess2020bringing}, a decentralized virtual machine enabling cross-chain transfers of crypto assets through distributed key management. RenVM facilitated the transfer of non-Ethereum assets, such as Bitcoin, to the Ethereum network, unlocking new opportunities for \emph{decentralized finance} (DeFi) applications. Additionally, the tBTC project went live, allowing Bitcoin holders to mint ERC-20 tokens \cite{bauer2022erc} on Ethereum without relying on trusted intermediaries, marking a significant step in cross-chain asset management.

THORchain's mainnet launched in \textbf{\texttt{2021}}, introducing a decentralized liquidity network that enabled native cross-chain asset swaps (e.g., BTC, ETH) without the need for wrapped tokens or intermediary chains. Meanwhile, Cosmos launched its Inter-Blockchain Communication (IBC) protocol \cite{kim2022inter}, enabling seamless asset and data interoperability between blockchains and marking the maturation of the Cosmos ecosystem.

LayerZero Labs launched the LayerZero cross-chain communication protocol \cite{zarick2021layerzero} in \textbf{\texttt{2022}}, enabling smart contracts to transmit messages across different blockchains, while providing off-chain proof for cross-chain communication. LayerZero’s Omnichain protocol opened up new possibilities for cross-chain DeFi and NFT development. The same year, Anyswap rebranded as Multichain \cite{multichain}, expanding its cross-chain bridge capabilities to support a broader range of blockchain networks. Through its multi-chain architecture, Multichain became a critical infrastructure within the cross-chain DeFi ecosystem.

In \textbf{\texttt{2023}}, with advancements in ZK-Rollup technology \cite{lavaur2022enabling}, an increasing number of cross-chain bridges began adopting ZK-Rollups to enhance the security and efficiency of Layer-2 cross-chain transactions, driving further progress in blockchain interoperability and scalability.

By \textbf{\texttt{2024}}, the growing demand for cross-chain interoperability led to the standardization of cross-chain communication protocols. Protocols such as Polkadot’s XCMP and Cosmos’ IBC gained widespread adoption, promoting the further integration of multi-chain ecosystems.

\subsection{Relationship between Interoperability with Scalability}

Blockchain trilemma, also known as "\emph{Scalability Trilemma}" \cite{buterin2021sharding, hafid2020scaling, rebello2024survey}, is a concept that highlights the challenge of achieving a perfect balance among three critical characteristics: security, scalability, and decentralization. According to this trilemma, a blockchain can optimize only two of these properties simultaneously, often at the expense of the third. This is similar to the CAP theory \cite{CAP} of traditional distributed systems.
In practical terms, blockchain systems tend to focus on optimizing one or two of these aspects while making trade-offs with the third. For instance, Bitcoin \cite{nakamoto2008peer} prioritizes decentralization and security but has limitations in scalability, leading to slower transaction times and higher fees. Some permissioned blockchains \cite{polge2021permissioned} may prioritize security and scalability but at the cost of decentralization, relying on fewer, trusted nodes to process transactions; Layer-2 solutions \cite{rebello2024survey, sguanci2021layer} or alternative consensus mechanisms \cite{xu2023survey} aim to improve scalability while attempting to maintain security and decentralization, though achieving all three at high levels remains a significant challenge. Therefore, balancing and even achieving these three characteristics is particularly important for the future development of the blockchain to adapt to more complex and large-scale scenarios.

Exploring interoperability, as a pathway for computational offloading, is an effective approach that not only mitigates compromises in decentralization but also achieves a more balanced trade-off within the “blockchain trilemma”. Consequently, we propose that: \emph{“Interoperability is an essential prerequisite for enabling service scalability”}.

$\mathcal{CCI}$ facilitates the seamless flow of assets across ecosystems, alleviating lock-in effects and promoting greater economic equity among users. Additionally, interoperability eliminates data and value silos, enhancing the collaborative synergy within blockchain communities while reducing data redundancy costs. For example, a token holder on one blockchain may participate in decentralized autonomous organization (DAO) voting on another \cite{fan2020multav}. Interoperability further allows applications to deploy across multiple chains, enabling data sharing and asset transfer beyond the constraints of any single chain, thereby extending user reach and market potential. It also enhances system adaptability, allowing applications to align with evolving blockchain platforms and protocols to meet diverse user demands and support ecosystem-wide scalability. Moreover, interoperability can improve the security of specific blockchains by anchoring less secure chains to more secure ones through mechanisms like sidechains \cite{singh2020sidechain} or timestamping \cite{tas2023interchain}, thereby enabling the regular creation of security checkpoints.

\begin{figure}[!ht] 
\small
\centering
\includegraphics[width=3.5in]{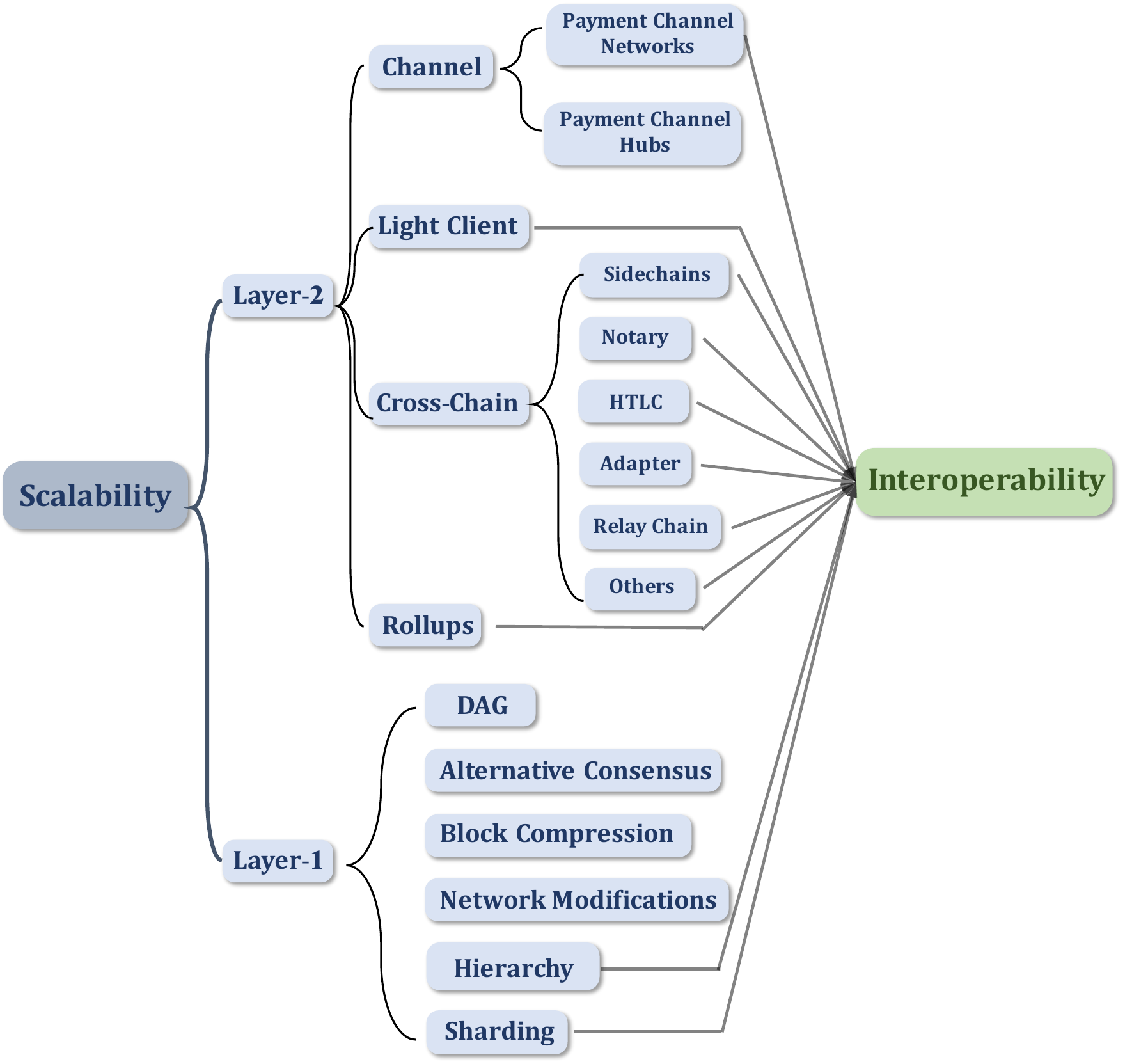} 
   \vspace{-2mm}
\caption{The technical correlation between interoperability and scalability.}
\label{scalawithIIC}
   \vspace{-2mm}
\end{figure}

As illustrated in Fig. \ref{scalawithIIC}, our analysis of scalability technologies across Layer-1 and Layer-2 \cite{sguanci2021layer,rebello2024survey} reveals that the technological foundations of interoperability are thoroughly embedded within scalability frameworks. 
A primary benefit of Layer-1 solutions is their foundational and holistic approach to improving blockchain performance. By addressing the root causes of performance issues through techniques such as sharding and hierarchy, Layer-1 solutions can lead to more sustainable improvements in interoperability and scalability. Conversely, Layer-2 solutions enhance blockchain functionality by adding supplementary layers or protocols atop existing blockchains, leaving the core blockchain protocols unchanged. This makes Layer-2 solutions generally easier to implement and adopt, offering greater flexibility in terms of interoperability and broader applicability.

On the whole, we contend that interoperability is indispensable for achieving scalability. This perspective is supported by a decade of extensive academic research \cite{rebello2024survey,belchior2021survey,wang2023exploring,ren2023interoperability} and industry backing \cite{connext,coindesk}, with many stakeholders viewing interoperability as a crucial enabler for large-scale adoption \cite{belchior2023brief}.

\section{Building Block and Preliminary}

In this section, we present the essential knowledge required to understand this survey. We begin by defining blockchain and interoperability. Following that, we present the definitions of security and the modes related to blockchain interoperability. Tab. \ref{SymbolDescription} describes the symbols commonly used in this paper.

\begin{table}[!h]
 \renewcommand{\arraystretch}{1.25}
 \caption{Symbol Description}
    \vspace{-2mm}
 \label{SymbolDescription}
 \centering
     \scriptsize
 \begin{tabular}{cc}
  \rowcolor{blue!6}
  \bottomrule
  \textbf{Symbol}                         & \textbf{Description}     \\   
  \midrule
         $\mathscr{S}$  &   Source Chain\\
        $\mathscr{T}$   &   Target Chain \\
        $\mathcal{L}_{\mathscr{S}}$ & the Ledger of $\mathscr{S}$ \\
        $\mathcal{L}_{\mathscr{T}}$ & the Ledger of $\mathscr{T}$ \\
        $\TxCC$ & the Cross-Chain Transaction \\
         $\mathcal{CCI}$ & Cross-Chain Interoperability \\
         TTP & Trusted Third Party \\

  \toprule
 \end{tabular}
    \vspace{-2mm}
\end{table}

\subsection{A Primer on Blockchain}

\textbf{Blockchain Basic.} Blockchain is a public ledger technology powering cryptocurrencies such as Bitcoin \cite{nakamoto2008peer} and Ethereum \cite{buterin2013ethereum}. It operates as a decentralized data structure, enabling dApps \cite{raval2016decentralized}, smart contracts \cite{liangwei}, and the recording of all network transactions. In this system, each node independently maintains a ledger copy, verifies peers, and can initiate, validate, and confirm transactions without a \emph{Trusted Third Party} (TTP). This decentralized framework strengthens security and resilience, minimizing single points of failure and the risk of data tampering.

\begin{nameddefinition}[\textbf{Blockchain Structures}] \label{blockchain_def}
\emph{A blockchain includes the following data structures:}
\begin{itemize}[itemsep=0.5pt]
\item \emph{\textbf{Transaction.} Which is the process of transferring cryptocurrency. It must specify the sender, recipient, transaction amount, and the sender’s signature;}
\item \emph{\textbf{Block.} Which consists of two components: the header and the body. The former contains a set of transactions, while the latter records the hash of the previous block, version, nonce, the root of the Merkle tree, etc;}
\item \emph{\textbf{Chain.} Which is a sequence of blocks, where each block contains the previous block's hash, serving as a pointer to link them together, forming a chain. New blocks are always appended to the chain.}
\end{itemize}
\end{nameddefinition} 
For simplicity, we use "\emph{blockchain}" and "\emph{chain}" interchangeably throughout this survey.

 \begin{figure}[!ht] 
\centering
\includegraphics[width=3in]{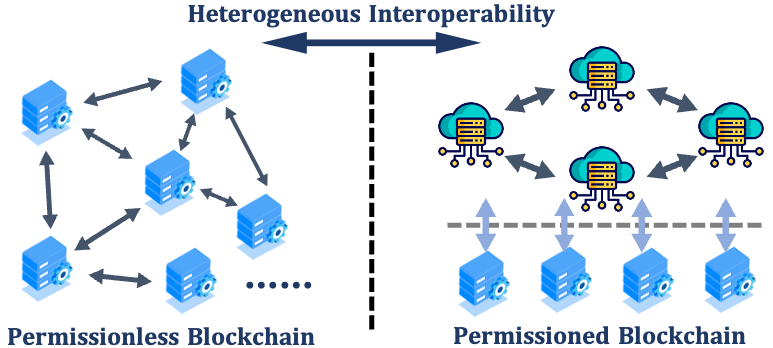} 
   \vspace{-2mm}
\caption{Network architectures for different types of blockchains.}
\label{blockchainnetwork}
   \vspace{-2mm}
\end{figure} 
\textbf{Blockchain Types.} Blockchain can be categorized into two types based on network architecture and authorization requirements (see Fig. \ref{blockchainnetwork}). When considering interactions between two chains of the same type, we define this as homogeneous interoperability, whereas interactions between different types of blockchains are referred to as heterogeneous interoperability. Typically, the latter is backwards compatible with the former.
\begin{nameddefinition}[\textbf{Permissionless Blockchain}] \label{permissionless}
\emph{In a permissionless blockchain system:}
\begin{itemize}[itemsep=0.5pt]
\item \emph{No identity authentication is required;}
\item \emph{Any node can join, send transactions, participate in consensus, or leave at any time;}
\item \emph{At any given moment, the number of participating nodes is subject to variation and cannot be reliably predicted.}
\end{itemize}
\end{nameddefinition}


\begin{nameddefinition}[\textbf{Permissioned Blockchain}] \label{permissioned}
\emph{A permissioned blockchain system is jointly managed and maintained by multiple organizations or institutions, where only authenticated nodes are permitted to join the network, read data, and execute transactions.}
\end{nameddefinition} 
Different from permissionless blockchain, to maintain consistency among replicated data on different nodes, the permissioned blockchain should employ a State Machine Replication (SMR) algorithm \cite{SMR}, ensuring nodes agree on the order of incoming transactions to maintain identical copies of the distributed ledger.

\textbf{Distributed Ledger Model.} In the following context, the terms $blockchain$ and $distributed$ $ledger$ are used interchangeably. Regarding interoperability, we consider the interaction between the source chain $\mathscr{S}$ and the target chain $\mathscr{T}$, which may involve distinct consensus participants and different consensus protocols. Let $\mathcal{L}$ represent a ledger, with $\mathcal{L}_{\mathscr{S}}$ and $\mathcal{L}_{\mathscr{T}}$ corresponding to the ledgers of $\mathscr{S}$ and $\mathscr{T}$, respectively. The state of a ledger is defined as a dynamically evolving sequence of transactions, denoted by $\langle \mathtt{Tx_{1}}, \ldots, \mathtt{Tx_{n}} \rangle$.
We assume the ledger state evolves in discrete rounds, indexed by natural numbers $r \in \mathbb{N}$. Thus, $\mathcal{L}^{P}[r]$ represents the state of ledger $\mathcal{L}$ at round $r$, which is defined as the state after applying all transactions recorded in $\mathcal{L}$ since round $r-1$, according to the perspective of some party $P$. Hence, $\Tx \in \mathcal{L}^{P}[r]$ can be denoted as a transaction $\Tx$ has been included in $\mathcal{L}$ as position $r$.

To maintain cross-chain protocol security, as a premise, either a singular $\mathscr{S}$ or $\mathscr{T}$ must exhibit the following properties \cite{backbone2015}:

\begin{nameddefinition}[\textbf{Robust Distributed Ledger}] \label{RDL}
\emph{We say that a robust distributed ledger must meet the following properties:}
\begin{itemize}[itemsep=0.5pt]
\item \emph{\textbf{Persistence.} For any two honest parties $P_{1}$ and $P_{2}$, if they adopt respective ledgers $\mathcal{L}^{P_{1}}[r_{1}]$ and $\mathcal{L}^{P_{2}}[r_{2}]$ at round $r_{1}$ and $r_{2}$ respectively, where $r_{1} \leq r_{2}$. It holds that $\mathcal{L}^{P_{1}}[r_{1}]$ $\preceq$ $\mathcal{L}^{P_{2}}[r_{2}]$.}
\item \emph{\textbf{Liveness.} After the environment submits a valid $\Tx$, any honest parties $P$ will report $\Tx \in \mathcal{L}^{P}[r']$ at round $r'$ after $t$ round.}

\end{itemize}
\end{nameddefinition} 

$Persistence$ guarantees that confirmed the cross-chain transaction $\TxCC$ is irreversible, and $liveness$ ensures the eventual inclusion of all valid $\TxCC$. These properties are ensured when the in-chain consensus adheres to the specified requirements \cite{backbone2015}: \emph{Common Prefix, Chain Quality, and Chain Growth}.

\textbf{Transaction Model.} When a transaction $\Tx$ is included in a ledger $\mathcal{L}$, it modifies the ledger’s state by specifying a set of operations that must be executed and agreed upon by consensus participants $P_{1},...,P_{n}$. The nature of these operations is system-dependent and can vary from simple transfers to the execution of complex programs \cite{wood2014secure}. For the sake of generality, we do not distinguish between different transaction models, i.e. $UTXO$ \cite{nakamoto2008peer} and the $account$-$based$ model \cite{wood2014secure}.

\subsection{Blockchain Interoperability Definition}

  \begin{table*}[!htbp]
 \caption{Comparative Definitions of Interoperability in Different Literature}
 \centering
 \scriptsize
 \setlength{\tabcolsep}{11.2pt}
 \renewcommand{\arraystretch}{1.5}
 \label{table1}
 \begin{threeparttable}    
  \begin{tabular}{p{0.35cm}p{2.05cm}p{12.5cm}}
   \rowcolor{blue!6}
   \bottomrule
   
\rotatebox{0}{\textsl{\textbf{Year}}}  & \rotatebox{0}{\textsl{\textbf{Proposer}}}&   \multicolumn{1}{c}{\rotatebox{0}{\textsl{\textbf{Core Defination}}}}   \\
   \hline

1996 &  Wegner et al. \cite{wegner1996interoperability} &  It refers to the capability of multiple software components to work together, even when there are variations in programming language, interface, and execution environment. \\ \hline

  2006 & Vernadat et al. \cite{vernadat2006interoperable} &  It denotes the capability of two or more systems to either offer services to one another or receive services, while efficiently leveraging a shared exchange for mutual benefit.\\ \hline
  
  2016  & Buterin et al. \cite{buchman2016tendermint}  & It involves three main operations: \ding{172} transferring assets between platforms; \ding{173} implementing payment-versus-payment and payment-versus-delivery models; \ding{174} retrieving information from one blockchain within another.\\  \hline

  2019 &  Pillai et al. \cite{pillai2019blockchain}  & It is designed not to directly alter the state of other blockchains. Instead, its purpose is to initiate specific functionalities on the other system, which are expected to carry out operations within their own network.  \\ \hline
  
 2019  &   Yaga et al. \cite{yaga2019blockchain}    & Its atomic transaction execution extends across multiple blockchains, enabling data recorded on one chain to be accessible, verifiable, and referenced by potentially external transactions in a semantically consistent manner. \\  \hline

2021 & Belchior et al. \cite{belchior2021survey}     &  The capability of $\mathscr{S}$ to modify the state of $\mathscr{T}$, facilitated by inter- or intra-cross-chain transactions, spanning both homogeneous and heterogeneous blockchain systems.  \\  \hline

2023 &  Wang et al. \cite{wang2023exploring}    & It refers to the capability to accurately execute asset transfers across a mix of homogeneous and heterogeneous blockchain systems while preserving the foundational design principles of each system. \\ \hline

2023 &  Ren et al. \cite{ren2023interoperability} &  It refers to the flexibility to transfer assets, share data, and execute smart contracts across public, private, and permissioned blockchains without altering their underlying systems. \\

   \toprule
  \end{tabular}
  
   
 \end{threeparttable}    
 \label{table_compare_inter}
\end{table*}

In the early development of computer science, numerous descriptions of interoperability were introduced \cite{wegner1996interoperability,geraci1991ieee,vernadat2006interoperable}. Interoperability is defined by the Institute of Electrical and Electronics Engineers (IEEE) as follows: \emph{"The ability of two or more systems or components to exchange information and to use the information that has been exchanged"} \cite{geraci1991ieee}. Vernadat et al. \cite{vernadat2006interoperable} expanded upon this definition from a systems perspective, and these conceptual frameworks have been effectively integrated into the discourse on blockchain interoperability. Although the technologies facilitating blockchain interoperability remain nascent and lack standardization, several representative definitions have emerged \cite{buchman2016tendermint,pillai2019blockchain,yaga2019blockchain,belchior2021survey,wang2023exploring,ren2023interoperability}. As illustrated in Tab. \ref{table_compare_inter}, these definitions underscore various dimensions of blockchain interoperability. We assert that blockchain interoperability, also referred to as cross-chain interoperability, \emph{which refers to the ability of blockchain networks to facilitate mutual interaction with one another by exchanging assets, data, or both, primarily manifests through Data Interoperability, Functional Interoperability, and Value Interoperability}. Data interoperability emphasizes cross-chain data access and transmission among disparate blockchains. Value interoperability pertains to asset exchanges and transfers across different blockchains. Functional interoperability seeks to integrate functionalities between various blockchains, such as enabling cross-chain calls to smart contracts. However, it is essential to recognize that each blockchain system operates as an isolated entity, and interoperability is an ancillary feature of the system. Consequently, when introducing new functionalities to a blockchain, it is imperative not to undermine its foundational role as a decentralized ledger system.

\subsection{A Generic Interoperability Paradigm}

Let us define a more generic interoperability paradigm that can signify the transfer of goods, assets, or objects between $\mathscr{S}$ and $\mathscr{T}$. We assume that an operation $\mathcal{O}_{\mathscr{S}}$ runs on $\mathscr{S}$, and an operation $\mathcal{O}_{\mathscr{T}}$ runs on $\mathscr{T}$. Operations can influence the blockchain state in two distinctive ways: \ding{172} by writing transactions to the blockchain; \ding{173} by halting interaction with the blockchain. These assumptions align with the CCC protocol model proposed in \cite{zamyatin2021sok}.

\begin{nameddefinition}[\textbf{A Generic Interoperability Paradigm}] \label{generic_inter}
\emph{The generic paradigm is constructed by the following phases:}
\begin{itemize}[itemsep=0.5pt]
\item \emph{\textbf{a) Setup.} The primary task during the setup phase is to establish the relevant information for both the parameterized $\mathscr{S}$ and $\mathscr{T}$, and to define the application-level specifications for interoperability to facilitate the initialization of cross-chain communication. For instance, in the case of digital asset exchanges, this involves specifying the asset types to be exchanged (e.g., Tokens or NFTs), the valuation standards (e.g., based on ERC-20 \cite{rahimian2021tokenhook} or ERC-721 \cite{cabot2022improving}), time constraints, and any additional conditions;}
\item \emph{\textbf{b) Commit on Source Chain $\mathscr{S}$.} Upon successful setup, a publicly verifiable commitment to execute a $\TxCC$ is submitted on $\mathscr{S}$. Specifically, $\mathcal{O}_{\mathscr{S}}$ writes the transaction to $\mathcal{L}_{\mathscr{S}}$. Based on the persistence and liveness of $\mathcal{L}_{\mathscr{S}}$ (as described in Def. \ref{RDL}), all honest participants in $\mathscr{S}$ will determine that the transaction has ultimately reached a stable state;}
\item \emph{\textbf{c) Verify.} The commitment made by $\mathcal{O}_{\mathscr{S}}$ on $\mathscr{S}$ is verified by $\mathcal{O}_{\mathscr{T}}$ (or $\mathcal{O}_{\mathscr{T}}$ receives the proof from $\mathcal{O}_{\mathscr{S}}$). Based on the persistence and liveness of $\mathcal{L}_{\mathscr{S}}$, the verification will succeed once the transaction stabilizes on $\mathcal{L}_{\mathscr{S}}$;}
\item \emph{\textbf{d.1) Commit on Target Chain $\mathscr{T}$.} Following successful verification, a commitment that is publicly verifiable to execute a $\TxCC$ is submitted on $\mathscr{T}$. $\mathcal{O}_{\mathscr{T}}$ writes the transaction to $\mathcal{L}_{\mathscr{T}}$. Based on the persistence and liveness of $\mathcal{L}_{\mathscr{T}}$, all honest participants in $\mathscr{T}$ will ascertain that the transaction has ultimately reached a stable state;}
\item \emph{\textbf{d.2) Abort.} If the verification fails, or if $\mathcal{O}_{\mathscr{T}}$ is unable to fulfill the commitment execution on $\mathscr{T}$, the protocol will execute an abort operation on $\mathscr{S}$ to revert the modifications to their original state, ensuring atomicity.}
\end{itemize}
\end{nameddefinition}

It is noteworthy that some asset exchange interoperability protocols follow a Two-Phase Commit (2PC) \cite{rahimian2021tokenhook} design, allowing phase $b)$ and $d.1)$ to be executed concurrently. Phase $d.2)$ is not peremptory, indicating that once $commit$ has been executed, abort is no longer an option \cite{zamyatin2021sok}.

\subsection{Security Definition of Blockchain Interoperability}

\begin{figure}[ht]
	\hfill
 \subfigure[]{
		\includegraphics[width=0.19\textwidth]{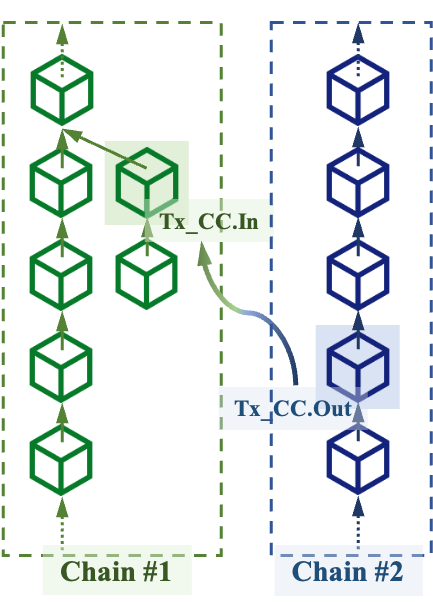}
		\label{crosschain_failure1}
	}
	\hfill
	\subfigure[]{
		\includegraphics[width=0.19\textwidth]{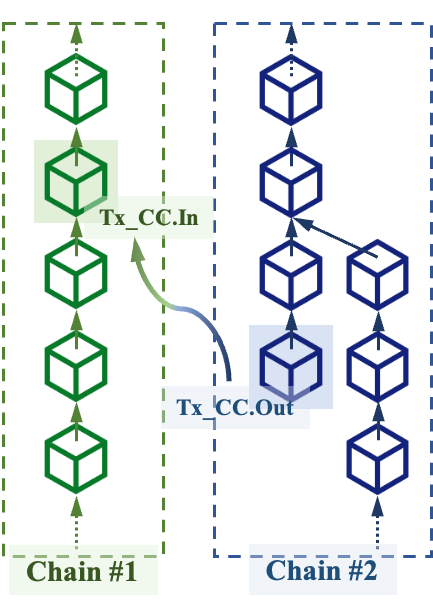}	 
    \label{crosschain_failure2}
    }
\vspace{-1mm}	
 \caption{Cases studies on cross-chain atomicity transfer failure. (a) $\mathtt{Tx\_CC.In}$ is not on the longest chain of Chain \#1; (b) $\mathtt{Tx\_CC.Out}$ is not on the longest chain of Chain \#2.}
	\label{crosschain_failure}
\vspace{-2mm}
\end{figure}

First of all, let's describe an example of an interoperability atomicity transmission failure. Let $\mathtt{Tx\_CC}$ denote a cross-chain transaction, where $\mathtt{Tx\_CC.In}$ represents the input impacting the ledger state of the transaction originator, and $\mathtt{Tx\_CC.Out}$ represents the output affecting the ledger state of the transaction recipient. The atomicity failure of $\mathtt{Tx\_CC}$ can be categorized into two scenarios, as illustrated in Fig. \ref{crosschain_failure}. In the first scenario (Fig. \ref{crosschain_failure1}), due to intentional or unintentional forks on Chain \#1, $\mathtt{Tx\_CC.In}$ fails to be executed, while $\mathtt{Tx\_CC.Out}$ successfully takes effect on the longest chain of Chain \#2. This situation introduces a potential double-spending risk for $\mathtt{Tx\_CC}$, necessitating measures to prevent such occurrences.
In the second scenario (Fig. \ref{crosschain_failure2}), although $\mathtt{Tx\_CC.Out}$ is forked and fails to take effect, this does not pose a double-spending risk. Instead, $\mathtt{Tx\_CC.Out}$ can be rewritten on the longest chain of Chain \#2, thereby mitigating the atomicity failure of $\mathtt{Tx\_CC}$ in this case.
Therefore, the challenge of blockchain interoperability stems from the need for atomic synchronization of transactions across two or multiple chains, e.g., in an atomic swap, a transaction $\mathtt{Tx\_CC.In}$ on Chain \#1 succeeds if and only if $\mathtt{Tx\_CC.Out}$ was previously posted on Chain \#2.

A necessary guarantee for a secure $\mathcal{CCI}$ protocol is \emph{\textbf{atomicity}}. Referring to \cite{zamyatin2021sok,scaffino2023glimpse}, we articulate in a weak and a strong variant. For $\mathscr{S}$ and $\mathscr{T}$, each with respective underlying ledgers $\mathcal{L}_{\mathscr{S}}$ and $\mathcal{L}_{\mathscr{T}}$, the goal of $\mathcal{CCI}$ can be described as the synchronization of processes $\#\mathtt{In}$ and $\#\mathtt{Out}$ such that $\#\mathtt{Out}$ writes $\TxCC.
{\mathtt{Out}}$ to $\mathcal{L}_{\mathscr{T}}$ if and only if $\#\mathtt{In}$ has written $\TxCC.{\mathtt{In}}$ to $\mathcal{L}_{\mathscr{S}}$. 
From $persistence$ and $liveness$ of $\mathcal{L}$ (Def. \ref{RDL}), it follows that eventually $\#\mathtt{In}$ writes $\TxCC.{\mathtt{In}}$ in $\mathcal{L}_{\mathscr{S}}$ and $\#\mathtt{Out}$ becomes aware of and verifies $\TxCC.{\mathtt{Out}}$ in $\mathcal{L}_{\mathscr{T}}$.
Hence, a secure $\mathcal{CCI}$ protocol must exhibit the following properties:

\begin{nameddefinition}[\textbf{The Security of $\mathcal{CCI}$}] \label{security cross-chain}
\emph{For both blockchain $\mathscr{S}$ and $\mathscr{T}$ with ledgers $\mathcal{L}_{\mathscr{S}}$ and $\mathcal{L}_{\mathscr{T}}$, each of which satisfies persistence and liveness required for a robust distributed ledger in Def. \ref{RDL}. Consider two processes, $\#\mathtt{In}$ on $\mathscr{S}$ and $\#\mathtt{Out}$ on $\mathscr{T}$, with to-be-synchronized transactions $\TxCC.{\mathtt{In}}$ and $\TxCC.{\mathtt{Out}}$.
 A $\mathcal{CCI}$ protocol is secure if the following properties can be satisfied:}
\begin{itemize}[itemsep=0.5pt]
\item \emph{\textbf{Weak Atomicity.} A valid $\TxCC.{\mathtt{In}}$ is reported stable on $\mathcal{L}_{\mathscr{T}}$ only if $\TxCC.{\mathtt{Out}}$ has be reported stable on $\mathcal{L}_{\mathscr{S}}$, i.e.: $\TxCC.{\mathtt{In}} \in \mathcal{L}_{\mathscr{T}} \Longrightarrow \TxCC.{\mathtt{Out}} \in \mathcal{L}_{\mathscr{S}}$.}
\item \emph{\textbf{Strong Atomicity.} There are no outcomes in which $\TxCC.{\mathtt{In}}$ is reported stable on $\mathcal{L}_{\mathscr{T}}$ but $\TxCC.{\mathtt{Out}}$ is not stable on $\mathcal{L}_{\mathscr{S}}$, or $\TxCC.{\mathtt{Out}}$ is reported stable on $\mathcal{L}_{\mathscr{S}}$ but $\TxCC.{\mathtt{In}}$ is not stable on $\mathcal{L}_{\mathscr{T}}$, i.e.: }
{
\begin{equation*}\label{strong ato}
\begin{aligned}
& \neg((\TxCC.{\mathtt{In}} \in \mathcal{L}_{\mathscr{T}} \wedge \TxCC.{\mathtt{Out}} \notin \mathcal{L}_{\mathscr{S}}) \vee \\
& (\TxCC.{\mathtt{Out}}\in \mathcal{L}_{\mathscr{S}} \wedge \TxCC.{\mathtt{In}} \notin \mathcal{L}_{\mathscr{T}})).
\end{aligned}
\end{equation*}
}
\end{itemize}

\end{nameddefinition}

The former ensures that $\TxCC.{\mathtt{Out}}$ appears on $\mathcal{L}_{\mathscr{T}}$ \underline{only if} $\TxCC.{\mathtt{In}}$ has been already written into $\mathcal{L}_{\mathscr{S}}$. The latter ensures that $\TxCC.{\mathtt{Out}}$ appears on $\mathcal{L}_{\mathscr{T}}$ \underline{if and only if} $\TxCC.{\mathtt{In}}$ has been already written into $\mathcal{L}_{\mathscr{S}}$.

\subsection{Interoperability Modes}
We propose three interoperability modes based on the existing works of literature \cite{wang2023exploring, belchior2023you,augusto2024sok}. The choice of interoperability mode determines the required protocol architecture, with each configuration delivering its own specific security guarantees.

\textbf{Asset Swap.} Asset swap refers to the exchange of different assets between two separate blockchains through an agreed-upon protocol. This typically occurs when users want to exchange one asset for another, such as swapping Bitcoin for tokens on Ethereum. Without migrating the assets to another blockchain, this exchange happens via decentralized cross-chain protocols (e.g., atomic swaps \cite{herlihy2018atomic} with HTLC \cite{koens2019assessing} and adaptor signatures \cite{thyagarajan2022universal}). In this process, each party retains its assets on its respective blockchains, but they achieve an equal-value swap.

\textbf{Asset Migration.} Asset migration refers to moving the asset from one blockchain to another, which encompasses locking or burning the asset in $\mathcal{L}_{\mathscr{S}}$ and creating or minting a representation of that asset in $\mathcal{L}_{\mathscr{T}}$. Once the asset is locked in $\mathcal{L}_{\mathscr{S}}$, the verification process is carried out in $\mathcal{L}_{\mathscr{T}}$. This verification can be achieved by replicating the consensus mechanism of $\mathscr{S}$ on $\mathscr{T}$ \cite{frauenthaler2020eth,ciobotaru2022accountable} or by employing proof-based mechanisms such as zero-knowledge proofs \cite{xie2022zkbridge, westerkamp2020zkrelay, garoffolo2020zendoo}.

\textbf{Data Transfer.} Data transfer focuses on the transfer of information, such as transaction histories or the state of smart contracts, and extends the concept of interoperability. Information written in one chain can be transferred or replicated to another chain, typically accompanied by proofs, such as the payload of a blockchain view \cite{belchior2022can}. Blockchain gateways are frequently employed to support this process, functioning via gateway-to-gateway protocols. \cite{pedreira2023trustable}. Examples include coordinating and managing decentralized autonomous organizations (DAOs) governance and actions across chains.

\subsection{Trust Model of Interoperability}

Zamyatin et al. \cite{zamyatin2021sok} have demonstrated that in an asynchronous setting, $\mathcal{CCI}$ is fundamentally impossible without a TTP. Therefore, the trust model is a crucial element that must be addressed when discussing interoperability solutions, and it is typically categorized into the following three types.

\textbf{TTP.} The simplest method of cross-chain verification relies on a TTP to verify state changes across chains during interoperability execution. TTP-based solutions are typically realized through external validators or consensus committees. External validators outsource the cross-chain verification process to a trusted custodian that is independent of both $\mathscr{S}$ and $\mathscr{T}$, bypassing the need for on-chain validation. These external validators may be static or dynamic and are often incentivized to act honestly by staking assets on relevant blockchains. Alternatively, consensus committees, composed of members from either $\mathscr{S}$ or $\mathscr{T}$, can handle verification. The committee members reach a consensus on the ledger’s state through mechanisms such as $BFT$  \cite{zhang2023bft} or \emph{the longest chain rule} \cite{pass2017fruitchains}. Misbehaviour by committee members can be viewed as a failure of the chain itself. In practice, external validators can be implemented via multi-signature contracts, requiring a set of signatures from the validators. The vote of the consensus committee is implemented through smart contracts, ensuring that committee members agree upon the execution outcome.

\textbf{Synchrony.} This model does not rely on TTP but assumes synchronized communication between participants and derives security from cryptographic hardness assumptions by using locking mechanisms. Such protocols are often referred to as non-custodial protocols, as they avoid transferring asset custody to a TTP. In the worst-case scenario, a failure would result in permanently locking funds rather than providing any financial gain to a third party. In practice, this model is realized through technologies such as HTLC, adaptors, time-lock puzzles \cite{agrawalr2024time}, and verifiable delay functions (VDFs) \cite{boneh2018verifiable}, often in combination with smart contracts.

\textbf{Hybrid.} In cases where a party crashes or the synchrony assumption fails. i.e., when a predefined timeout is exceeded, the watchtower is employed to enforce commitments \cite{avarikioti2020cerberus}. This structure was first introduced and applied to off-chain payment channels \cite{gudgeon2019sok}, which can help channel users monitor the blockchain online in real-time and perform specific actions on behalf of users when needed. It is particularly useful in atomic swaps utilizing HTLC, where one party crashes after the secret in the hash lock has been revealed. Additionally, we refer to the model that incorporates both TTP and synchrony as the Hybrid model.

\subsection{Interoperability Layers}

\begin{figure}[!ht] 
\small
\centering
\includegraphics[width=3.5in]{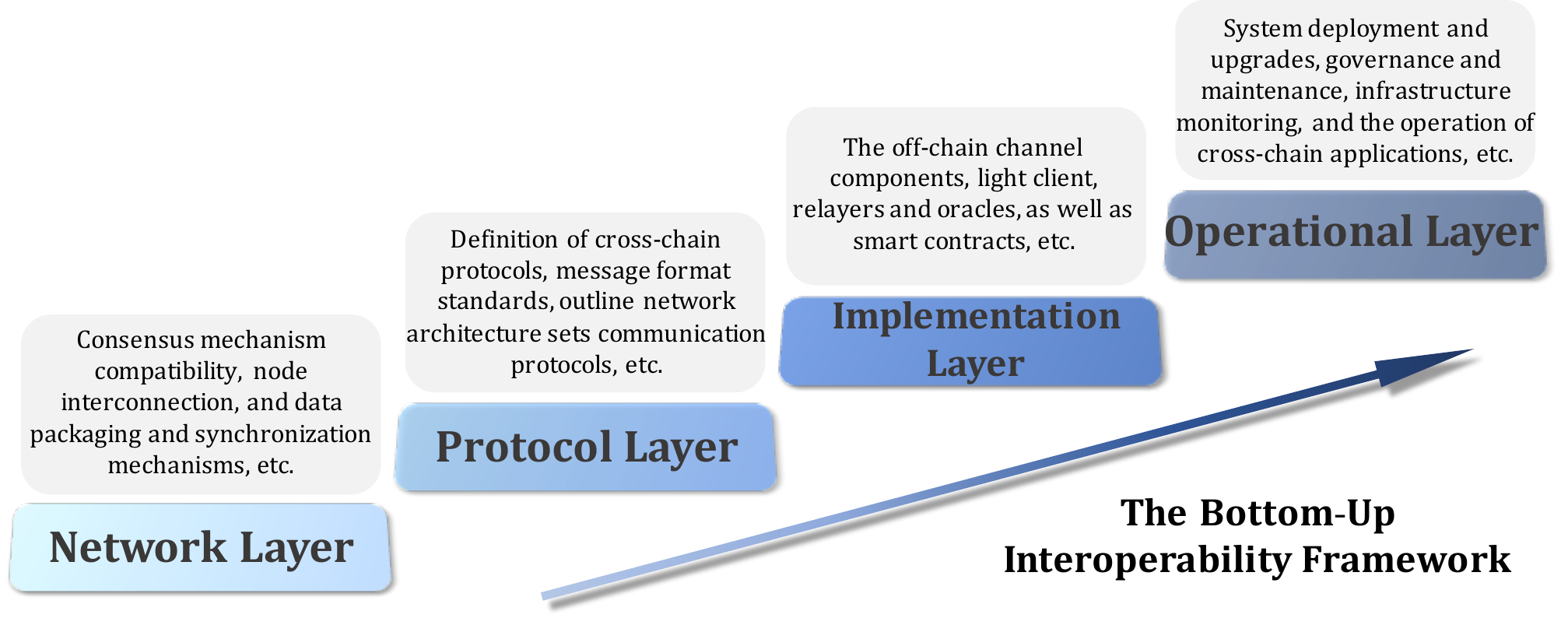} 
   \vspace{-2mm}
\caption{Interoperability layers.}
\label{CCIlayer fig}
   \vspace{-2mm}
\end{figure}

From a security perspective, interoperability solutions can be categorized into multiple layers, as shown in Fig. \ref{CCIlayer fig}. This layered classification is supported by existing literature \cite{augusto2024sok, crosschainRiskFramework}.

\emph{Network Layer} serves as the foundation, focusing on the underlying logic of interoperability solutions, such as the validity and compatibility of consensus rules, methods of node interconnection, local data packaging, and synchronization mechanisms. This layer is critical in distinguishing between homogeneous and heterogeneous solutions.
\emph{Protocol Layer} addresses the architectural decisions required for constructing interoperability protocols. This includes defining various types of participants, their roles and responsibilities, as well as ensuring security, performance, standardizing message formats, etc.
\emph{Implementation Layer} involves the development of complex on-chain and off-chain components, while accounting for diverse programming languages, smart contract standards, oracles, etc.
Finally, \emph{Operational Layer} covers system deployment, updates, maintenance, regulatory oversight, governance, the operation of external validators, and the management of dApps in cross-chain contexts, etc.
Most solutions span at least one or two layers. The following sections concentrate on the key layers targeted by each solution.

\section{Existing Solutions} \label{totalsolution}

\begin{figure*}[!ht] 
\small
\centering
\includegraphics[width=4.6in]{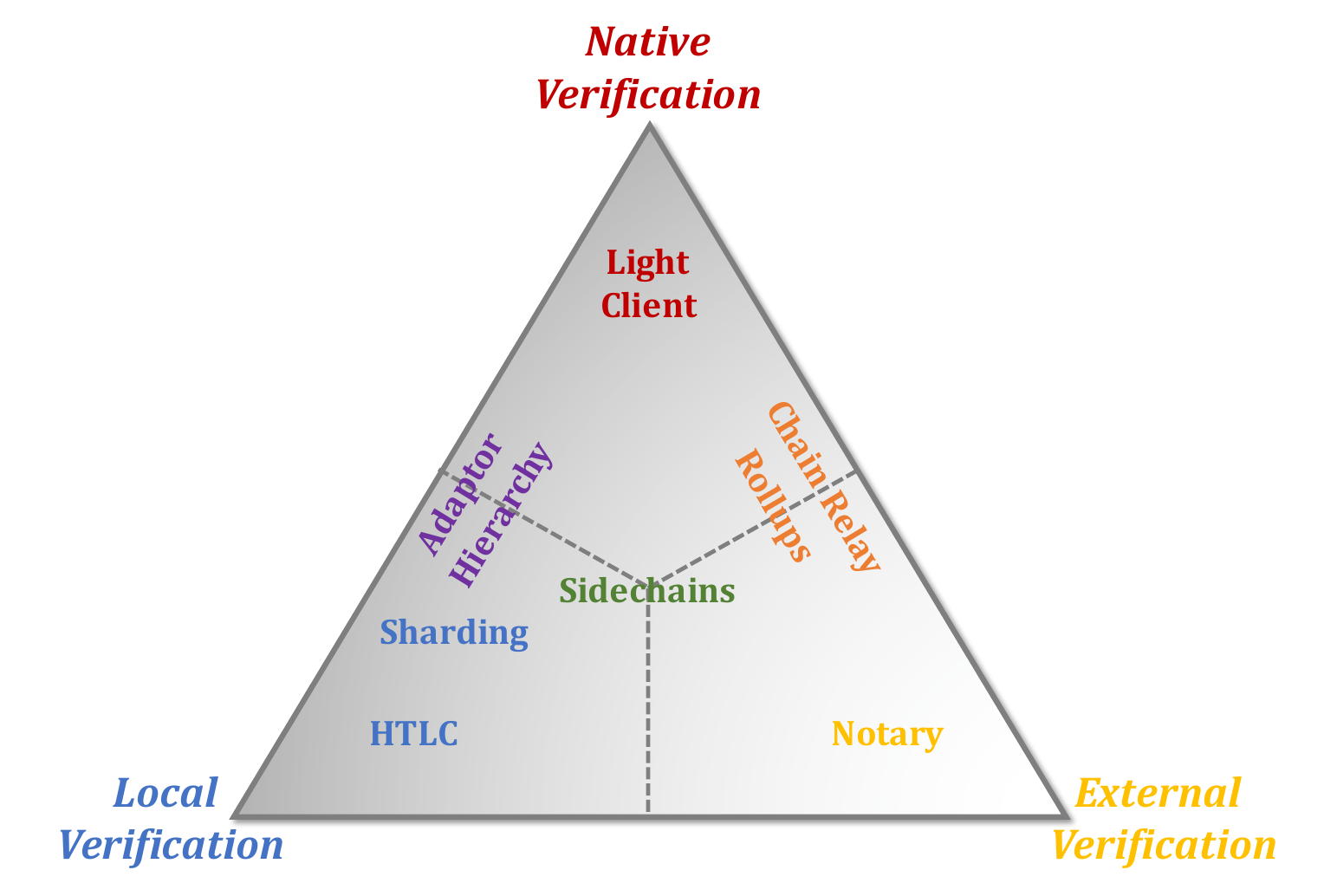} 
   \vspace{-2mm}
\caption{Technology classification triangle (The darker the \textcolor{gray!80}{gray color}, the stronger the trustlessness).}
\label{verificationfig}
   \vspace{-2mm}
\end{figure*}

This section presents concrete solutions. Categorizing blockchain interoperability has been a persistent challenge. In reality, each classification may overlap, indicating there is no universally fixed categorization \cite{wang2023exploring}. Our approach focuses on interoperability verification, dividing it into native verification, local verification, and external verification. The specific technologies are outlined in Fig. \ref{verificationfig}.

\begin{itemize}[itemsep=0.5pt]

\item  \emph{Native verification} refers to cross-chain transactions or states verified directly through the blockchain's consensus mechanism and rules, without reliance on third parties. Here, the verification is entirely performed on-chain by the nodes of either $\mathscr{S}$ or $\mathscr{T}$ , with security ensured by the blockchain's inherent model. For example, one blockchain might synchronize the state or block headers of another using a light client and verify based on the consensus of the counterpart chain. This method, relying on internal rules and consensus algorithms, is considered decentralized and trust-minimized.
\item \emph{External verification} involves cross-chain transaction or state verification by an external third party (often witnesses or validators) that does not directly belong to $\mathscr{S}$ or $\mathscr{T}$. This process relies on independent intermediaries or validator networks (e.g., MPC network, TEE network, Multi-signature group, or Oracles \cite{caldarelli2022overview}), which can be centralized or decentralized, ensuring transaction integrity.
\item \emph{Local verification} pertains to operations or transactions on a specific chain, verified directly by the chain’s local nodes, without requiring external chain data. For example, in state channels, participants verify each other's transactions during execution and settlement. This method applies to intra-chain transactions and is generally used to secure smart contracts, state transitions, or transaction execution on-chain. Given the opposing economic interests of the transacting parties, the potential for collusion is effectively eliminated.

\end{itemize}

Consider three trustless verification mechanisms with respective security metrics: source chain security, $M_{1}$; target chain security, $M_{2}$; and external verifier security, $M_{3}$, where external verification introduces an additional security assumption. Thus, the security metric for $\mathcal{CCI}$ is approximated as $M=\textsf{Min}(M_{1},M_{2},M_{3})$ under external verification, $M=M_{1} \oplus M_{2}$ under local verification (assuming fully opposing transacting parties), and $M=\textsf{Max}(M_{1}, M_{2})$ under native verification. Generally, $M_{3}$ represents the weakest link, often criticized in external verification despite its higher efficiency.

Chain relay integrates both native and external verification mechanisms. Typically, chain relay achieves native verification through its consensus, while occasionally utilizing external validators, such as intermediary networks, specific nodes, or Oracles to assist in verifying cross-chain transactions. 
In contrast, rollups process and compress large volumes of transactions on Layer-2, subsequently submitting the aggregated results to the Layer-1 main chain to ensure data integrity and state consistency. Rollups mandate that Layer-2 inherit the security properties of Layer-1, such that only the state finalized on Layer-1 is accepted as authoritative. As a result, rollups effectively combine native and external verification mechanisms.
Sidechains, on the other hand, can be designed with centralized, consortium-based, or SPV-based anchoring methods, positioning them centrally within Fig. \ref{verificationfig}. The following sections delve into each technology's unique characteristics, providing a detailed analysis of their respective technology’s unique characteristics.

\subsection{Atomic Swaps}

\textbf{Atomic swaps} is a type of contract that facilitates decentralized cryptocurrency exchanges \cite{herlihy2018atomic}. In this context, the term "atomic" implies that the transfer of ownership of one asset inherently triggers the transfer of ownership of another, satisfying the $atomicity$ property defined in Def. \ref{security cross-chain}. This concept was first introduced by TierNolan on the Bitcointalk forum in 2013 \cite{atomic_swap2013}. 
For four years, atomic swaps remained largely theoretical. Until 2017, when Charlie Lee, the founder of Litecoin \cite{yu2024bitcoin}, tweeted about successfully performing a cross-chain atomic swap between LTC and BTC, exchanging 10 LTC for 0.1167 BTC. Since that event, numerous decentralized exchange platforms and independent traders have adopted the technology for cryptocurrency trading \cite{atomic_swap2020}. Additionally, specialized cryptocurrency wallets, such as Atomic Wallet \cite{suratkar2020cryptocurrency} and Liquality, have been developed to facilitate cross-chain atomic swaps.

Atomic swaps must maintain $fungibility$, meaning that observers of the ledger (aside from the transacting parties) should not be able to distinguish between transfers executed as part of an atomic swap and standard asset transfers on the same ledger.
Currently, cross-chain atomic swaps require a minimum of four transactions, although some solutions attempt to reduce the number of transactions to two \cite{succinct_atomic_swap}, but it will increase the real-time online requirements for the exchanging parties. The most commonly used atomic swap technologies include hash time-lock contracts (HTLC) \cite{poon2016bitcoin}, and adaptor signatures \cite{aumayr2020generalized}.
While some methods \cite{bentov2019tesseract,lind2016teechan,wen2024mecury} propose deferring atomic swap functionality to \emph{Trusted Execution Environments} (TEEs) \cite{sabt2015trusted}, such solutions require all users to possess a TEE, which is impractical. Furthermore, recent research has revealed significant vulnerabilities in TEEs \cite{chen2019sgxpectre,van2019tale}.
We next describe HTLC and adaptor signature techniques in detail.

  \begin{table*}[!htbp]
 \caption{Comparative Analysis of Interoperability solutions based on HTLC}
 \centering
 \scriptsize
 \setlength{\tabcolsep}{13.2pt}
 \renewcommand{\arraystretch}{1.5}
 \label{table1}
 \begin{threeparttable}    
  \begin{tabular}{>{\columncolor{blue!6}}p{1.5cm}p{1.2cm}p{0.8cm}p{0.2cm}p{0.2cm}p{8.85cm}}
   \bottomrule
   
\rotatebox{0}{\textsl{\textbf{Reference}}}  & \rotatebox{60}{\textsl{\textbf{Technique}}}   &  \rotatebox{60}{\textsl{\makecell{\textbf{Trust}\\\textbf{Model}}}}   &  \rotatebox{60}{\textsl{\textbf{Privacy}}}\tnote{\ding{172}}     &   \rotatebox{60}{\textsl{\textbf{Generic}}}         & \rotatebox{0}{\textsl{\textbf{Summary of Advantages}}}  \\
   \hline

LN \cite{poon2016bitcoin}  &  HTLC &   Synchrony  & $\usym{2713}$ &  \ding{108} & It Creates a network of micropayment channels that enables bitcoin scalability,
micropayments down to the satoshi, and near-instant transactions.   \\

  CheaPay \cite{zhang2019cheapay} & CHTLC &  Synchrony  & $\usym{2717}$  &   \ding{119} & It examines the issue of payment routing in PCNs through an optimization lens, intending to minimize the transaction fee associated with a payment path. \\

AMHLs \cite{malavolta2018anonymous}  & Multi-hop lock &  Synchrony  &  $\usym{2713}$  &  \ding{119}  &   It serves as a versatile primitive, applicable beyond multi-hop payments in PCNs, and illustrates how this primitive can be leveraged to achieve $\mathcal{CCI}$ within PCNs.  \\

Deshpande et al. \cite{deshpande2020privacy} & HTLC and Schnorr signatures  & Hybrid  &   $\usym{2713}$  &   \ding{119}  &
It introduces the primitive of atomic release of secrets (ARS), which facilitates the atomic exchange of pre-agreed secrets in transactions, and illustrates how ARS can be applied to build privacy-protecting atomic swaps. \\

MAD-HTLC \cite{tsabary2021mad}   & HTLC-Spec &  Synchrony & $\usym{2717}$ &  \ding{119} & 
A new approach is proposed that harnesses miner rationality to secure smart contracts, and it is employed to design MAD-HTLC, which implements the HTLC-Spec.  \\

Cross Channel\cite{guo2023cross}    &  HTLC, zk-SNARK &  Hybrid  &  $\usym{2713}$  &  \ding{119}  & It is the first off-chain channel that supports cross-chain services, effectively reducing the high latency inherent in asynchronous networks, and delivering both strong security and practical utility. \\

zkCross \cite{guo2024zkcross}  & HTLC and zk-Rollup   &  Hybrid  &  $\usym{2713}$     & \ding{108}  &  It overcome three important challenges in cross-chain privacy-preserving auditing, namely Cross-chain Linkability Exposure, Incompatibility of Privacy and Auditing, and Full Auditing Inefficiency. \\

   \toprule
  \end{tabular}
  
  \begin{tablenotes}
   \footnotesize
   \item[\ding{172}] Privacy involves safeguarding the confidentiality of the identities of the sender and receiver, payment amount, and payment path within $\mathcal{CCI}$.
   
  \end{tablenotes} 
 \end{threeparttable}    
 \label{HTLC table}
\end{table*}

\begin{figure}[!ht] 
\small
\centering
\includegraphics[width=3.3in]{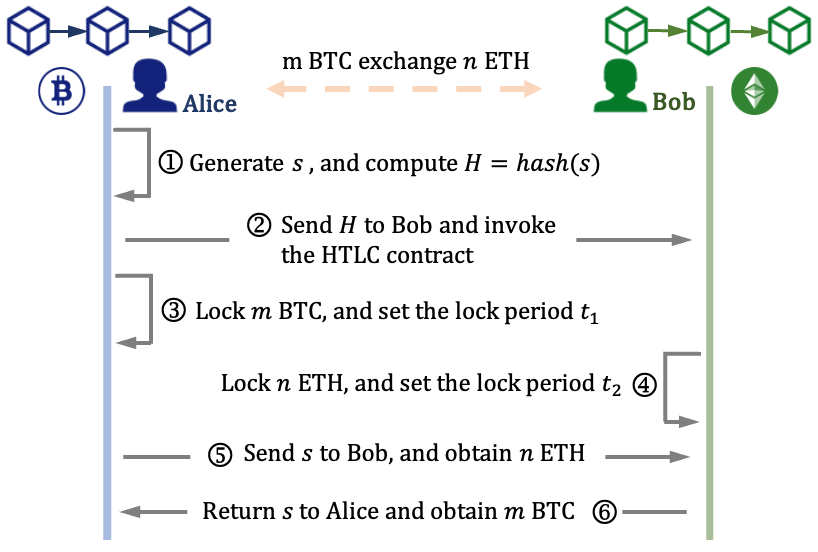} 
   \vspace{-2mm}
\caption{HTLC interaction between two different blockchains.}
\label{hash locking fig}
   \vspace{-2mm}
\end{figure}

\subsubsection{\textbf{HTLC-Based Atomic Swaps}}
HTLC was originally proposed to enable cross-chain transactions in DEXs \cite{koens2019assessing} and serves as a core technology for atomic swaps \cite{herlihy2018atomic}. It facilitates conditional payments across different blockchains through programmable logic and asset collateralization, with a notable application being \emph{payment channel networks} (PCNs) \cite{papadis2020blockchain}. The concept of HTLC is derived from sequential game theory \cite{escardo2011sequential}, where users on the same or different blockchains make decisions in sequence, based on the order of time. These decisions form the basis of a game-theoretic approach to achieving cross-chain asset swaps via collateralized transactions. 
The core components of HTLC are time-lock and hash-lock. A time-lock ensures that both parties to a transaction must submit their respective actions within a predefined time frame for the transaction to be valid. The commitment for this transaction is void if the time expires, and each party retains their assets. Conversely, a hash-lock involves setting a hash function, where a party can prove their commitment by revealing the pre-image $s$ that generates the hash value $\mathcal{H}=hash(s)$. If the corresponding hash value $\mathcal{H}$ is verified, the commitment remains valid; otherwise, it expires. 
HTLC allows for asset exchanges between distinct blockchain systems, ensuring that while the total quantity of assets on each blockchain remains unchanged, the ownership of these assets can be swapped, facilitating cross-chain asset exchange but not actual asset migration. Tab. \ref{HTLC table} summarizes some of the HTLC solutions that we identified in the literature.

\textbf{Technological Process of HTLC.} For example, as shown in Fig. \ref{hash locking fig}, consider the scenario where Alice, operating within the Bitcoin network, wishes to exchange $m$ BTC for $n$ ETH held by Bob in the Ethereum network: \ding{172} Alice generates a random secret $s$ and computes its hash value $\mathcal{H}=hash(s)$; \ding{173} Alice sends $\mathcal{H}$ to Bob and invokes the HTLC contract in the Ethereum network; \ding{174} Alice then locks $m$ BTC in the Bitcoin network through a locking contract, setting a time limit $t_{1}$. This contract promises Bob that he can obtain the $m$ BTC if he provides the pre-image $s$ of $\mathcal{H}$ within the time limit $t_{1}$; \ding{175} Upon learning that Alice has locked $m$ BTC, Bob locks $n$ ETH in the Ethereum network under a similar locking contract with a time limit $t_{2}$ (where $t_{2}<t_{1}$), promising Alice that she can claim the $n$ ETH if she provides the pre-image $s$ within time $t_{2}$; \ding{176} Alice then sends $s$ to Bob and unlocks the contract on Ethereum to receive $n$ ETH. If she fails to unlock within $t_{2}$, the system returns the $n$ ETH to Bob; \ding{177} Bob, upon receiving $s$, submits it to the Bitcoin network to unlock the $m$ BTC. If the contract is not unlocked within $t_{1}$, the system returns the $m$ BTC to Alice.

\textbf{Limitations of HTLC.} HTLC-based atomic swaps are deployed in practice \cite{koens2019assessing,tsabary2021mad} and have a wide range of applications \cite{adams2024layer,malavolta2018anonymous,wood2016polkadot}. Despite their advantages, these methods exhibit intrinsic limitations that undermine their utility, which we summarize below:

\begin{itemize}[itemsep=0.5pt]
\item \emph{Compatibility of the Hash Function.} Both $\mathscr{S}$ and $\mathscr{T}$ must support compatible hash functions within their scripting languages, and each ledger must represent the hash function using the same number of bits. Otherwise, atomicity may be compromised, as one ledger might not allow sufficiently large pre-images \cite{thyagarajan2022universal}. Beyond atomicity, using the same hash value across both ledgers also raises privacy concerns, as observers could link two HTLCs as part of the same swap. Finally, a fundamental issue arises in that many cryptocurrencies, such as Monero \cite{lai2019omniring}, Ripple \cite{schwartz2014ripple}, or Zcash \cite{sasson2014zerocash} (with shielded addresses), do not support HTLC contract computation in their scripting languages.

\item \emph{Limitations of Time-Lock.} To facilitate this feature, both ledgers must include support for time-lock functionality in their respective scripting languages. However, adding time-lock conflicts with privacy protection for several reasons: \ding{172} It makes time-locked transactions easier to distinguish from transactions without time restrictions \cite{monerotimelockwoes}; \ding{173} It may interfere with other privacy-enhancing operations already in place on the ledger \cite{Timelockedtransactionoutputs}; \ding{174} Even if it is possible to implement time locks in a privacy-preserving manner, it significantly increases computational and storage costs for the ledger \cite{ImplementConfidentiallyTimelockedFunds,monerotimelockwoes}; \ding{175} Both parties involved in the transaction may be exposed to price speculation during the waiting period, such as front-running attacks \cite{daian2020flash}. Therefore, in such cases, designing privacy-focused cryptocurrencies requires avoiding time-locked assets as a design principle \cite{thyagarajan2022universal}.

\item \emph{Single-Asset Swap.} The swap is limited to two parties and does not support multiparty exchanges. In addition, given the significant value differences among cryptocurrencies, current atomic swaps are typically restricted to small values of $m$ (or $n$) to match swap offers (e.g. $m$ BTC by $n$ ETH). In practice, there are users, such as market makers or exchanges, who hold diversified portfolios across multiple ledgers. If multi-asset swaps were possible, they could leverage several of their assets to match swap offers more efficiently.

\end{itemize}

\subsubsection{\textbf{Adaptor Signature-Based Atomic Swaps}}

Adaptor signature \cite{aumayr2020generalized} allows users to create a pre-signature for a message $\mathcal{M}$, which, on its own, is not valid. However, it can be transformed into a valid signature once the user reveals a specific secret value. Fig. \ref{adaptorsigfig} provides the formal definition of adaptor signatures. As a promising cryptographic primitive, it not only addresses several limitations of HTLCs, but have also found applications in areas such as DeFi, payment channel networks, multi-party signature protocols, and privacy-enhancing transactions. Recent research has investigated its use in multi-party atomic swap scenarios.

\begin{figure}[htbp]
\centering
\fbox{
\begin{minipage}[b]{0.9\linewidth}
\centering
\footnotesize
\underline{Adaptor Signature $\Pi_{\mathsf{AS}}$}
\begin{description}
\item An adaptor signature scheme $\Pi_{\mathsf{AS}}$ w.r.t a hard relation $\mathcal{R}$ and a signature scheme $\Pi_{\mathsf{DS}}=(\mathsf{KGen},\mathsf{Sign},\mathsf{Vf})$ consists of algorithms $(\mathsf{pSign},\mathsf{Adapt},\mathsf{pVf},\mathsf{Ext})$ defined as:
\item[$\vartriangleright \hat{\sigma} \leftarrow \mathsf{pSign}(sk,\mathcal{M},\mathcal{S})$:]The pre-sign algorithm takes as input a secret key $sk$, message $\mathcal{M} \in \{0,1\}^{*}$ and statement $\mathcal{S}\in L_{\mathcal{R}}$, outputs a pre-signature $\hat{\sigma}$.

\item[$\vartriangleright 0/1 \leftarrow \mathsf{pVf}(pk,\mathcal{M},\mathcal{S}, \hat{\sigma})$:]The pre-verify algorithm takes as input a public key $pk$, message $\mathcal{M} \in \{0,1\}^{*}$, statement $\mathcal{S}\in L_{\mathcal{R}}$ and pre-signature $\hat{\sigma}$, outputs a bit $b$.

\item[$\vartriangleright \sigma \leftarrow \mathsf{Adapt}(\hat{\sigma},\mathcal{W})$:]The adapt algorithm takes as input a pre-signature $\hat{\sigma}$ and witness $\mathcal{W}$, outputs a signature $\sigma$.

\item[$\vartriangleright \mathcal{W} \leftarrow \mathsf{Ext}(\sigma,\hat{\sigma},\mathcal{S})$:]The extract algorithm takes as input a signature $\sigma$, pre-signature $\hat{\sigma}$ and statement $\mathcal{S}\in L_{\mathcal{R}}$, outputs a witness $\mathcal{W}$ such that $(\mathcal{S}, \mathcal{W})\in \mathcal{R}$, or $\bot$.

\end{description}
\end{minipage}
}
\caption{A generic adaptor signature scheme.}
\label{adaptorsigfig}
\end{figure}

\begin{figure*}[!ht] 
\small
\centering
\includegraphics[width=4.5in]{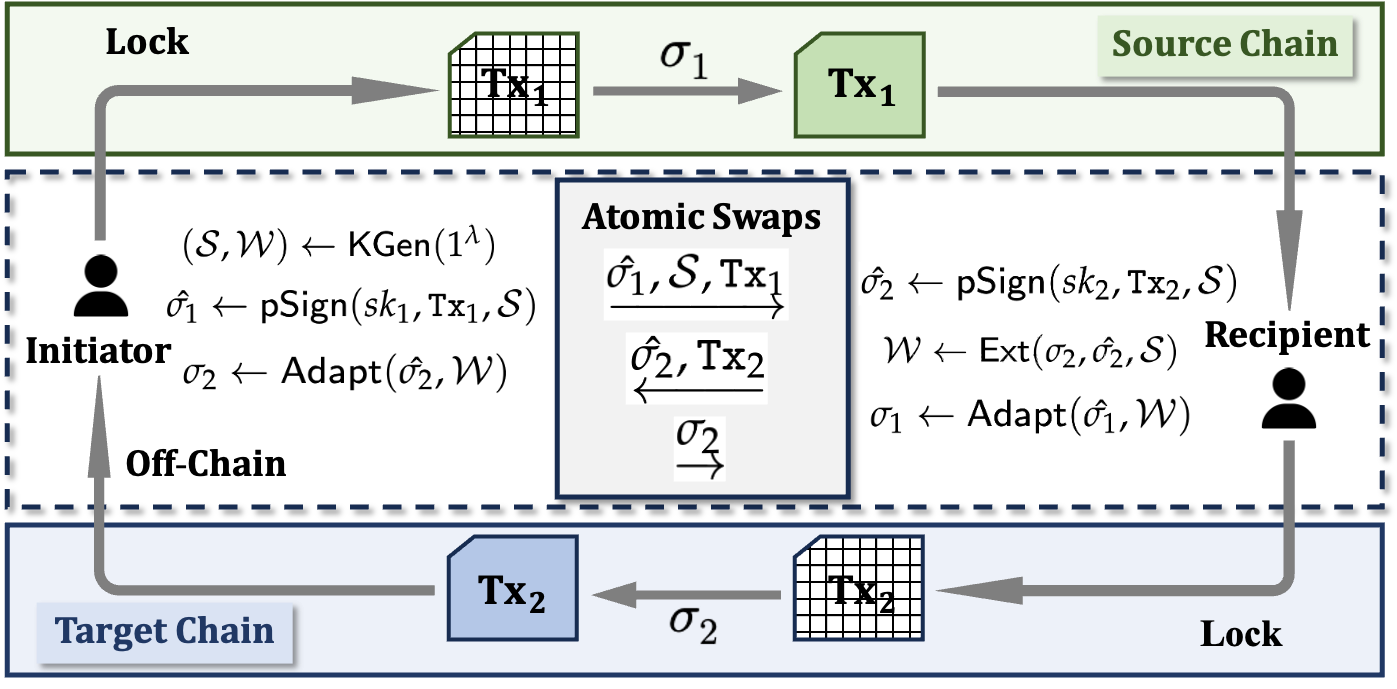} 
   \vspace{-2mm}
\caption{Atomic Swaps based on adapter signature.}
\label{adaptor_label}
   \vspace{-2mm}
\end{figure*}

Atomic swap protocol based on adaptor signature involves the interaction between an initiator on the source chain and a recipient on the target chain to exchange assets $\Tx_{1}$ and $\Tx_{2}$, as illustrated in Fig. \ref{adaptor_label}. To ensure fairness, both parties apply time locks to the assets involved, primarily to provide the recipient with sufficient time to complete the transaction and prevent the initiator from claiming both assets.
The protocol begins with the initiator generating a hard relation $(\mathcal{S}, \mathcal{W})\leftarrow \mathsf{KGen}(1^{\lambda})$, and using the statement $\mathcal{S}$ to produce a pre-signature $\hat{\sigma}_{1}$ for the transaction $\Tx_{1}$ (i.e., transferring $\Tx_{1}$ to the recipient). The pre-signature is then sent to the recipient, who verifies its correctness. Upon successful verification, the recipient uses the same statement $\mathcal{S}$ to generate a pre-signature $\hat{\sigma}_{2}$ for the transaction $\Tx_{2}$ (i.e., transferring $\Tx_{2}$ to the initiator), and returns it to the initiator.
The initiator verifies $\hat{\sigma}_{2}$ and then adapts it into a full signature $\sigma_2$ using the witness $\mathcal{W}$. The initiator then broadcasts $\sigma_2$ on-chain to claim $\Tx_{2}$. Observing this, the recipient extracts $\mathcal{W}$ from $\hat{\sigma}_1$ and $\sigma_1$, adapts $\hat{\sigma}_1$ into a full signature $\sigma_1$, and broadcasts $\sigma_1$ on-chain to claim $\Tx_{1}$. This completes the atomic and fair exchange process.

In Tab. \ref{adaptortab}, we provide a detailed analysis of the strengths, weaknesses, and suitability of various protocols.

\textbf{Compared with HTLC.} Atomic swaps utilizing adaptor signatures offer the following key advantages:

\begin{itemize}[itemsep=0.5pt]
\item \emph{Reduced Lock Time and On-Chain State.} It eliminates the reliance on on-chain scripts like time-lock and hash-lock used in "secret-hash" swaps, thereby reducing the time assets remain locked.

\item \emph{Higher Off-Chain Efficiency.} The primary interactions in adaptor signature schemes occur off-chain, with only the final state requiring on-chain confirmation. This makes cross-chain transactions more lightweight and reduces complexity and transaction fees, especially in Mulit-path Payment \cite{liu2023lightpay} and frequent cross-chain operations scenarios. 

\item \emph{Enhanced Privacy.} While HTLC necessitates using the same hash value across chains, adaptor signatures decouple transactions, thereby minimizing the exposure of publicly visible information on-chain. 

\item \emph{Support for Multi-Party and Multi-Chain Scenarios.} By incorporating multi-signatures or other cryptographic primitives, adaptor signatures provide greater flexibility for multi-party, multi-chain, and multi-asset atomic swaps \cite{thyagarajan2022universal,you2024multi}, making them more scalable for complex cross-chain transaction environments.

\end{itemize}

\textbf{Open Issues of Adaptor Signature.} Despite the growing applicability of adaptor signatures, several unresolved issues or challenges remain. In multi-party atomic swap scenarios, mitigating collusion attacks continues to be a significant open problem. 
Possible solutions include employing reputation systems, mandating participants to furnish deposits or collateral, or utilizing advanced cryptographic methods like threshold signatures or MPC. Nevertheless, these approaches may add significant complexity and overhead to the protocols.
 While some multi-party atomic swap protocols \cite{thyagarajan2022universal,you2024multi} have made substantial progress in cross-chain asset exchanges, it is crucial to acknowledge the limitations and assumptions of these protocols.
In PipeSwap \cite{ni2024pipeswap}, it was pointed out that the scheme in \cite{thyagarajan2022universal} is vulnerable to $double$-$claiming$ attacks, which are relatively easy to execute and can naturally extend to other scriptless cross-chain swap protocols and PCNs.
 Future research should address these challenges and enhance the protocols' robustness in real-world scenarios.

\begin{table*}[!htbp]
 \caption{Comparative Analysis of Interoperability solutions based on Adaptor Signature}
 \centering
 \scriptsize
 \setlength{\tabcolsep}{13.2pt}
 \renewcommand{\arraystretch}{1.5}

 \begin{threeparttable}    
  \begin{tabular}{>{\columncolor{blue!6}}p{0.8cm}p{1.2cm}|p{0.6cm}|p{7.45cm}|p{3.15cm}}
   \rowcolor{blue!6}
   \bottomrule
   
\textsl{\textbf{Reference}}  &   \textsl{\textbf{Object}}  &   \textsl{\textbf{Generic}}  & \textsl{\textbf{Strong Points}}  &  \textsl{\textbf{Weak Points}} \\
   \hline
   
\cite{deshpande2020privacy}  &  \multirow{2}{*}{Two-Party}   &  \multirow{2}{*}{\ding{109}} &  It represents an atomic secret release scheme that is built upon the combination of adapter signatures and the Schnorr signature algorithm.    & \multirow{2}{*}{\makecell{They are limited to two-party\\ and do not consider\\ the challenges of\\ multi-party atomic\\ swaps scenario.}} \\  
\cline{1-1} \cline{4-4}
\cite{klamti2022post}                  &  &  &   It is an enhanced two-party adapter signature scheme grounded in quantum-secure coding theory problems.    &   \\
   \hline

\cite{kajita2024generalized}     &   \multirow{3}{*}{Multi-Party}   &  \multirow{3}{*}{\ding{119}}  & It proposes a generalized adaptor signature scheme for N parties, enabling secure multi-party transactions. & \multirow{4}{*}{\makecell{They often focus on specific\\ scenarios or address a limited\\ set of potential attack vectors.\\ 
Potential vulnerabilities linked\\ to time-lock puzzle mechanisms\\ and the computational burden of\\ managing multiple blockchains\\ still require resolution.}} \\   
\cline{1-1} \cline{4-4} 
\cite{chen2024privacy}                           &    &  & It proposes a privacy-preserving multi-party cross-chain transaction protocol based on a novel pre-adaptor signature scheme.  & \\
\cline{1-1} \cline{4-4}
\cite{ji2023threshold}                           &    &  & It proposes the concept of threshold adaptor signatures for enhancing the security and fault tolerance of multi-party swaps. &  \\
 \cline{1-4}
 
\cite{thyagarajan2022universal}                         &  \multirow{2}{*}{\makecell{Multi-Party\\Non-custodial\\Multi-Asset\\Universal}}  & \multirow{2}{*}{\ding{108}} & It establishes a complete framework for general atomic swaps, incorporating adapter signatures and time-lock puzzle techniques to optimize practicality.  &  \\
\cline{1-1} \cline{4-5}
\cite{you2024multi}                          &              &  & \multicolumn{2}{c}{\makecell{It represents the first fully scalable off-chain atomic swap protocol, supporting multiple participants\\ (of any number), while ensuring zero overhead for the local blockchain, without the dependency\\ on smart contracts or trusted third parties.}}  \\

   \toprule
  \end{tabular}

 \end{threeparttable}    
 \label{adaptortab}
\end{table*}

 \subsection{Notary-based Token Swap Bridges}

 \begin{figure}[!ht] 
\small
\centering
\includegraphics[width=3.5in]{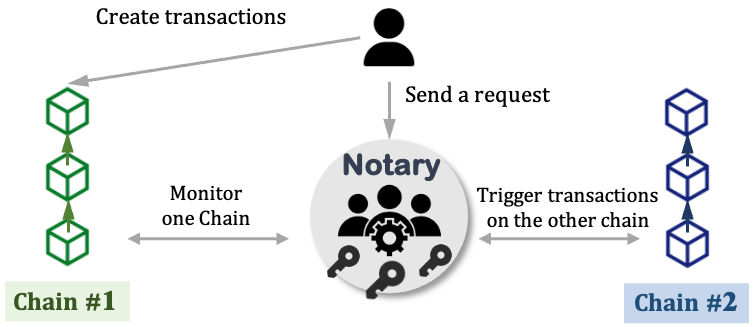} 
   \vspace{-2mm}
\caption{Workflow of a notary-based scheme via a TTP.}
\label{notaryfig}
   \vspace{-2mm}
\end{figure} 

Due to the independence of different blockchains in cross-chain transactions, they cannot directly understand or verify the state changes on each other's chain. Therefore, a trusted intermediary, known as a "\textbf{Notary}", is introduced to act as a bridge between the two chains. Notaries are widely used for their efficiency and ease of implementation \cite{interledger,yin2022bool,wang2023cross}. A typical notary workflow can be described as follows (Fig. \ref{notaryfig}):

 \begin{itemize}[itemsep=0.5pt]
\item \emph{Initiating Transactions.} The user initiates a transaction or event on $\mathcal{L}_{\mathscr{S}}$ (e.g., locking a certain amount of tokens).

\item \emph{Notary Verification on $\mathscr{S}$.} The notary monitors the transaction on the $\mathscr{S}$ and verifies whether the transaction has been successfully executed. Verification usually involves checking whether the transaction has been confirmed by the consensus mechanism of $\mathscr{S}$.
\item \emph{State Notification to $\mathscr{T}$.}  Once the notary confirms the event or transaction on $\mathcal{L}_{\mathscr{S}}$, it notifies $\mathscr{T}$, indicating that the event has occurred. For example, the notary can issue proof on $\mathscr{T}$ to indicate that assets on $\mathscr{S}$ have been locked.
\item \emph{Execution of Cross-Chain Operation.} Based on the notary's proof, $\mathscr{T}$ executes the corresponding operation on-chain, such as releasing an equivalent amount of assets or triggering a cross-chain smart contract call.

\end{itemize} 

\textbf{Notary Evolution.} Variants of this mechanism include centralized notary schemes and decentralized notary schemes.
Tian et al. \cite{tian2021enabling} designed a decentralized notary scheme for executing atomic swaps in cryptocurrency exchange protocols. This protocol involves a verification committee (a group of notaries) responsible for inspecting and verifying transactions, with a notary election mechanism mitigating the risk of a single point of failure. Similarly, RenVM \cite{you2024multi} employs a Byzantine Fault-Tolerant network combined with Secure Multi-Party Computation (SMPC) to facilitate cross-chain asset transfers, replacing centralized custodianship with a decentralized, trustless custodian model. 
 
 While distributed collective signatures enhance the decentralization of notary groups, this method does not eliminate the issues of trust and incentives for notaries. As a result, some researchers have turned to reputation metrics to address trust issues associated with notaries. Xiong et al. \cite{xiong2022notary} improved the reliability of notaries by refining the internal selection process of the notary group and integrating collateral pools with a reputation-driven incentive system. Niu et al. \cite{niu2023nft} introduced an enhanced reputation value model that ensures notary reliability while reducing the risk of over-centralization. Zhao et al. \cite{zhao2022dynamic} developed a reputation-based notary election mechanism using an advanced PageRank algorithm, effectively preventing malicious nodes from becoming notaries. Similarly, Sun et al. \cite{sun2022decentralized} adopted a reputation-based election method, randomly selecting notaries from high-reputation candidates to handle cross-chain transactions, while updating reputation values to restrict malicious behavior by notaries. Hu et al. \cite{hu2024blockchain} introduced reputation decay and dynamic window mechanisms to prevent inactive malicious notaries from regaining reputation over time. In contrast to these approaches, Bool Network \cite{yin2022bool} is a secure notary platform that uses Ring VRF and TEEs to hide the notary group, reducing trust conflicts.

 \textbf{Limitations of Notary.} 
Notary-based cross-chain technology, valued for its simplicity and flexibility, is theoretically compatible with most heterogeneous blockchain interoperability needs. However, its reliance on external notary entities introduces a trust assumption that undermines the core principles of decentralization and trustlessness in blockchain systems. This has become a major obstacle to its broader adoption.
In practice, the approach is mainly used in low-frequency cross-chain scenarios, such as asset transfers or cross-chain smart contract calls, where accuracy is critical but real-time performance is not. The system’s security and correctness depend entirely on the notary’s proper behavior.
Moreover, its dependence on off-chain entities introduces regulatory and compliance risks. If notaries are influenced by legal or policy constraints, the system may lose its neutrality and global accessibility. These limitations make notary-based solutions more suitable for short-term, domain-specific applications rather than fully autonomous and secure cross-chain ecosystems.
As a result, researchers are increasingly exploring alternative cross-chain mechanisms to balance decentralization with performance better.

 \subsection{Light Client}
Interoperability verification is typically achieved by running full nodes or employing \textbf{light clients} with linear storage overhead, which scales with the length of $\mathscr{S}$. The core concept of the light client was first introduced by Satoshi Nakamoto in his original whitepaper \cite{nakamoto2008peer} as an SPV solution.

A Light Client\cite{chatzigiannis2022sok} is a type of node in blockchain networks that, compared to full nodes, aims to provide fundamental verification and interaction capabilities with significantly lower resource and storage requirements. As a result, this technology serves as a low-cost alternative for node implementation and can act as a bridge for data verification and communication in blockchain interoperability mechanisms.
Specifically, a light client leverages the consensus mechanism of the source chain to ensure the authenticity of data, while functioning as a verification module on the target chain to validate the legitimacy of transactions from the source chain. In scenarios requiring cross-chain synchronization of account states (e.g., balances or assets), the light client enables efficient state synchronization by verifying block headers and associated state proofs. As a result, the security of light clients often depends on the robustness of $\mathcal{L}_{\mathscr{S}}$ (see Def. \ref{RDL} for details).

\begin{figure*}[!ht] 
\small
\centering
\includegraphics[width=5.0in]{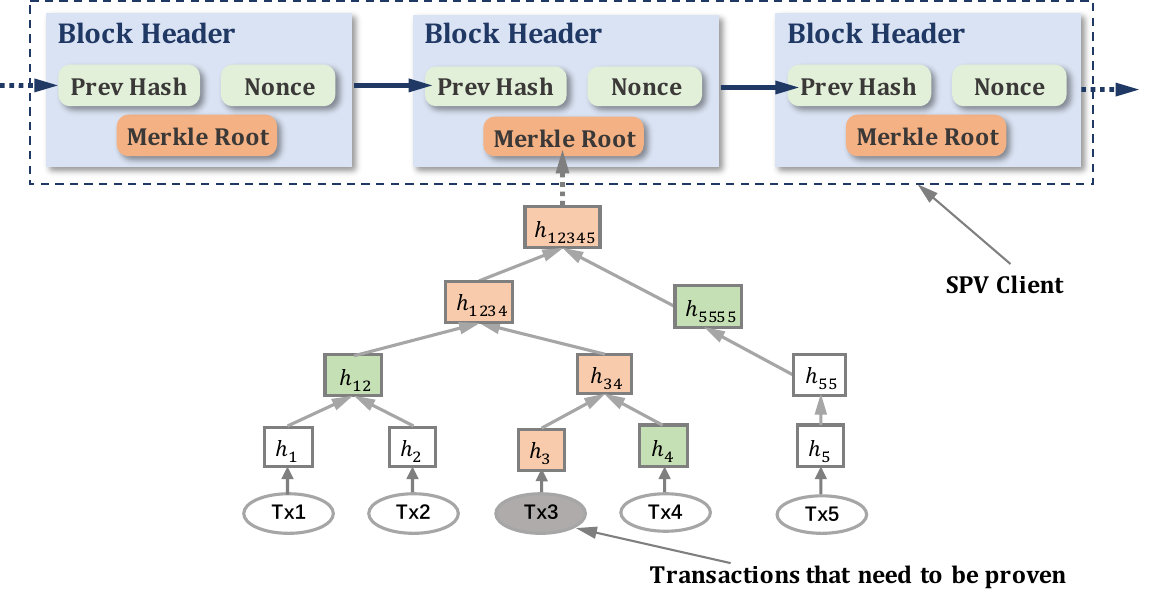} 
   \vspace{-2mm}
\caption{Simplified Payment Verification. \textcolor{orange!75}{The orange items} in the Merkle Tree constitute the proof of Tx3. \textcolor{darkgreen}{The green item} can be computed and validated against the Merkle Root.}
\label{SPVfig}
   \vspace{-2mm}
\end{figure*}

 \textbf{Simplified Payment Verification (SPV).} SPV operates by utilizing Merkle proofs, a critical component that allows light nodes to verify whether a transaction is included in a block without downloading the entire block (see Fig. \ref{SPVfig}). Specifically, a transaction’s Merkle proof consists of its Merkle path and the root of the Merkle tree. The Merkle path is a collection of sibling nodes along the path from the transaction to the root of the tree. By verifying the Merkle path, a light node can confirm that the transaction indeed exists within a specific block, thus participating in the blockchain network without maintaining the entire chain’s data.



  \begin{table*}[!htbp]
 \caption{Comparison of Light Client Solutions.}
 \centering
 \scriptsize
 \setlength{\tabcolsep}{9.2pt}
 \renewcommand{\arraystretch}{1.5}
 \label{table1}
 \begin{threeparttable}    
  \begin{tabular}{>{\columncolor{blue!6}}cccccccc}
     \rowcolor{blue!6}
   \bottomrule
   
  \rotatebox{0}{\textsl{\makecell{\textbf{Concrete}\\\textbf{Type}}}}\tnote{\ding{172}}      &  \rotatebox{0}{\textsl{\textbf{References}}} &\rotatebox{0}{\textsl{\makecell{\textbf{Information}\\\textbf{Relayed}}}}    &  \rotatebox{0}{\textsl{\makecell{\textbf{Backward}\\\textbf{Compatibility}}}}\tnote{\ding{173}}    &  \rotatebox{0}{\textsl{\makecell{\textbf{Storage}\\\textbf{Overlhead}}}} & 
 \rotatebox{0}{\textsl{\makecell{\textbf{No Trusted}\\\textbf{Setup}}}}
 &\rotatebox{0}{\textsl{\makecell{\textbf{Upfront}\\\textbf{Mining Secure}}}}\tnote{\ding{174}}  &
 \rotatebox{0}{\textsl{\makecell{\textbf{Communication}\\\textbf{Complexity}}}}\tnote{\ding{175}}  \\
   \hline

LC  & \cite{btcrelay}\cite{westerkamp2020zkrelay}\cite{frauenthaler2020eth}  &  Linear & $\usym{2713}$  & Linear  & $\usym{2713}$   &  $\usym{2713}$  &  $\mathcal{O}(C)$ \\

 \hline

 SLC & \cite{kiayias2016proofs}\cite{kiayias2020non}\cite{bunz2020flyclient}  & Logarithmic & $\usym{2717}$  & Logarithmic & $\usym{2713}$  &  $\usym{2713}$ & $\mathcal{O}(k\cdot\mathsf{polylog}(C))$ \\

 \hline

  ZK  & \cite{bonneau2020coda} \cite{xie2022zkbridge} \cite{belchior2024harmonia}  & Linear & $\usym{2713}$  & Constant & $\usym{2717}$  &  $\usym{2713}$ &  $\mathcal{O}(1)$  \\

 \hline

  \multirow{2}{*}{SSPV}\tnote{\ding{176}}   & \cite{bartoletti2018bitml}\cite{statelessspv}   & \multirow{2}{*}{Constant} & \multirow{2}{*}{$\usym{2713}$}  & \multirow{2}{*}{Constant} & \multirow{2}{*}{$\usym{2713}$}  &  $\usym{2717}$  &\multirow{3}{*}{$\mathcal{O}(k)$} \\
\cline{2-2} \cline{7-7}

   & \cite{scaffino2023glimpse} &  &  &  & &  $\usym{2713}$&\\

 \cline{1-7}

PSLC  & \cite{aumayr2024blink}&  Constant & $\usym{2713}$  & Constant & $\usym{2713}$  &  $\usym{2713}$&\\

   \toprule
  \end{tabular}
  
  \begin{tablenotes}
   \scriptsize
   \item[\ding{172}] $Abbreviation$: Light Client (LC), Super-Light Client (SLC), Zero-Knowledge Based (ZK), Stateless SPV (SSPV), Provably Secure Light Client (PSLC).
   \item[\ding{173}] Super-light Clients with logarithmic complexity were proposed \cite{kiayias2016proofs,kiayias2020non,bunz2020flyclient}, but they either require constant PoW difficulty \cite{kiayias2020non} or an hard fork in Bitcoin \cite{bunz2020flyclient}, and are thus not backward compatible.
   \item[\ding{174}] By knowing the transaction to be verified in advance, a malicious prover can exploit the fact that users on $\mathscr{S}$ cannot ensure that the proof corresponds to the correct suffix of the chain. The prover can pre-construct a forged subchain. Since there is no backward time constraint on executing an upfront mining attack, the attacker will eventually succeed in finding a sufficient number of forged blocks, regardless of their mining power or the need to bribe any miners \cite{scaffino2023glimpse}.
   \item[\ding{175}] Let $C$ denote the lifetime of the system (informally, the length of $\mathscr{S}$ or $\mathscr{T}$) and $k$ denote the security parameter. According to Def. \ref{RDL} with the Bitcoin Backbone model \cite{backbone2015}, $k$ is the \emph{common prefix} parameter, which is constant for a protocol execution, albeit with the trade-off of logarithmically increasing the probability of failure in the lifetime of the system.
   \item[\ding{176}] In SSPV, users provide proof $\pi$ to the smart contract with quasi-Turing completeness hosted on $\mathscr{T}$, convincing it that a transaction has appeared on PoW-based $\mathscr{S}$. This proof consists of the block header containing the transaction, the Merkle inclusion proof of the transaction within the block, and $n$ subsequent confirmation block headers. The smart contract subsequently validates the Merkle proof and ensures that each of the $n+1$ block headers constitutes a legitimate subchain of its parent chain, and ensures that all headers contain sufficient PoW, meaning their hash values are less than the predetermined target.
   
  \end{tablenotes} 
 \end{threeparttable}    
 \label{lightclienttablecompare}
\end{table*}

\textbf{Light Client Evolution.} While SPV light clients save storage space (Bitcoin's block headers are around 80 bytes compared to the full block size of about 1MB), they still require processing a large amount of data proportional to the chain’s length. For Bitcoin, this amounts to approximately 60MB of storage, while for Ethereum, it requires around 4GB. To reduce this storage burden, various optimizations have emerged. The first succinct construction was the interactive \emph{Proofs of Proof-of-Work} (PoPoW) protocol \cite{kiayias2016proofs}, which achieves polylogarithmic communication costs. INPoPoW \cite{kiayias2020non} removed the need for interactivity and provided security and succinctness for 1/2 adversaries under optimistic conditions. This was later optimized \cite{karantias2020compact} and further improved to more practical solutions \cite{daveas2020gas}, with backward compatibility ensured through redesigns \cite{kiayias2021velvet}. In later work, the optimistic environment constraint was addressed, enabling succinctness for all adversaries, with security guarantees for up to 1/3 threshold adversaries \cite{kiayias2021mining}. Another alternative, FlyClient \cite{bunz2020flyclient}, was proposed to provide security and succinctness for 1/2 adversaries, adding support for variable difficulty adjustments.

More recently, universal (recursive) zero-knowledge (ZK) technologies have been employed to construct light clients with constant communication overhead \cite{bonneau2020coda,xie2022zkbridge,belchior2024harmonia}. 
For example, DendrETH \cite{belchior2024harmonia} is a decentralized and efficient ZK proof-based light client, which mitigates security problems by lowering the attack surface by relying on the properties of ZK proofs. 
However, these methods incur high computational costs and require a trusted setup for key generation and verification. To develop a constant communication light client without the need for a trusted setup, the concept of \emph{stateless SPV} (SSPV) was proposed by Prestwich   \cite{howto} and implemented by Summa \cite{Summa}. 
Recently, Barbára et al. \cite{barbara2022bxtb} implemented, and for the first time formalized, stateless SPV within the BxTB cross-chain exchange. 
However, Scaffino et al. revealed that this construction is vulnerable to upfront mining attacks, rendering it insecure \cite{scaffino2023glimpse}. They proposed a new protocol named Glimpse \cite{scaffino2023glimpse}, which builds on the stateless SPV idea by introducing high-entropy transactions to prove that the provided chain segment is "fresh" and not pre-mined. Aumayr et al. \cite{aumayr2024blink} refined the problem of PoPoW and proved that Blink has optimal communication cost, constructing the first provably secure Optimal Proof of Proof-of-Work without a trusted initialization Setting. A comparative summary of various light clients for $\mathcal{CCI}$ verification is presented in Tab. \ref{lightclienttablecompare}.

\textbf{Open Issues of Light Client.} Research focus has shifted from SLC and SSPV to PSLC (see Tab. \ref{lightclienttablecompare}), with an emphasis on reducing cross-chain verification size and communication complexity without compromising security. However, no solution has been developed that fully satisfies functional, security, and efficiency properties while remaining practical for clients and minimizing overhead for consensus participants or full nodes \cite{chatzigiannis2022sok}. Existing studies have not sufficiently addressed the inefficiencies associated with frequent offline phases of light clients. Even for light clients with efficient bootstrapping protocols, frequent offline periods may still be inefficient due to the time lag in synchronizing with the blockchain state. 
Future research may explore the delegation of certain computational tasks of light clients to participants on the source chain, such as consensus nodes or full nodes. By introducing appropriate incentive mechanisms, this approach can ensure the feasibility and reliability of such delegation. Not only does this strategy significantly reduce the storage and computational burden on light clients, but it also improves overall system efficiency, offering a more optimized solution for cross-chain verification and state synchronization.


 \subsection{Sidechains with Wrapped Assets}
     \begin{figure*}
\centering
\subfigure[Two-Way Peg in Sidechains]{
\includegraphics[width=3.5in]{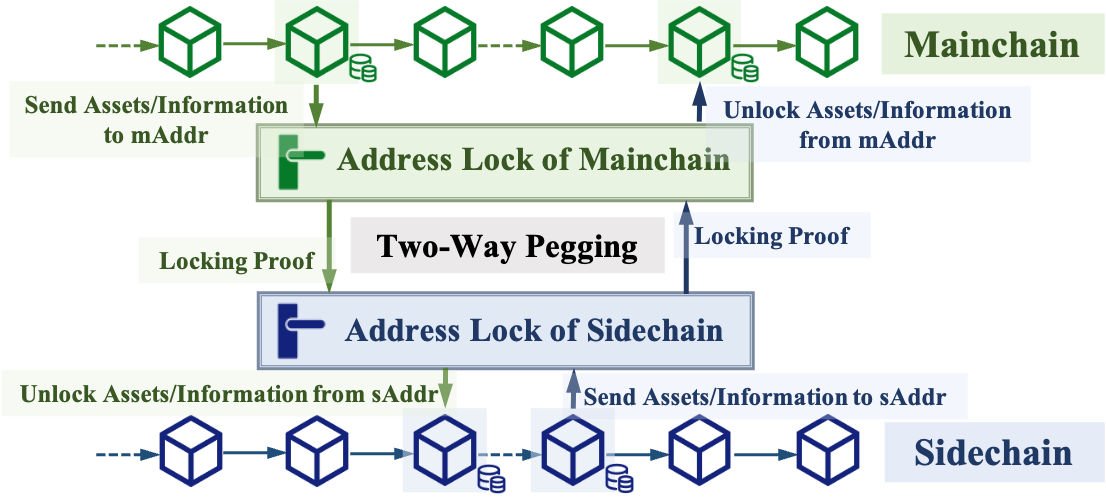}
}
\quad
\subfigure[Two forms of Sidechains]{
\includegraphics[width=3.1in]{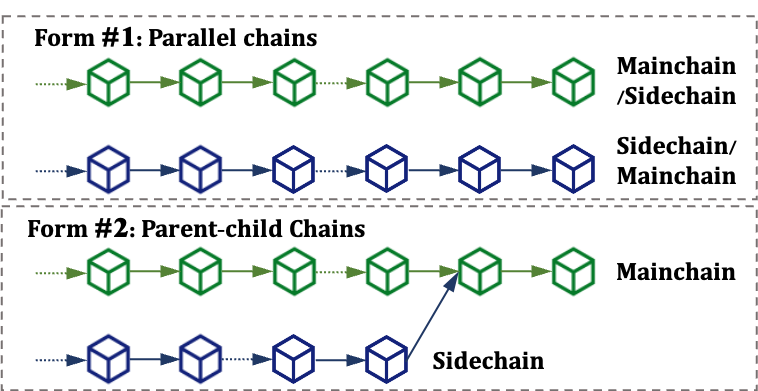}
}
\caption{Sidechains construction and types.}
\label{sidechainintroduction}
\vspace{-2mm}
\end{figure*}

\textbf{Sidechains}, also known as pegged sidechains, is a cross-chain technique that facilitates blockchain interoperability by supporting bidirectional transfers between blockchains \cite{sidechainpeg2014}. In addition to enhancing interoperability, sidechains contribute to the scalability and upgradability of blockchains \cite{possidechain}. They enable blockchains to offload transactions, executing them on sidechains, thus promoting scalability. Furthermore, new functionalities can be explored by bootstrapping sidechains from the mainchain. 

\textbf{Two-way peg} mechanisms can be categorized as centralized or federated. In a centralized two-way peg, a TTP performs token locking, which offers speed and simplicity but introduces a single point of failure and centralization \cite{singh2020sidechain}. In contrast, the federated two-way peg distributes control among a group of notaries, thereby mitigating issues of centralization and single points of failure \cite{sidechainpeg2014,dilley2016strong}. Depending on the mode of implementation, two-way Pegs can be implemented as following five modes: Single Custodian, Consortium, SPV, Driving Chain and Hybrid, with specific descriptions and comparison referenced in Tab. \ref{sidechain_mode_table}.

 Fig. \ref{sidechainintroduction}(a) illustrates the bidirectional transfers facilitated by a two-way peg. To transfer assets from the mainchain to the sidechain, users send assets to an external address associated with a consortium \cite{sidechainpeg2014}, which acts as an intermediary for locking and unlocking assets or information. After a specified transaction time commitment, the consortium releases equivalent assets on the sidechain. As depicted in Fig. \ref{sidechainintroduction}(b), sidechains are generally divided into two types: parallel chains and parent-child chains, where the mainchain serves as the parent in the latter.
 
 \textbf{Wrapped Assets} in sidechains are digital representations of underlying assets on other chains. They are deposited (wrapped) on $\mathscr{S}$ when the corresponding original tokens have been locked on $\mathscr{T}$, and then they are destroyed or withdrawn (unwrapped) to redeem the original ones.
 This category encompasses assets issued on sidechains and collateralized on parent chains, such as Liquid \cite{dilley2016strong} tokens L-BTC wrapped by BTC. It also includes wrapped tokens, such as WBTC on Ethereum \cite{ethereum} wrapped by BTC.

  \begin{table*}[!htbp]
 \caption{Performance Comparison of Different Two-way Pegs Mode for Sidechains Implementations.}
 \centering
 \scriptsize
 \setlength{\tabcolsep}{13.2pt}
 \renewcommand{\arraystretch}{1.5}
 \label{table1}
 \begin{threeparttable}    
  \begin{tabular}{>{\columncolor{blue!6}}cccccc}
   \bottomrule
   
\rotatebox{0}{\textsl{\textbf{Benchmark}}}   & \rotatebox{10}{\textsl{\textbf{Single Custodian}}}    &   \rotatebox{10}{\textsl{\textbf{Consortium}}}  &   \rotatebox{10}{\textsl{\textbf{SPV}}}  &   \rotatebox{10}{\textsl{\textbf{Driving Chain}}}   &   \rotatebox{10}{\textsl{\textbf{Hybrid}}}   \\
   \hline


\emph{Realization Approach}\tnote{\ding{172}}  &  Central Exchange  &   Multi-Party Signature & Soft Fork  &  Soft Fork  & Soft Fork \\

\emph{Implementation Difficulty}\tnote{\ding{173}}  &  \ding{109}  & \ding{109}  & \ding{119} & \ding{119} & \ding{108} \\

\emph{Security Strength}\tnote{\ding{174}}  &   \ding{109}  & \ding{119}  & \ding{119} & \ding{119} & \ding{108}   \\

\emph{Degree of Centralization} &  \ding{108}  & \ding{119}  & \ding{109} & \ding{119} & \ding{119} \\

\emph{Interoperability Efficiency} &  \ding{108}  & \ding{108}  & \ding{109} & \ding{119} & \ding{119}  \\

\emph{Trust Model} & TTP  & TTP  &  Synchrony &   TTP  &  Hybrid   \\

\emph{Typical Case}   &  Liquid \cite{nick2020liquid}   &  Cumulus \cite{gai2021cumulus}   &  Pegged Sidechains \cite{back2014enabling}  &   Drivechain \cite{sztorc2015drivechain}   &  IBC \cite{kwon2019cosmos}    \\

   \toprule
  \end{tabular}
  
  \begin{tablenotes}
   \scriptsize
   \item[\ding{172}] Driving Chain allows miners from the mainchain to control the sidechain. By involving mainchain miners in the consensus process of the sidechain, this mechanism ensures the security of the sidechain. 
Hybrid model combines characteristics from the aforementioned models to achieve higher security, decentralization, and flexibility. Typically, Hybrid model selects different mechanisms based on specific needs, such as a combination of Consortium with SPV. The security of these three models—Consortium, SPV, and Driving chain—relies on the longest-chain rule \cite{pass2017fruitchains}, so their implementation often requires a soft fork \cite{shahsavari2019theoretical}.

\item[\ding{173}] Sort order (from low to high): Single-Custodian\ding{226}Consortium\ding{226}SPV\ding{226}Driving Chain\ding{226}Hybrid. The first two are relatively straightforward, as they do not require complex cross-chain protocols or smart contract mechanisms. SPV relies on light client verification. The latter two demand deep modifications to blockchain infrastructure, making them more challenging to implement.

\item[\ding{174}] Sort order (from low to high): Single-Custodian\ding{226}Consortium\ding{226}Driving Chain\ding{226}SPV\ding{226}Hybrid. The security of TTP-based models is relatively weak, and in SPV, if the light client is maliciously compromised (e.g., eclipse attack) \cite{alangot2020decentralized}, it could result in faulty cross-chain verification. In Driving Chain, insufficient economic incentives for miners may still pose security risks. Hybrid model, which combines the strengths of other approaches, offers the highest level of security.

  \end{tablenotes} 
 \end{threeparttable}    
 \label{sidechain_mode_table}
\end{table*}

\begin{table*}[!ht]
 \caption{Comparison among Different Sidechains Constructions.}
 \centering
 \scriptsize
 \setlength{\tabcolsep}{11.2pt}
 \renewcommand{\arraystretch}{1.5}
 \label{table1}
 \begin{threeparttable}    
  \begin{tabular}{>{\columncolor{blue!6}}lcccccc}
   \rowcolor{blue!6}
   \bottomrule
\rotatebox{0}{\textsl{\textbf{Schemes}}}  &\multicolumn{3}{c}{\textsl{\textbf{Universality}}}  & \textbf{\textsl{\textbf{PS}}}\tnote{\ding{175}} & \textbf{\textsl{\textbf{CC}}}\tnote{\ding{175}}  & \textbf{\textsl{\textbf{SM}}}  \\
   \cline{2-4}  & \textbf{AR}\tnote{\ding{172}} & \textbf{VD}\tnote{\ding{173}} & \textbf{H}\tnote{\ding{174}} &   &  \\
   \hline
   
   \textbf{BTCRelay} \cite{btcrelay} & PoW Chains & No & $\usym{2717}$  & $\mathcal{O}(C)$  &  $\mathcal{O}(C)$ & SPV \\
   
   \textbf{zkRelay} \cite{westerkamp2020zkrelay}  & PoW Chains & No & $\usym{2717}$  & $\mathcal{O}(C)$  &  $\mathcal{O}(C)$ &  SPV \\
   
   \textbf{ETHRelay} \cite{frauenthaler2020eth} & PoS Chains & No &  $\usym{2717}$& $\mathcal{O}(C)$  &  $\mathcal{O}(C)$   & SPV  \\
   
   \textbf{Drivechains} \cite{drivechain}  & PoW Chains & No &  $\usym{2717}$ &     $\mathcal{O}(C)$  &  $\mathcal{O}(C)$ & Driving Chain    \\
   
   \textbf{SEPoW} \cite{sepow}  & PoW Chains & No & $\usym{2717}$   &  $\mathcal{O}(\mathsf{log}(C))$  &   $\mathcal{O}(\mathsf{log}(C))$  &  Driving Chain  \\  
   
   \textbf{FlyClient} \cite{bunz2020flyclient}  & PoW Chains &  Yes & $\usym{2717}$  & $\mathcal{O}(k\cdot\mathsf{polylog}(C))$ & $\mathcal{O}(k\cdot\mathsf{polylog}(C))$   & SPV\\
   
   \textbf{Txchain} \cite{txchain}  & PoW Chains &  Yes & $\usym{2717}$  & $\mathcal{O}(k\cdot\mathsf{polylog}(C))$ & $\mathcal{O}(k\cdot\mathsf{polylog}(C))$    & SPV \\
   
   \textbf{PoW Sidchains} \cite{kiayias2020proof}  & PoW Chains &   No  & $\usym{2717}$ & $\mathcal{O}(\mathsf{log}(C))$   &  $\mathcal{O}(\mathsf{log}(C))$ & Consortium \\
   
   \textbf{PoS Sidechains} \cite{possidechain}  & PoS Chains & No & $\usym{2717}$ &  $\mathcal{O}(S)$   &   $\mathcal{O}(S)$  & Consortium \\
   
   \textbf{Yin et al.}  \cite{yin2021sidechains}  & PoW or PoS Chains & Yes &  $\usym{2717}$ & $\mathcal{O}(S)$   &   $\mathcal{O}(S)$ & Consortium \\
   
   \textbf{PSSC} \cite{deng2023pssc} & Chains with SNARK & N/A & $\usym{2713}$   &  $\mathcal{O}(1)$  & N/A & Hybrid \\

   \textbf{Zendoo} \cite{garoffolo2020zendoo} & Chains with zk-SNARK &   N/A  &   $\usym{2713}$ &  $\mathcal{O}(1)$  & N/A  & Hybrid  \\
    
   \textbf{Cumulus} \cite{gai2021cumulus} & Chains with customized SC   & N/A &  $\usym{2713}$  &  $\mathcal{O}(1)$  & N/A  & Consortium \\
   
   \textbf{USSC} \cite{li2024ussc}   &  PoW or PoS Chains  & Yes &  $\usym{2713}$ &  $\mathcal{O}(S)$   &   $\mathcal{O}(S)$ & Consortium  \\
   
   \textbf{Ge-Go} \cite{yin2023sidechains} & PoW or PoS Chains & Yes &   $\usym{2717}$ &  $\mathcal{O}(1)$   &   $\mathcal{O}(1)$   & Consortium  \\
   
   \textbf{Glimpse} \cite{scaffino2023glimpse} & PoW Chains & Yes &   $\usym{2713}$ &  $\mathcal{O}(k)$   &   $\mathcal{O}(k)$   & Hybrid \\

   \toprule
  \end{tabular}
  
  \begin{tablenotes}
   \footnotesize
   \item[$\bigstar$] $Abbreviation$. AR: Applicability Range; VD: Variable Difficulties; H: Heterogeneous; PS: Proof Size; CC: Computation Cost; SM: Sidechains Mode.
   \item[\ding{172}] In the parent-child chains form, only the mainchain serves as the study object of applicability range. This is because the sidechain is bootstrapped from the mainchain and has customizability. Additionally, we consider the mainchain with basic payment functionality. Here SC means smart contract.
   \item[\ding{173}] This refers to whether the sidchains construction can be applied to various sidechains with variable difficulties.
    \item[\ding{174}] Which means whether a sidechains construction supports the interoperability between heterogeneous systems; for instance, PoW-based chains communicate with PoS-based chains. 
    \item[\ding{175}] Let $C$ denote the lifetime of the system (informally, the length of the mainchain or sidechain) and $k$ denote the \emph{common prefix} parameter. $S$ is the validation set (or committee) size.

\end{tablenotes} 
\end{threeparttable}    
\label{tablesidechaincompare}
\end{table*}

\textbf{Sidechains Evolution.} The current researches on sidechains primarily focus on three key technical dimensions: universality, performance, and security. We provide a detailed comparison of these research works across key metrics such as universality, proof size, and computational cast, as summarized in Tab. \ref{tablesidechaincompare}.

\emph{Universality.} To enhance the universality of sidechain constructions, several researchers have proposed innovative approaches. Kiayias et al. \cite{kiayias2020proof} introduced a PoW sidechain architecture applicable to blockchain systems using PoW consensus. Similarly, Gaži et al. \cite{possidechain} presented a construction designed for PoS blockchains. Westerkamp et al. \cite{westerkamp2020zkrelay} developed zkRelay, another sidechain framework compatible with PoW systems. Additionally, Yin et al. \cite{yin2021sidechains} proposed two distinct sidechain architectures: one optimized for speed within PoS blockchains and another for efficiency in PoW systems. Compared to earlier studies \cite{drivechain,btcrelay,sztorc2015drivechain}, these approaches expanded the applicability of sidechain solutions, extending beyond specific blockchain systems to support a broader range of blockchain consensus mechanisms. However, these solutions still necessitate forking of the mainchain, introducing potential security vulnerabilities, and face limitations when enabling interoperability between heterogeneous blockchains. 
To further improve universality, Zendoo \cite{garoffolo2020zendoo} employs zk-SNARKs \cite{chen2022review} to enable secure communication between a mainchain and multiple sidechains without relying on trusted intermediaries, making it compatible with various blockchain consensus models. The protocol remains susceptible to security risks despite these advancements due to potential forks. PSSC \cite{deng2023pssc} leverages SNARK technology to construct a general sidechain architecture tailored for IoT environments, which supports heterogeneous chains while maintaining constant storage size. However, the complex script language design poses challenges for practical application.

\emph{Efficiency.} Some researchers have focused on improving the efficiency of sidechains. In systems like BTCRelay \cite{btcrelay}, zkRelay \cite{westerkamp2020zkrelay}, and ETHRelay \cite{frauenthaler2020eth}, cross-chain proofs consist of block headers that increase linearly with the length of the chain, resulting in substantial storage and communication overhead for the nodes in sidechains. To reduce the size of these proofs, Kiayias et al. \cite{kiayias2016proofs} introduced PoPoW, a cryptographic primitive that generates succinct proofs of transactions occurring in PoW blockchains. In this approach, the complexity of the proof is sublinear to the length of the PoW blockchain. To further minimize proof sizes, Kiayias et al. \cite{kiayias2020non} proposed NIPoPoW, which scales logarithmically with the length of the blockchain. However, NIPoPoW is limited to blockchains with fixed block difficulty.
 FlyClient \cite{bunz2020flyclient}, and Txchain \cite{txchain} improved upon NIPoPoW, achieving smaller proof sizes with logarithmic complexity, but these protocols require continuous PoW difficulty or Bitcoin hard forks, making them not backward-compatible.
  Additionally, other works \cite{xie2022zkbridge,teutsch2019retrofitting} leverage zk-SNARKs to reduce the size of the proofs, ensuring that the proof size remains constant, regardless of the length of the blockchain.

\emph{Security.} The security of sidechain constructions largely depends on the level of decentralization and the fulfillment of three key security properties: $persistence$, $liveness$ (Def. \ref{RDL}), and $atomicity$ (Def. \ref{security cross-chain}). Dilley et al. \cite{dilley2016strong} introduced a federated model that facilitates cross-chain asset transfers among various blockchains. This model employs trusted federated boards to manage assets, allowing transfers only when most board members approve, thereby mitigating centralization risks. Kiayias et al. \cite{kiayias2020proof} proposed the first decentralized sidechain framework specifically for PoW blockchains. Similarly, Gaži et al. \cite{possidechain} developed the first formal framework for PoS sidechain constructions, offering rigorous security processing and validation. Other works, like Cosmos \cite{kwon2019cosmos}, Polkadot \cite{wood2016polkadot}, and Liquid \cite{nick2020liquid}, have also made improvements to cross-chain verification. Their validation relies on trusted committees or federations or is left unspecified, lacking formal security definitions.

\textbf{Compared with Light Client.} The primary technical distinction between sidechains and light clients lies in their respective verification targets. Sidechains verify the nodes on either the main or sidechain, while light clients verify transactions for lightweight nodes. A sidechain is an independent chain capable of supporting high-frequency transactions, providing a smart contract execution environment \cite{liangwei}, and performing consensus mechanisms \cite{xu2023survey}. In contrast, a light client serves as a lightweight verification system that does not participate in consensus but merely validates the correctness of transactions.

\begin{figure*}[!ht] 
\small
\centering
\includegraphics[width=5.0in]{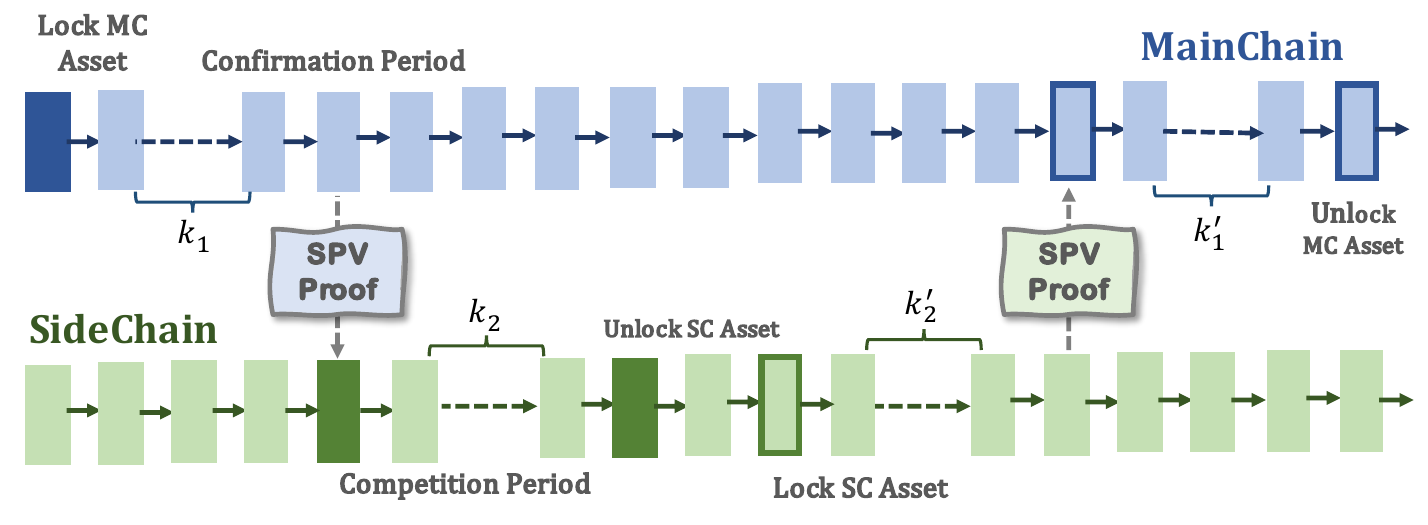} 
   \vspace{-2mm}
\caption{Conventional wrapped assets transfer for SPV sidechains from $\mathscr{S}$ to $\mathscr{T}$ and back again.}
\label{sidechainSPV}
   \vspace{-2mm}
\end{figure*}

However, both technologies share common ground in that they can utilize the lightweight verification mechanism derived from SPV. As illustrated in Fig. \ref{sidechainSPV}, we present a simplified wrapped asset transfer process based on SPV sidechains as an example, which can be simplified as following steps: 
\ding{172} Lock the assets of $\mathscr{S}$;
\ding{173} Wait for a confirmation period on $\mathcal{L}_{\mathscr{S}}$ ($k_{1}$ blocks) to ensure sufficient proof of work, which helps resist DoS attacks;
\ding{174} After the confirmation period, the user creates a minting transaction on $\mathcal{L}_{\mathscr{T}}$ with SPV proof of lock transaction in $\mathscr{S}$. The assets of $\mathscr{T}$ remain locked during a competition period;
\ding{175} During the competition period, which prevents double-spending, other users can provide an updated SPV proof to invalidate the minting transaction of $\mathscr{T}$ if the mainchain assets are moved. This is called a reorganization proof;
\ding{176} After the competition period ($k_{2}$ blocks\footnote{It is possible that $k_{1}\neq k_{2}$ due to differing blockchain parameters, such as variations in block generation time or network synchrony.}), the tokens of $\mathscr{T}$ are minted and can circulate;
\ding{177} To withdraw assets to $\mathcal{L}_{\mathscr{S}}$, and repeat the above steps.

Thus, it becomes evident that all verification mechanisms supporting light clients can be directly or indirectly applied to sidechains, a concept further explained in the Glimpse \cite{scaffino2023glimpse} protocol.

\textbf{Open Issues of Sidechains.}
Sidechains enables the sharing of states between mainchain and sidechain, allowing users to securely lock tokens on one and utilize them on the other chain. This facilitates higher transaction frequency and faster instant transactions on the sidechain \cite{pillai2022cross}. However, frequent token transfers between the mainchain and sidechain introduce additional security risks, particularly with regard to fraudulent transfers. This increases the complexity of interface design and may further lead to resource centralization among miners.
In certain scenarios, cross-chain asset interoperability is typically achieved through the use of wrapped assets, wherein a trusted entity locks the original tokens and issues equivalent wrapped tokens for use on the sidechain. However, this approach poses centralized trust risks, as the operation of wrapped tokens relies on a centralized authority. Moreover, wrapped tokens may face economic challenges due to the following reasons:

\begin{itemize}[itemsep=0.5pt]

\item  \emph{Value Parity.} The system must ensure that the value of the wrapped tokens remains consistent with the original tokens, as any deviation could lead to market instability.
\item \emph{Secure Custody.} The locked original assets must be securely held by the trusted entity. If the custodian fails, there is a risk of theft or loss of the assets.
\item \emph{Exchange Rate Stability.} The exchange rate between the wrapped tokens and the original tokens must remain stable over time; otherwise, users’ trust in the wrapped tokens could be eroded.
\item \emph{Liquidity Pressure.} Users can redeem wrapped tokens at any time, which may create liquidity stress on the custodied asset pool. This risk is heightened during periods of market volatility or mass redemptions \cite{caldarelli2021wrapping}.

\end{itemize}

Moreover, ensuring the persistence and liveness of interactions between mainchains and sidechains in asynchronous network environments presents a significant challenge. In other words, it is essential to guarantee that valid asynchronous cross-chain transactions are executed correctly, eventually recorded on-chain, and confirmed by a sufficient number of subsequent blocks to achieve stability. This introduces new technical difficulties for sidechain construction\cite{yang2024asyncsc}.
For instance, key open challenges include designing cross-chain interaction models that remain robust under asynchronous conditions—especially for resource-constrained nodes with intermittent connectivity—and ensuring the ordering consistency of cross-chain transactions to prevent conflicts and state inconsistencies arising from out-of-order execution.

\subsection{Chain Relay-based Swap Bridges}

To overcome the limitations of the mainchain, some sidechain solutions have evolved into \textbf{chain relay} mechanism. Chain relay combines the strengths of notary schemes and sidechain solutions: on one hand, chain relay adopts the intermediary approach from notary mechanisms, allowing compatibility with diverse heterogeneous chains without modifying $\mathscr{S}$; on the other, by using a third-party chain as an intermediary, chain relay can act as “sidechains” for multiple chains, thereby ensuring decentralization and trust in cross-chain processes.

Fundamentally, chain relay operates through a light client within a smart contract. Off-chain untrusted relayers continuously transfer block headers from $\mathscr{S}$ to $\mathscr{T}$. To prevent malicious relayers from submitting invalid block headers, smart contracts ensure correct relay operations via two safeguards: \ding{172} internal verification of block headers through a partially replicated consensus mechanism of $\mathscr{S}$, and \ding{173} enhanced system stability through fork management.

The concept of relay chains originated with BTC Relay \cite{btcrelay} and has been widely implemented in interoperability protocols. XCLAIM \cite{zamyatin2019xclaim} leverages BTC Relay to achieve trustless atomic swaps between Bitcoin and Ethereum, introducing a cryptocurrency collateral mechanism to enable multi-party asset exchange and redemption requests. Westerkamp et al. \cite{westerkamp2020zkrelay} proposed zkRelay, which supports batch block header processing and uses zkSNARKs for on-chain and off-chain verification, ensuring fixed verification costs. Verilay \cite{westerkamp2022verilay}, the first relay solution for PoS blockchains, deployed on Ethereum 2.0, validates the PoS protocol by generating final blocks and provides methods to retrieve validator public keys. Tesseract \cite{bentov2019tesseract} employs a TEE as a relay to enable secure real-time cryptocurrency exchanges and support cross-chain transactions and asset tokenization.

\textbf{Compared with sidechains.} Some researchers \cite{belchior2021survey,wang2023exploring,ou2022overview} classify chain relay and sidechains together as the coordinating interoperable technologies due to their reliance on light client validation mechanisms. However, we assert that they have fundamental differences: \emph{sidechains are homogeneous extensions of primary chains, while chain relay connects distinct chains (either homogeneous or heterogeneous) to facilitate asset transfer and message exchange}. See Tab. \ref{sidechainsrelaycompare} for further differences.

  \begin{table}[!htbp]
 \caption{Peimary Differences Between Sidechains and Chain Relay.}
 \centering
 \scriptsize
 \setlength{\tabcolsep}{11pt}
 \renewcommand{\arraystretch}{1.5}
 \label{table1}
 \begin{threeparttable}    
  \begin{tabular}{>{\columncolor{blue!6}}p{1.35cm}p{2.6cm}p{2.6cm}}
   \bottomrule
   
\rotatebox{0}{\textsl{\textbf{Benchmark}}}   & \multicolumn{1}{c}{\rotatebox{0}{\textsl{\textbf{Sidechains}}}}   &     \multicolumn{1}{c}{\rotatebox{0}{\textsl{\textbf{Chain Relay}}}}  \\
   \hline

\textbf{Core Functionality}  & Extending Mainchain  &  Cross-Chain Transfer  \\

\textbf{Subordination Relationship}  & Sidechain Subordinate to the Mainchain  & Chains Operate Independently \\

\textbf{Processing Method}  & Synchronize Block Header &  No Need to Synchronize Block Header  \\

\textbf{Transaction Speed}  & Relatively Fast & Relatively Slow \\

\textbf{Security} & Reliant on Mainchain  & Based on Each Chain\\

   \toprule
  \end{tabular}

 \end{threeparttable}    
 \label{sidechainsrelaycompare}
\end{table}

Traditional classifications \cite{ou2022overview} also include distributed private key control, which enhances chain relay security by redundantly distributing private keys to verifiers. Although not an independent cross-chain approach, distributed private key control improves the security of notary and relay chain solutions.

\textbf{Limitations.} Chain relay enables validation of any transaction on $\mathscr{S}$, achieving partial decentralization and atomicity and defending against double-spending attacks, yet they remain vulnerable to MEV attacks and do not provide transaction privacy protection. Additionally, chain relay protocols are costly to operate, with limited cross-chain efficiency, particularly in terms of time. Overall, chain relay solutions have advantages in heterogeneity; however, only a few relay chains are currently operational, and significant node subsidies through incentive mechanisms are required.

\subsection{Miscellaneous Interoperability Solutions}

\subsubsection{\textbf{Rollups}} Rollups \cite{gorzny2024rollup} is a sidechains solution that batches processes transactions from a source chain and executes them on an external chain. It can be categorized into Optimistic Rollups (e.g., Arbitrum \cite{kalodner2018arbitrum} and Optimism \cite{optimism}) and Zero-Knowledge Rollups (e.g., ZkSync \cite{jainexploring} and Loopring \cite{wang2018loopring}), with each differing in trust model and proof mechanism.

\begin{itemize}[itemsep=0.5pt]
\item \emph{Optimistic Rollups} \cite{khalil2024parole} operate under the assumption that all transactions are valid, submitting results directly to the main chain. A "challenge period" mechanism is only triggered when a dispute arises, requiring fraud-proof submission. This model improves processing speed but may introduce delays in cross-chain environments due to the challenge period.

\item \emph{ZK-Rollups} \cite{yamamoto2023examination} generate a ZK proof for each batch of transactions, ensuring data correctness and consistency when submitted to the main chain. ZK-Rollups hold a significant advantage in cross-chain operations, as their instant verification enhances the security and efficiency of cross-chain transfers.
\end{itemize}

Rollups can batch multiple transactions across chains into a single cross-chain operation, reducing fees and increasing data transmission efficiency. Additionally, Rollups' proof generation inherently provides data consistency, minimizing the trust cost for asset migration across chains and reducing potential security risks during cross-chain interactions. Typically, Rollups are funded through native bridges (e.g., Polygon's PoS bridge and zkEVM bridge) \cite{ilisei2024analyzing}, which serve as Layer-2 onboarding pathways for technologies like Starkware \cite{starkware} and ZkSync \cite{jainexploring}. The latter enhances Layer-1 scalability by parallelizing instances of EVM circuit execution.

\subsubsection{\textbf{Burn-and-Mint Style Protocol}} 
In the field of blockchain interoperability, the burn-to-claim protocol proposed by Pillai et al. \cite{pillai2021burn,pillai2020burn} is a notable burn-and-mint mechanism. This protocol facilitates cross-chain asset transfers through a two-step process: locking and burning the asset on $\mathscr{S}$, followed by minting an equivalent asset on $\mathscr{T}$.

The protocol effectively mitigates the risk of double-spending attacks, as assets are irreversibly burned before they are claimed on $\mathscr{T}$. Additionally, it preserves trustlessness, as no TTP is required to manage the transaction. However, the author does not provide proof of the protocol's inherent resistance to Maximal Extractable Value (MEV) attacks \cite{pillai2021burn}. Malicious miners may exploit their control over transaction execution order within a block to gain an unfair advantage, which could undermine the fairness of asset exchange rates.

\emph{Limitations.} While the Burn-to-Claim protocol facilitates cross-chain asset transfers, it exposes transaction details and involves a complex recovery process in the event of a failed transfer, meaning the atomicity of the protocol is not always guaranteed \cite{pillai2023formal}. Moreover, current burn-and-mint designs often rely on APIs or centralized gateways, compromising the principles of decentralized trust and security.

\subsubsection{\textbf{Hierarchical Blockchain}} A hierarchical architecture is a design approach that separates blockchain consensus tasks into multiple layers, aimed at enhancing the performance, scalability, and security of blockchain systems. However, this naturally creates interoperability requirements between the underlying and upper-layer blockchains. 

Some protocols \cite{kogias2016enhancing,abraham2016solida,pass2016hybrid,gilad2017algorand} employ the underlying blockchain using consensus mechanisms like PoW or PoS to prevent double-spending attacks. After selecting a set number of nodes, these nodes undergo identity verification, followed by a chain consensus algorithm to generate an upper-layer blockchain. Other protocols \cite{li2020scalable,wu2025dbpbft,guo2024hierarchical,deng2024distributed} select optimal leaders to serve as committee members for the upper-layer chain based on the underlying committee and reputation mechanisms, making these committees responsible for $\mathcal{CCI}$ tasks (e.g. based on 2PC \cite{chen2024parallel,rahimian2021tokenhook}).

Through coordination of high-level consensus, $\mathcal{CCI}$ between upper-layer and lower-layer blockchains can be more efficient, ensuring global consistency by sharing the underlying consensus. Additionally, state synchronization and information transmission between lower-layers are simplified and secured, preventing incompatibilities between different consensus protocols.

\subsubsection{\textbf{Sharding}} Sharding \cite{luu2016secure} is a scalability technique designed to enable blockchain networks to process more transactions concurrently. The core idea of sharding is to partition nodes into smaller committees (shards, each maintaining a separate chain). Each shard manages a disjoint subset of the overall blockchain state, performing intra-shard consensus independently and processing different transactions in parallel. However, each shard still stores the entire state ledger. To optimize storage efficiency, the authors of OmniLedger \cite{kokoris2018omniledger} proposed \emph{state sharding}, where each shard is responsible for storing and managing only a subset of the ledger data. Several other state-of-the-art state sharding protocols have since been introduced, including Monoxide \cite{wang2019monoxide}, BrokerChain \cite{huang2022brokerchain}, and Pyramid \cite{hong2023prophet}.

\emph{State sharding} is one of the most challenging aspects of implementing sharding. In the context of state sharding, verifying cross-shard transactions becomes particularly complex, as nodes in different shards store distinct portions of the ledger. Thus, mechanisms must be developed to facilitate the transfer of transactions or ledger state exchanges across shards \cite{sonnino2020replay}. Cross-shard transactions involve two or more shards, requiring coordination between them, which introduces the need for interoperability between different chains or ledgers. To handle cross-shard transactions, protocols such as OmniLedger \cite{kokoris2018omniledger}, RapidChain \cite{zamani2018rapidchain}, and ChainSpace \cite{al2017chainspace} use 2PC protocol. Additionally, Monoxide \cite{wang2019monoxide} introduces a relay transaction mechanism to ensure the atomic finality of cross-shard transactions. In these mechanisms, the makespan of cross-shard transactions tends to be higher than that of intra-shard transactions. Furthermore, a high proportion of cross-shard transactions increases the complexity of the sharded blockchain system, potentially degrading system performance. Some of the latest research \cite{huang2024account,lin2024spiralshard} focuses on addressing these challenges.


\section{Representative platforms in Industry}
Interoperability platforms are technological frameworks designed to enable seamless collaboration among diverse applications, devices, and systems.
In the industry, a primary function of interoperability platforms is their ability to support communication across different blockchain protocols and accommodate various data formats. Another key function is standardization. by adhering to industry standards and protocols, these platforms ensure reliability and trust in system interactions. 
Standards like SWIFT in banking \cite{robinson2023global}, OPC UA in industrial big data, and HL7 standards in health IT \cite{cruz2018using} illustrate how interoperability platforms facilitate standardized communication across sectors.     

Given that blockchain interoperability is a crucial practical aspect of the modern decentralized economy, numerous industry platforms provide such services. We present several notable platforms and categorize them broadly into two types: interoperability based on permissionless blockchains and based on permissioned blockchains. As described in Def. \ref{permissionless} and Def. \ref{permissioned}, permissionless and permissioned blockchains serve different purposes and exhibit distinct characteristics, which in turn influence their interoperability requirements.

\subsection{Interoperability for Permissionless Blockchains}

Interoperability among permissionless blockchains typically focuses on the transfer of assets across networks. Mechanisms such as atomic swaps, relays, and sidechains enable the direct exchange of cryptocurrencies between different permissionless blockchains without the need for intermediaries.
Below, we present an in-depth overview of several leading platforms, with a focus on their comparative analysis as shown in Tab. \ref{tab:interledger_polkadot}.

\begin{table*}[!htbp]
\caption{Analysis and Comparison of Different Permissionless Blockchain Interoperability Platforms}
\centering
\scriptsize
\setlength{\tabcolsep}{10pt}
\renewcommand{\arraystretch}{1.5}
\label{tab:interledger_polkadot}
\begin{threeparttable}
\begin{tabular}{>{\columncolor{blue!6}}p{1.8cm}p{1.9cm}p{1.9cm}p{1.9cm}p{1.9cm}p{1.9cm}p{1.9cm}}
\bottomrule

\textsl{\textbf{Benchmark}} & \textsl{\textbf{Interledger}} & \textsl{\textbf{Polkadot}}  &  \textsl{\textbf{LayerZero}} & \textsl{\textbf{RSK}}  &  \textsl{\textbf{HyperService}} & \textsl{\textbf{Cosmos}}  \\
\hline

\textbf{Design Goal} & Cross-ledger payment protocol & Multi-chain interoperability platform & Cross-chain messaging protocol & Bitcoin-based smart contract platform & Cross-chain programmability & Decentralized blockchain interoperability \\
\hline

\textbf{Core Mechanism} & Hash lock, escrow & Parachain, relay chain & Light client, oracle & Sidechain & HSL, cross-chain gateway & IBC, Hub-Zone architecture \\
\hline

\textbf{Consensus Model} & Depends on native ledger consensus & NPoS (Nominated PoS) & Native blockchain consensus & PoW (merged mining) & NSB-based coordination & Tendermint-BFT \\
\hline

\textbf{Interoperation Method} & Connector & XCMP protocol & Cross-chain messaging & Two-way peg lock & Gateway routing & Zone, IBC protocol \\
\hline

\textbf{Main Use Cases} & Cross-ledger payment, distributed payment gateway & Multi-chain dApps, DeFi interoperability & Cross-chain asset/NFT transfer & DeFi, cross-border payment, smart contracts & dApps, NFT transfer & Multi-chain DApps, DeFi, NFT cross-chain \\
\hline

\textbf{Data Transfer Mechanism} & ILP packet transfer & Parachain XCMP messaging & Packet-based payment transfer & Two-way peg escrow account & HyperBridge middleware & IBC cross-chain messaging \\
\hline

\textbf{Cross-chain Capability} & Strong (multi-ledger compatible) & Strong (multi-chain support) & Strong (supports major chains) & Weak (Bitcoin-focused) & Medium & Strong (Zone extensible) \\
\hline

\textbf{Security Mechanism} & Escrow + hash lock & Relay chain shared security & Relay + oracle dual validation & Merged mining with Bitcoin security & NSB-based & Tendermint + IBC verification \\
\hline

\textbf{Execution Speed} & Fast & Efficient & Efficient (packet optimized) & Slow & Fast & Fast \\
\hline

\textbf{Native Token} & None & DOT & None & R-BTC & HSP & ATOM \\
\hline

\textbf{High-level Protocols} & Supported (sharded payment, price query) & Extended via parachains & Supports multi-protocol interaction & EVM-compatible, supports Solidity & EVM/WASM-compatible & Supports IBC and smart contract calls \\

\toprule
\end{tabular}
\end{threeparttable}
\end{table*}

\subsubsection{\textbf{Interledger}}
Interledger protocol (ILP) \cite{interledger} is one of the most classical notary-based interoperability platforms \cite{hope2016interledger} based on Hash-locking. Its primary function is to facilitate the transfer of bundled payments across different payment networks or ledgers. ILP\footnote{\url{https://github.com/interledger}} creates a system that connects transacting parties, enabling two distinct systems to exchange currencies via third-party connectors\footnote{Connectors, which act as intermediaries forwarding ILP data packets between the sender and receiver, can generate revenue through currency conversion fees, subscription charges, or other mechanisms.} without requiring mutual trust.

ILPv4, the simplified version\footnote{\url{https://github.com/interledger/interledger.org-v4}} of ILP, is optimized for routing numerous low-value data packets, commonly referred to as "penny swaps". This version can be integrated with any type of ledger, including those not originally designed for interoperability. Moreover, it is designed to function in conjunction with various higher-level protocols, which implement features ranging from quoting prices to sending larger sums using chunked payments. The precondition for implementing ILP is the concept of escrow. Escrow refers to a process where the sender creates an escrowed transaction, putting assets under conditional hold without transferring ownership. 
Custodial transactions are governed by preimage conditions, which permit any party with knowledge of the condition to confirm or revoke the transaction. The sender may also impose a time lock, preventing any modification or deletion of the transaction during the lock period. Upon expiration of the time lock, the custodial transaction is automatically invalidated.
\begin{figure}[!ht] 
\small
\centering
\includegraphics[width=3.5in]{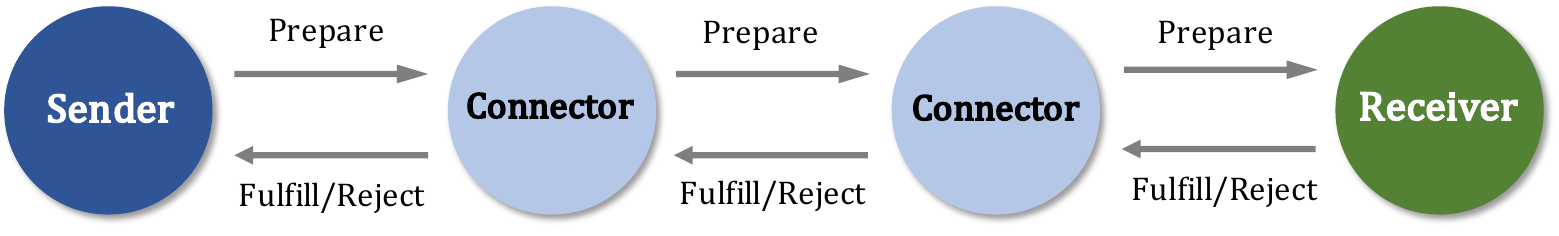} 
   \vspace{-2mm}
\caption{Interledger Multi-Hop transaction schematic.}
\label{interledger fig}
   \vspace{-2mm}
\end{figure}

Let us use Fig. \ref{interledger fig}, along with a simple example, to illustrate the entire process \cite{interledger_process} of an atomic transaction via ILP for detail (Scenario assuming one sender $\mathcal{S}$, one receiver $\mathcal{R}$ and two connectors $\mathcal{C}_{1}$, $\mathcal{C}_{2}$): \ding{172} $\mathcal{S}$ and $\mathcal{R}$ agree on the hashlock $\mathcal{H}$. The preimage $\mathcal{P}$ is only known to $\mathcal{R}$; \ding{173} Sender prepares the transfer to $\mathcal{C}_{1}$ by creating and funding an HTLC on $\mathscr{S}$ with $\mathcal{H}$; \ding{174} $\mathcal{C}_{1}$ prepares a transfer to $\mathcal{C}_{2}$ via their shared payment channel, also using $\mathcal{H}$; \ding{175} $\mathcal{C}_{2}$ prepares a transfer to $\mathcal{R}$ on their shared trustline using $\mathcal{H}$; \ding{176} If $\mathcal{R}$ produces the preimage $\mathcal{P}$ before the transfer timeout, $\mathcal{C}_{2}$ will "pay" $\mathcal{R}$ by increasing his balance on their trustline; \ding{177} If $\mathcal{C}_{2}$ sends $\mathcal{P}$ to $\mathcal{C}_{1}$ before their transfer times out, $\mathcal{C}_{1}$ will send a signed claim to pay $\mathcal{C}_{2}$; \ding{178} If $\mathcal{C}_{1}$ submits $\mathcal{P}$ to $\mathscr{S}$ before the timeout, the transfer will be executed and $\mathcal{S}$ will receive the proof $\mathcal{P}$ that $\mathcal{R}$ was paid.

The protocol fundamentally relies on the formulation of address rules for each account and the definition of a standardized cross-chain messaging format. Transactions are completed exclusively when consensus is reached among all participants. Connectors, functioning as trustless intermediaries, can be operated by any party with access to two or more ledgers, ensuring secure transaction execution.

\subsubsection{\textbf{Cosmos}}

\begin{figure}[!ht] 
\small
\centering
\includegraphics[width=3.4in]{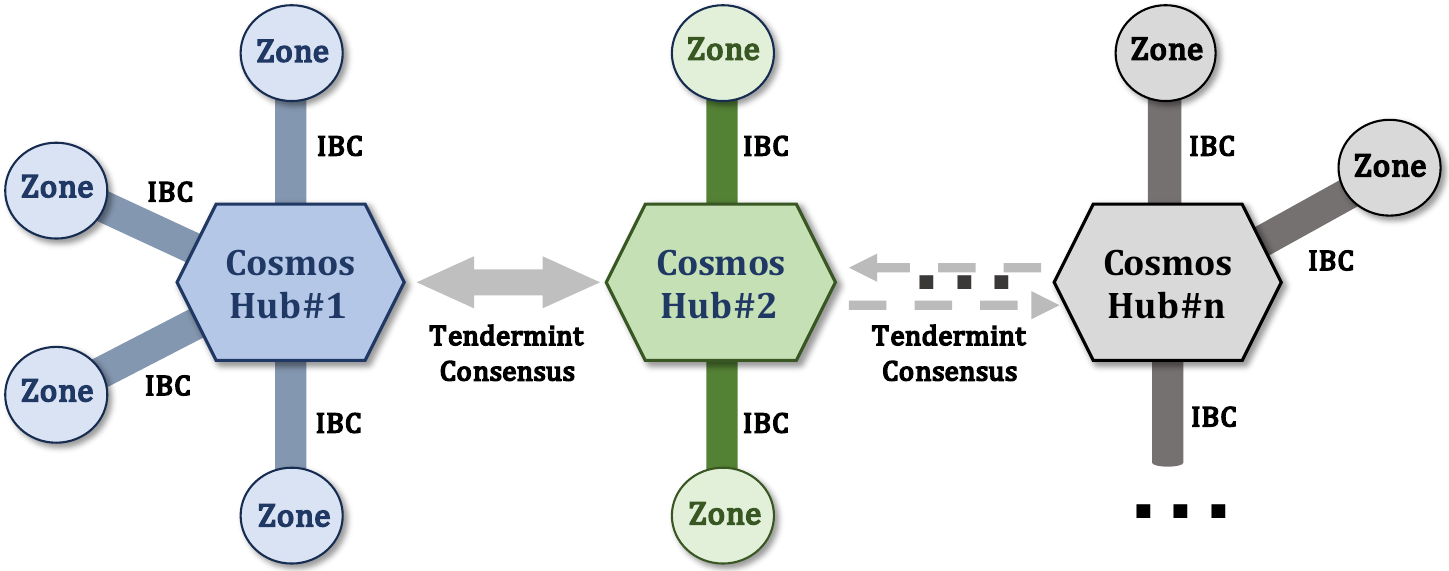} 
   \vspace{-2mm}
\caption{Cosmos architecture.}
\label{cosmos fig}
   \vspace{-2mm}
\end{figure}

The concept of Cosmos\footnote{\url{https://github.com/cosmos/cosmos}} was first introduced by Jae Kwon in 2017 \cite{kwon2018network,kwon2019cosmos}, who is also the founder of Tendermint \cite{kwon2014tendermint}. Kwon proposed two novel concepts in Cosmos: the Hub and the Zone. The Hub functions as a relay chain that handles cross-chain interactions, managing and coordinating communication between blockchains, while the Zones are parallel chains within the Cosmos ecosystem. Together, these Zones form a network of independent blockchains. The architecture of Cosmos is illustrated in Fig. \ref{cosmos fig}.

To facilitate interoperability among parallel chains, Cosmos introduced the Inter-Blockchain Communication (IBC)\footnote{\url{https://github.com/cosmos/ibc-go}} protocol. This protocol supports the transfer of various digital assets, ranging from cryptocurrencies to non-fungible tokens (NFTs), as well as cross-chain smart contracts. The Cosmos SDK\footnote{\url{https://github.com/cosmos/cosmos-sdk}}, by default, utilizes the Tendermint consensus engine \cite{buchman2016tendermint}, a proof-of-stake consensus algorithm, to secure the network. Tendermint’s instant finality enables the transmission of state and data across multiple heterogeneous chains. The Cosmos Hub adopts a decentralized governance mechanism, where network participants can stake ATOM (the native token of the Cosmos Hub) to become consensus validators and earn rewards. The more ATOM staked, the greater the validator’s voting power.

Currently, Cosmos has several application cases, such as serving as a Layer-2 scaling solution for Ethereum. Previously, Ethereum employed the Casper consensus protocol \cite{pititto2022gasper} as a Layer-1 scaling solution, aiming to transition Ethereum to a proof-of-stake (PoS) consensus mechanism. Cosmos, in its design, also made Ethereum-compatible with the Ethereum Virtual Machine (EVM), and its underlying blockchain, which uses a PoS protocol called Tendermint, is referred to as Ethermint.

As the IBC protocol matures and more blockchains join the Cosmos ecosystem, Cosmos gradually realizes its vision of becoming the “Internet of Blockchains”. In the future, Cosmos will further advance applications in DeFi, NFTs, DAOs, and other use cases, promoting the prosperity of the cross-chain ecosystem.

\subsubsection{\textbf{Polkadot}}

\begin{figure}[!ht] 
\small
\centering
\includegraphics[width=3.4in]{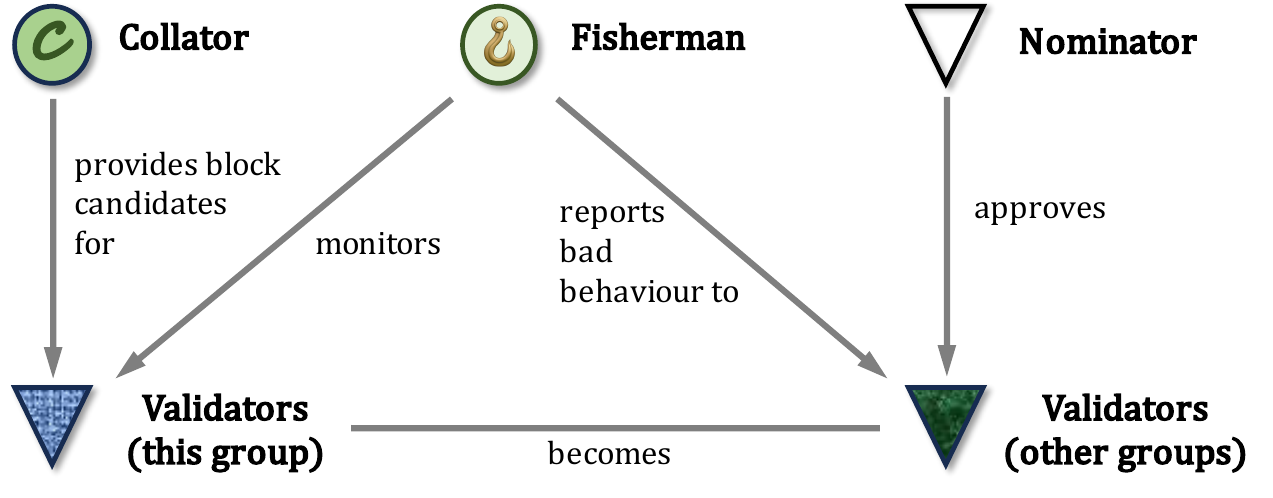} 
   \vspace{-2mm}
\caption{Participating roles of Polkadot.}
\label{polkadot fig}
   \vspace{-2mm}
\end{figure}

Polkadot \cite{wood2016polkadot,burdges2020overview} is a relay-chain network platform\footnote{\url{https://polkadot.com}; \url{https://wiki.polkadot.network/docs/build-guide}; \url{https://wiki.polkadot.network/docs/learn-bridges}} based on interchain protocols, allowing multiple independent chains to run in parallel or connect to other chains, such as Ethereum, through bridging. Polkadot categorizes the nodes in its network into four roles: Collator, Fisherman, Nominator, and Validator on the relay chain. The relationships among these roles are illustrated in Fig. \ref{polkadot fig} and the functions of each role can be expressed as:

\begin{itemize}[itemsep=0.5pt]
\item \emph{Collator} maintains a “full node” of a parallel chain, storing all essential information of the parallel chain and enabling transactions with other nodes on the chain. The primary task of the collator is to organize and execute on-chain transactions and submit them, along with ZK proofs, to the responsible Validators.
\item \emph{Validators} deploy the relay chain client, validating blocks submitted by collators, approving blocks produced by parallel chains, and executing the relay chain’s consensus before packaging blocks onto the chain.
\item \emph{Nominators} represent a group with staking interests. Their main responsibility is to select trustworthy validators and stake their assets with these validators. Validators are elected by nominators. By staking their assets, nominators trust the elected validators to maintain the network, and they receive rewards or penalties proportionate to those of the validators.
\item \emph{Fishermen} do not participate in block production on the relay or parallel chains. Their role is to monitor and report any malicious behavior by the other participants, earning a one-time reward for successful detection.
\end{itemize}

In the Polkadot network, cross-chain transactions are facilitated by a queuing mechanism, where the Merkle tree structure plays a critical role in ensuring data integrity. Transactions are routed from the exit queue of $\mathscr{S}$, through the relay chain, and into the entry queue of $\mathscr{T}$, with the relay chain maintaining records of the relayed transactions. The relay chain manages the queues and guarantees the atomicity of transactions. If any issues arise at any point in the process, the entire transaction is invalidated, and the relay chain assumes responsibility for validating and executing the transaction.
 When a cross-chain transaction is needed, $\mathscr{S}$ places the cross-chain transaction in its output queue alongside other transactions. The collator of $\mathscr{S}$ identifies the cross-chain transaction, packages it, and sends it to the validators, with Fishermen monitoring its legitimacy. Once validated, the cross-chain transaction is placed into the input queue of $\mathscr{T}$ and referenced in the relay chain. Finally, $\mathscr{T}$ executes the transactions in its input queue.

A key distinction between Cosmos and Polkadot pertains to the sovereignty of parallel chains, which varies significantly between the two network architectures. In Cosmos, parallel chains maintain autonomy in consensus, whereas Polkadot requires parallel chains to achieve global consensus through the relay chain to ensure shared security.

\subsubsection{\textbf{HyperService}}
HyperService\footnote{\url{https://github.com/HyperService-Consortium}} \cite{liu2019hyperservice} is the first platform designed to build and execute programmable dApps on heterogeneous blockchains. From a macro perspective, HyperService is built upon two key innovations: a programming framework for developers to write cross-chain dApps and a cryptographic protocol for securely implementing these dApps on blockchain networks. The programming framework introduces the Unified State Model (USM), a blockchain-agnostic and scalable model designed to describe cross-chain dApps. Additionally, HyperService introduces HSL, a high-level programming language tailored to the USM model for writing cross-chain dApps. These dApps, written in HSL, are then compiled into HyperService executable files and executed by the underlying cryptographic protocol.
\begin{figure}[!ht] 
\small
\centering
\includegraphics[width=3.5in]{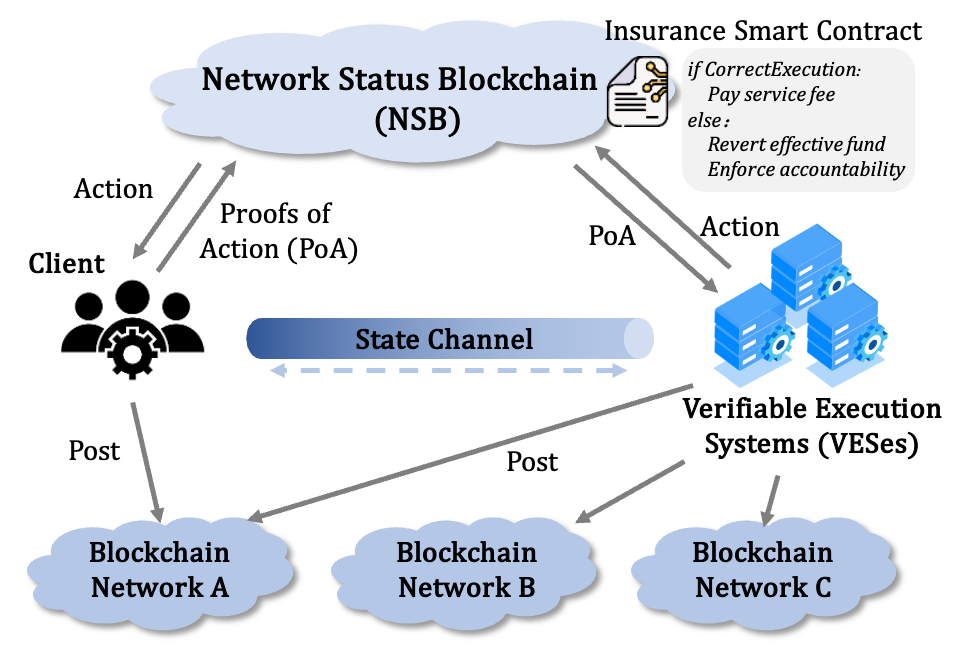} 
   \vspace{-2mm}
\caption{The architecture of HyperService.}
\label{hyperservice fig}
   \vspace{-2mm}
\end{figure}

As shown in Fig. \ref{hyperservice fig}, the HyperService architecture comprises three main components:

\begin{itemize}[itemsep=0.5pt]
\item \emph{Verifiable Execution Systemes (VESes):} Conceptually acting as a blockchain driver, the VESes compile high-level dApp programs provided by the client into executable blockchain transactions, which are runtime executables on HyperService. 
\item \emph{Network State Blockchain (NSB):} Designed as a "blockchain of blockchain", the NSB offers an objective and unified view of the dApp’s execution state. 
\item \emph{Insurance Smart Contract (ISC):} ISC arbitrates the correctness or violations of dApp executions, based on information provided by the NSB, without relying on trust. In cases of anomalies, the ISC rolls back all executed transactions, ensuring financial atomicity and holding malicious actors accountable.
\end{itemize}

HyperService introduces a groundbreaking paradigm for interoperability, streamlining the complexities of dApp development while ensuring atomicity and consistency in cross-chain operations within a secure, trustless environment. This platform holds profound significance for the future of blockchain applications, particularly in DeFi and cross-chain smart contract execution, where it paves the way for more seamless and secure interactions across diverse blockchain ecosystems.

\subsubsection{\textbf{LayerZero}}

\begin{figure}[!ht] 
\small
\centering
\includegraphics[width=3.5in]{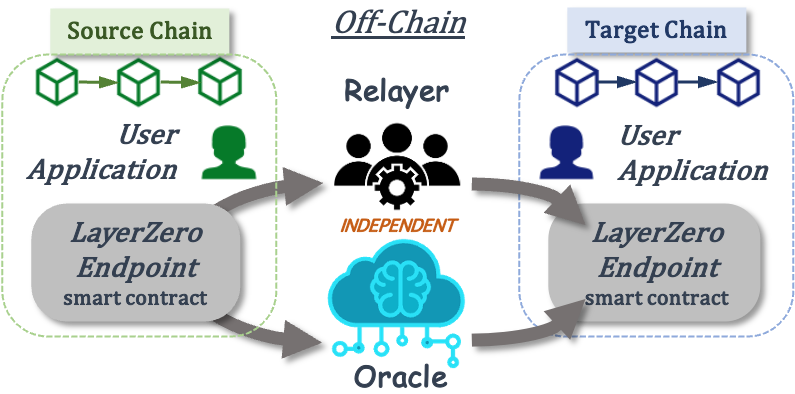} 
   \vspace{-2mm}
\caption{LayerZero upholds the integrity of $\mathcal{CCI}$ by mandating the corroboration of each transaction by two distinct entities \emph{Oracle} and \emph{Relayer}, ensuring its validity.}
\label{layerzerofig}
   \vspace{-2mm} 
\end{figure}

LayerZero\footnote{\url{https://github.com/layerzero-Labs}} \cite{zarick2021layerzero, layerzeroweb} is a decentralized cross-chain communication protocol that facilitates data transfer through endpoints on the source and target chains, along with an off-chain infrastructure consisting of oracles and relayers, as illustrated in Fig. \ref{layerzerofig}. This architecture allows LayerZero to transmit messages and state information, making it an efficient cross-chain bridging solution.

\begin{itemize}[itemsep=0.5pt]
\item \emph{Endpoints:} These are the foundational components of LayerZero on the blockchain, responsible for transmitting messages between different chains. They manage the reception and dispatch of on-chain data, ensuring that on-chain activities can seamlessly interact with the off-chain transmission mechanisms.

\item \emph{Oracle and Relayer:} The distinctive aspect of LayerZero lies in its use of two independent off-chain roles to achieve state synchronization. The oracle is tasked with retrieving data from $\mathscr{S}$ and transmitting it to $\mathscr{T}$, while the relayer is responsible for verifying the data's validity. This design separates data acquisition from verification, enhancing the protocol’s security.
\end{itemize}

Unlike trust-minimized native verification protocols, such as Polkadot’s XCMP and Cosmos’ IBC, LayerZero adopts a novel trust assumption: the oracle and relayer are assumed not to collude. This design increases the flexibility and efficiency of cross-chain communication, allowing developers to select a security model that balances trust assumptions with performance costs. However, it also implies a partial reliance on external verification, somewhat undermining the system’s trust-minimization properties.

LayerZero's trust model involves a partial reliance on the relayer, as it employs an implicit, on-demand state synchronization mechanism rather than the traditional explicit block header synchronization. While implicit synchronization is less costly, it necessitates a trade-off between trust and performance. By default, Chainlink is chosen as the oracle provider, with LayerZero itself serving as the relayer provider. Although these services can be substituted with user-defined solutions, the system inherently relies to some extent on social trust. Despite the associated trust risks, LayerZero’s flexibility and efficiency present significant potential for applications in $\mathcal{CCI}$.

\subsubsection{\textbf{RSK}}
RSK\footnote{\url{https://rootstock.io/}} (Rootstock) \cite{lerner2022rsk} is a sidechain platform enabling Bitcoin blockchain interoperability through a two-way peg mechanism. Its goal is to expand Bitcoin’s functionality to support smart contracts and dApps. RSK achieves these objectives through key technologies such as smart contracts, the two-way peg, and merged mining.

One of RSK's core features is its smart contract capability. RSK is compatible with the Ethereum Virtual Machine (EVM), allowing developers to use Ethereum’s development tools (such as the Solidity programming language) to build and deploy smart contracts on the RSK platform. This compatibility enables Ethereum applications and smart contracts to be ported to RSK, thereby expanding Bitcoin’s use cases.

A crucial technology behind RSK is merged mining, a concept first referenced in the well-known sharding consensus system Monoxide \cite{wang2019monoxide}. Merged mining\footnote{\url{https://github.com/rsksmart}} allows miners to utilize their computational power to secure both the Bitcoin and RSK networks by publishing blocks on RSK and earning additional fees with minimal extra cost. This process aligns RSK with the Bitcoin network, ensuring that RSK inherits the security of Bitcoin’s computational power.

\begin{figure}[!ht] 
\small
\centering
\includegraphics[width=3.5in]{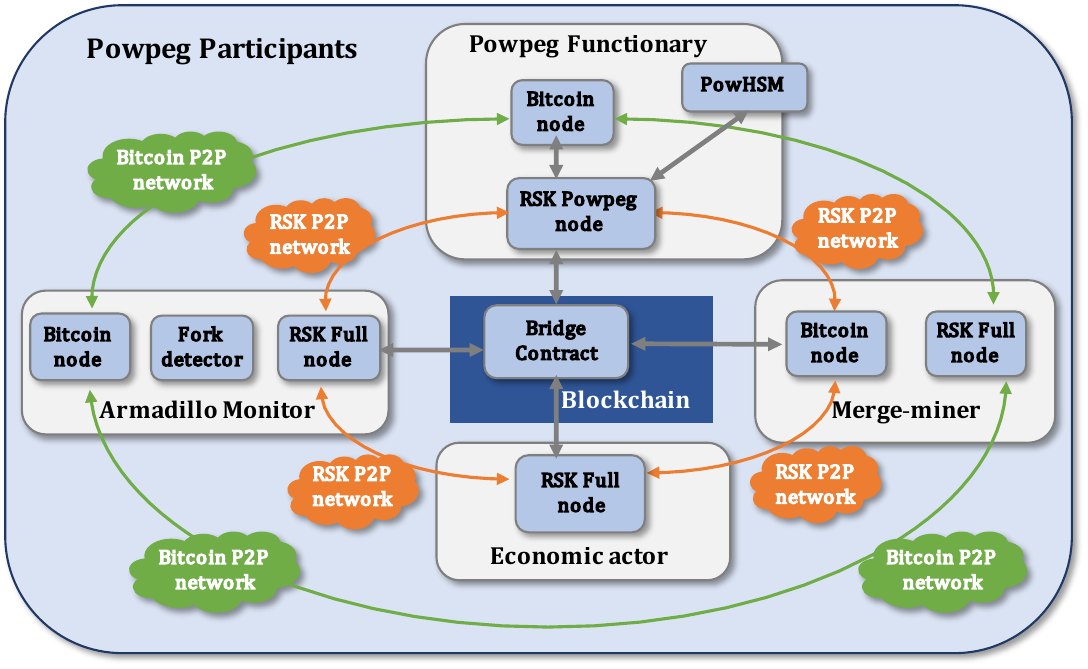} 
   \vspace{-2mm}
\caption{RSK architecture.}
\label{RSKfig}
   \vspace{-2mm}
\end{figure}

When users wish to transfer Bitcoin to RSK, they send BTC to a special multi-signature address, where the BTC is locked on the Bitcoin. In return, an equivalent amount of RBTC is generated on RSK. Users can then use RBTC for transactions or to execute smart contracts on RSK. When users want to convert RBTC back into Bitcoin, they can destroy the RBTC through the two-way peg mechanism, and the corresponding BTC will be unlocked from the multi-signature address on the Bitcoin mainchain. To ensure the security of this process, RSK employs three key components: the Bridge smart contract, Pegnatories, and the Armadillo monitor. Fig. \ref{RSKfig} illustrates the different components of the RSK architecture.

Despite offering smart contract functionality similar to Ethereum, RSK's ecosystem remains relatively small and faces strong competition from platforms such as Ethereum \cite{ethereum}, Polkadot \cite{wood2016polkadot}, and Cosmos \cite{kwon2019cosmos}. Additionally, the incentive structure for merge mining must be sufficiently attractive to miners, as inadequate participation could compromise the security of the RSK network.

 \subsection{Interoperability for Permissioned Blockchain}
 
 Permissioned blockchains restrict access to specific participants and are often utilized by enterprises for internal operations. These blockchains are designed to provide enhanced control, privacy, and efficiency. The interoperability of which primarily emphasizes the integration of various enterprise systems and applications. By implementing standardized protocols and data formats, different permissioned blockchains can effectively understand and process data seamlessly.
 
\textbf{Cactus}\footnote{\url{https://github.com/opentaps/cactus}} \cite{cactuswhite} is an open-source $\mathcal{CCI}$ platform designed to streamline communication between enterprise blockchains. Utilizing a flexible plugin-based architecture, it enables custom connectors for specific blockchain networks. Cactus interacts with different blockchains through "ledger connectors" and "validators", ensuring secure and consistent data transfer. It supports multiple consensus algorithms and multi-signature mechanisms, enhancing cross-chain security. The platform's modular design facilitates seamless integration with diverse permissioned blockchain systems.

\textbf{WeCross}\footnote{\url{https://github.com/WeBankBlockchain/WeCross}} \cite{wecrosswhite}, developed by WeBank, is an open-source cross-chain platform focused on efficient interoperability between consortium blockchains. It employs a four-component architecture (Zone, Router, Stub, and Resource) to manage flexible cross-chain connections and data exchange. WeCross supports 2PC and HTLC to ensure atomic and irreversible cross-chain transactions. The platform is compatible with major consortium blockchains like Hyperledger Fabric \cite{androulaki2018hyperledger} and FISCO BCOS \cite{li2023fisco}, enabling asset swaps and data access control. Using a smart contract-based framework, WeCross ensures transparency and security in cross-chain operations.

\textbf{FireFly}\footnote{\url{https://github.com/hyperledger/firefly}} \cite{firefly} is a blockchain interoperability platform tailored for enterprise use, focusing on the integration of both on-chain and off-chain data. Utilizing a microservices architecture, it creates a modular and scalable ecosystem that simplifies the management of smart contracts, digital assets, and external data sources. Unlike Cactus, FireFly emphasizes application development, offering a comprehensive software development kit for building DApps across multiple blockchains.

\textbf{Weaver}\footnote{\url{https://github.com/hyperledger-labs/weaver-dlt-interoperability}} \cite{weaver}, an open-source project under Hyperledger Labs, provides a $\mathcal{CCI}$ framework without relying on trusted intermediaries. It utilizes a "relay" and "driver" architecture, coordinated through smart contracts to synchronize states across blockchains. Weaver prioritizes compatibility with existing blockchain systems, avoiding modifications to underlying protocols, and employs a decentralized identity platform to protect user privacy and security. The platform supports cross-chain data sharing, asset transfers, and state validation, making it suitable for multi-permissioned blockchain environments.

\textbf{Cacti}\footnote{\url{https://github.com/hyperledger-cacti/cacti}} \cite{cacti} is a versatile interoperability platform that leverages the advanced technical capabilities of Cactus \cite{cactuswhite} and Weaver \cite{weaver}, a project from Hyperledger Labs. It offers a seamless integration path for users of both platforms. Unlike traditional approaches, Cacti does not force separate blockchain networks to merge into a single overarching chain. Nor does it require the creation of a new settlement chain or consensus protocol that other networks must adopt. Instead, Cacti enables independent networks to retain their decision-making autonomy while facilitating cross-network transactions as needed.

\section{Prospective and Intersecting Field Research Avenues}

In this section, we analyze several prospective and intersecting field research approaches to offer scholars new perspectives in cross-disciplinary areas.

\subsection{Redactable Blockchain with Interoperability}

The concept of redactable blockchain was initially proposed by Giuseppe et al. \cite{ateniese2017redactable}, with the aim of enabling controlled modifications to on-chain data, effectively overcoming the limitations of blockchain's immutability and providing a more flexible data storage paradigm. This includes scenarios such as removing inappropriate content, enhancing storage scalability, and complying with the "right to be forgotten" laws \cite{ye2023survey}. These solutions target issues like deleting illicit content for digital currency like Bitcoin \cite{deuber2019redactable}, revising vulnerable smart contracts and on-chain states for Ethereum \cite{tian2023accountable}, and editing on-chain data for the permissioned blockchain paltform\cite{manevich2021redacting}.

However, existing redactable blockchain proposals remain underdeveloped, with most efforts focusing on designing redaction policies for individual chains. These approaches face limitations when applied in $multi$-$chain$ environments. In particular, when dealing with rewriting across multiple heterogeneous or homogeneous blockchains, identifying and redacting relevant transactions becomes a critical challenge. If a cross-chain transaction is modified, the on-chain states of different blockchains are interdependent, meaning that rewriting a specific block or transaction in one blockchain may have direct or indirect impacts on the states of others. Consequently, the corresponding on-chain states must be redacted accordingly to maintain data consistency across blockchains. Failure to do so could lead to inconsistencies in dApps that span multiple chains, posing challenges in maintaining redaction consistency.

\begin{figure}[!ht] 
\small
\centering
\includegraphics[width=3.5in]{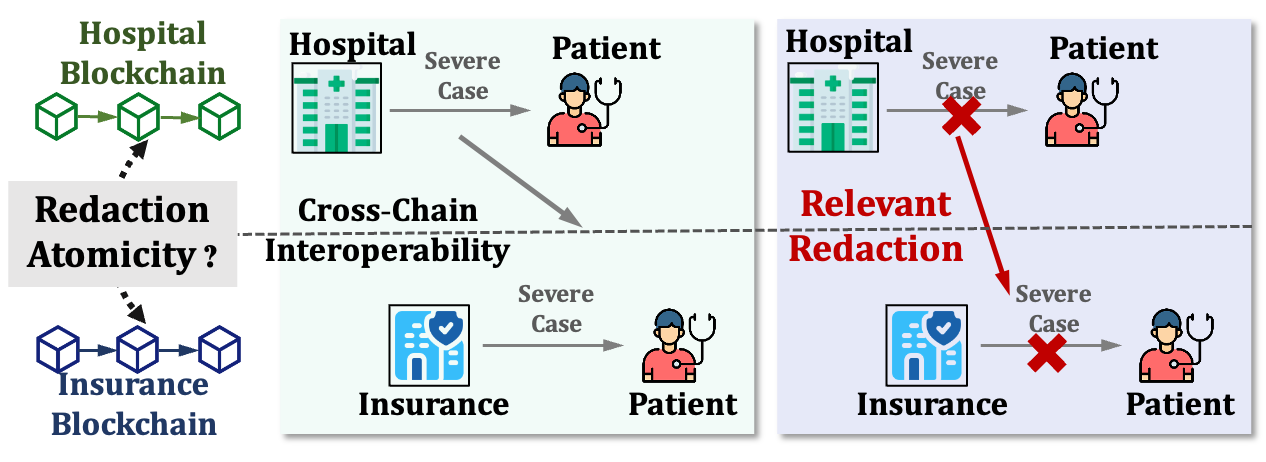} 
   \vspace{-2mm}
\caption{A linkage scene: Redaction cross two different blockchains.}
\label{redactionfig}
   \vspace{-2mm}
\end{figure}

\textbf{A Linkage Scene.} To illustrate the practical significance of combining redactable blockchain and $\mathcal{CCI}$, consider the following scenario involving two independent blockchains, as shown in Fig. \ref{redactionfig}: one manages medical records, while the other handles health insurance services. The hospital blockchain specializes in managing patient health records and treatment data, while the insurance blockchain deals with policy applications, claims, and related transactions. Suppose a physician in a medical institution incorrectly diagnoses a patient with a severe illness and uploads the erroneous information to the medical records blockchain. However, the physician fails to detect and correct the error in time. Subsequently, the patient submits an application for high-risk insurance coverage and compensation to the insurance blockchain based on the faulty medical records. Since the insurance system relies on data from the hospital blockchain, the policy and payout are approved. Without the support of redactable blockchain and interoperability technology, both the misdiagnosis record and erroneous insurance application would be permanently stored on both blockchains. This could lead to data bloat, impact the patient's future medical and insurance records, and potentially cause financial losses for the insurance company. Therefore, an appropriate cross-chain rewriting method is needed to remove the erroneous record from the hospital blockchain, followed by rewriting the related transactions on the insurance blockchain to synchronize the correction.

Another crucial consideration is maintaining atomicity during cross-chain redaction. If a transaction on a specific chain is fully redacted, resulting in a state change, any cross-chain transactions dependent on the previous state must be rewritten accordingly. Alternatively, if rewriting these cross-chain transactions is not possible, the states associated with the redacted transaction must be reverted. Ensuring atomic redaction presents a significant challenge.

Research in this area remains limited \cite{hu2023ivyredaction,du2024starcross}. Although Hu et al. \cite{hu2023ivyredaction} proposed the LvyRedaction, which can achieve atomicity and consistency in cross-chain editing, it requires middleware support and is limited to permissioned blockchains. In the future, a unified and robust solution will be essential to support the editing of transactions across different blockchains. This solution must include mechanisms for monitoring editing transactions, generating editing suggestions, and verifying editing proposals to ensure the atomicity, consistency, and auditability of the redaction and interoperability processes.

\subsection{Asynchronous Consensus with Interoperability}

Most current cross-chain technologies are based on network time assumptions to achieve \emph{global time}\footnote{The state evolution of two distinct blockchains may progress at different $time$ intervals. So a clock $\vartheta$ maps a given epoch on any ledger to the time on a global \cite{zamyatin2021sok} synchronized clock $\vartheta: s \rightarrow t$.} synchronization. 
Synchronous networks rely on the assumption that all messages are received within a specified time limit, denoted as $\Delta$, whereas partially synchronous networks function without a time constraint until a Global Stabilization Time (GST) event occurs, after which messages must be received within $\Delta$ \cite{kokoris2018omniledger}.
 However, these time-based assumptions lack robustness. As blockchain systems scale up, the workload for consensus increases, potentially preventing nodes from reaching global consistency. In addition, it may also destabilize the entire cross-chain system, ultimately leading to protocol failures.

To address these challenges, there is a need to explore interoperability technologies \cite{zhang2022arc,xie2024rac} that adopt asynchronous consensus. Sidechains, relay chains, and sharding depend on their respective consensus mechanisms, while the advantage of asynchronous consensus is that it does not rely on network performance for its protocol design. 
In 2001, Cachin et al. \cite{cachin2001secure} introduced the first asynchronous Byzantine atomic broadcast protocol, CKPS01. In 2016, Miller et al. \cite{miller2016honey} presented the first practically applicable asynchronous consensus protocol for blockchain environments—HoneyBadgerBFT. BEAT \cite{duan2018beat} employs a modular design to reduce consensus latency and improve throughput. Dumbo \cite{guo2020dumbo} is the first fully practical asynchronous BFT consensus protocol, which enhances HoneyBadger using provably reliable broadcast and multi-value Byzantine agreement.
\begin{figure}[!ht] 
\small
\centering
\includegraphics[width=3.5in]{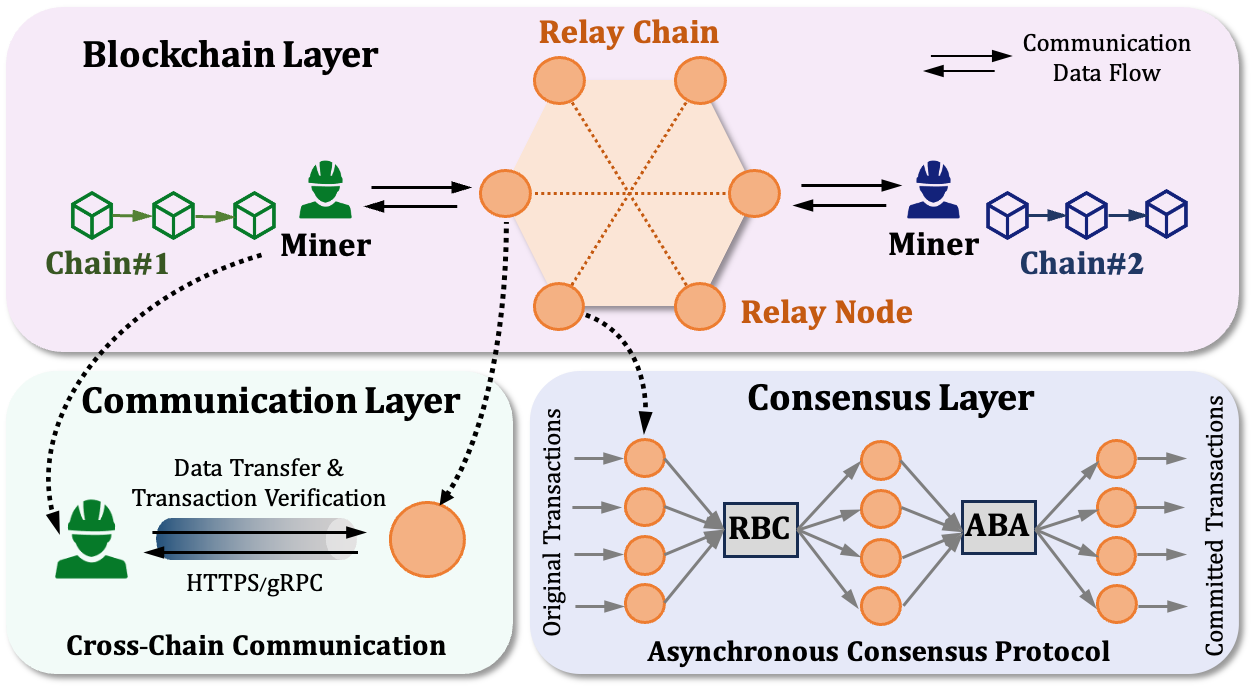} 
   \vspace{-4mm}
\caption{Potential case: an interoperability framework based on asynchronous consensus.}
\label{asyfig}
   \vspace{-2mm}
\end{figure} 

As illustrated in Fig. \ref{asyfig}, we propose a universal cross-chain framework based on asynchronous consensus and relay chain, with asynchronous consensus as the core component. Future research will focus on extending this framework to more protocols, incorporating asynchronous consensus theory and structure into cross-chain technology to further enhance system interoperability and scalability.

\subsection{Growing Web3 \& Metaverses Through Interoperability}


The underlying data storage of the metaverse and web3 relies on blockchain technology, and its development is intrinsically linked to interoperability \cite{yang2024interoperability}. This involves connecting different virtual worlds and allowing users to move seamlessly between them while maintaining ownership and functionality of digital assets. In the metaverse, interoperability means that users can transfer their digital identity, assets, and experiences across various virtual platforms without losing functionality or ownership \cite{bokolo2022exploring}.

To achieve cross-metaverse interoperability, two fundamental components \cite{li2023metaopera} must be seamlessly connected:

\begin{itemize}[itemsep=0.5pt]

\item  \emph{Identity.} In the metaverse, identity establishes the uniqueness of users and digital assets and is fundamental for linking user behaviors with assets. To enable interoperability, standardized identity markers are needed across different metaverses, encompassing user identities, assets, currencies, items, and their transfers. Although users may create multiple identities across various metaverses, each identity should be managed independently, similar to maintaining separate accounts on different social media platforms to safeguard privacy and retain control.

\item \emph{Objects.} Objects in the metaverse include digital assets, avatars, and interactive entities, each characterized by unique attributes such as gender, material, rendering effects, and functionality. These attributes can be generated through technologies like 3D and digital twins (DT) scanning or created directly by users. To support interoperability, objects with the same identity should maintain consistent attributes across metaverses, ensuring a seamless user experience and facilitating functionality and interaction across diverse environments.

\end{itemize}

Based on this foundation, a set of universal standards needs to be established to regulate cross-platform asset, identity, and data handling. Currently, various platforms use proprietary formats for data storage, leading to assets in one environment often being incompatible with others. Therefore, establishing universal standards will enhance asset compatibility between different virtual spaces.

\begin{itemize}[itemsep=0.5pt]

\item  \emph{Token Standards.} Standards such as ERC-721 \cite{cabot2022improving} and ERC-1155 \cite{yang2023non} lay the groundwork for the transfer and recognition of NFTs across different platforms. These standards allow digital assets to be recognized and utilized across multiple dApps within the same blockchain network, thus enhancing interoperability.

\item \emph{Cross-Platform Protocols.} There is a need to develop cross-platform protocols to define the methods of data exchange between different virtual environments. This may include standardized avatar formats, item specifications, and transaction mechanisms to ensure assets maintain consistent appearances and functionalities across various virtual worlds, providing users with a seamless experience.

\end{itemize}

Enhancing blockchain interoperability is fundamental to achieving interconnectedness in the metaverse, as it provides a decentralized and secure foundation for asset transfers, identity verification, and data exchange between virtual environments. By establishing standardized blockchain protocols, various metaverse platforms can facilitate interoperability, enabling users to freely access and trade digital assets. This capability fosters richer and more coherent experiences, ultimately contributing to a more integrated metaverse ecosystem. Through robust interoperability mechanisms, users can seamlessly navigate different virtual worlds while retaining ownership and functionality of their digital assets.

\section{Current Challenges}

The blockchain field faces numerous persistent challenges. Based on a review of interoperability literature from 2023 onward \cite{wang2023exploring,ren2023interoperability,belchior2023brief,augusto2024sok,li2024blockchain}, these challenges can be broadly categorized into four main areas: infrastructure, security, privacy, and scalability. We summarize these categories in Tab. \ref{challengestab} and provide a detailed breakdown of specific challenges within each category. Certain challenges, such as privacy protection in asset swaps and the prevention of double-spending, have been relatively well-addressed through existing technologies. However, issues like script compatibility, network interconnectivity, and architectural compatibility remain in the early stages of research and exploration, requiring further technological advancements.
Building on this categorization, we analyze current interoperability challenges from three broader perspectives: trustlessness, regulatory compliance, and knowledge frameworks. This broader analysis offers a multi-dimensional perspective on the challenges of interoperability. For a deeper understanding of specific challenges in privacy and security, readers are encouraged to consult \cite{augusto2024sok}.


  \begin{table*}[!htbp]
 \caption{Key Challenges Highlighted in Interoperability Reviews Since 2023}
 \centering
 \scriptsize
 \setlength{\tabcolsep}{11.2pt}
 \renewcommand{\arraystretch}{1.5}
 \label{table1}
 \begin{threeparttable}    
  \begin{tabular}{lccccccccc}
   \rowcolor{blue!6}
   \bottomrule
   
\rotatebox{0}{\textsl{\textbf{Category}}}  & \rotatebox{0}{\textsl{\textbf{Challenge}}}  &    \rotatebox{0}{\textsl{\textbf{Description}}} &  \rotatebox{0}{\textsl{\textbf{Progress}}}  &  \rotatebox{0}{\textsl{\textbf{Reference}}} \\
   \hline

\multirow{3}{*}{\rotatebox{60}{\textbf{Infrastructure}}} & Interoperable Architecture & e.g., Universal Modules, Booting Approaches, etc.  & \ding{119}  &  \cite{wang2023exploring}\\
 & Scripting Language & Need for Simpler and More Compatible Scripting Languages  & \ding{119}  &  \cite{ren2023interoperability}\\
  & Cryptographic Primitives & e.g. Cryptographic Sources Used in Adapter Signatures  & \ding{119}  &  \cite{belchior2023brief,wang2023exploring,ren2023interoperability}\\
&  Network Compatibility & e.g. More Universal API, Heterogeneous Networks in Sidechains  & \ding{109}  &  \cite{belchior2023brief}\\
   \hline

\multirow{3}{*}{\rotatebox{30}{\textbf{Security}}} & Monitoring & Enhanced Compliance and Legality Standards  & \ding{119}  &  \cite{augusto2024sok,belchior2023brief}\\
 & Trustless & Reliability in Synchronous Modes that Do Not Rely on TTP   & \ding{119}  &  \cite{li2024blockchain,ren2023interoperability}\\
 & Double-Spending Attack & Involve the Reliability of Consensus in $\mathscr{S}$ or $\mathscr{T}$  & \ding{108}  &  \cite{ren2023interoperability}\\
& Smart Contract Security & Contract Vulnerability Remediation, Governance, and Upgrades  & \ding{119}  &  \cite{li2024blockchain,augusto2024sok,ren2023interoperability,wang2023exploring}\\
   \hline

   \multirow{3}{*}{\rotatebox{30}{\textbf{Privacy}}} & Asset Swaps & Non-Distinguishability and Non-Linkability  & \ding{108}  &  \cite{belchior2023brief,augusto2024sok}\\
 & Asset Transfers & Anonymity and Confidentiality  & \ding{119}  &  \cite{belchior2023brief,augusto2024sok}\\
& Data Transfers & Unbreakability of Ciphertext and Security of Key Agreement  & \ding{109}  &  \cite{belchior2023brief,augusto2024sok}\\
&  Heterogeneous Systems  & Unlinkability and Anonymity Across Different Ledgers  & \ding{109}  &  \cite{ren2023interoperability}\\
   \hline

   \multirow{2}{*}{\rotatebox{30}{\textbf{Scalability}}} & Sharding & e.g. Coordinating State Sharding, Reducing Attacks on one Shard  & \ding{108}  &  \cite{wang2023exploring}\\
 & Layer-2: Rollups & Complex Contracts and Inefficiency of ZK Proofs  & \ding{119}  &  \cite{li2024blockchain}\\

   \toprule
  \end{tabular}
  
  \begin{tablenotes}
   \footnotesize
   \item[$\bigstar$] $Symbol$. \ding{108} - mostly addressed; \ding{119} - partially addressed; \ding{109} - unresolved or insufficiently addressed.
   
  \end{tablenotes} 
 \end{threeparttable}    
 \label{challengestab}
\end{table*}

\subsection{Trust Model Discrepancies} 
Blockchain networks generally function on varying trust models and security protocols. Bridging the differences in trust models across networks while preserving security and decentralization is a challenging task. Achieving this requires a thoughtful approach to consensus mechanisms, cryptographic methods, governance structures, etc.
Trustless cross-chain transactions aim to minimize dependence on third-party verifiers or intermediaries, typically employing mechanisms such as state channels \cite{negka2021blockchain} and hash-locked transfers \cite{hardjono2021blockchain}. These technologies require precise design and rigorous testing to ensure secure asset transfers across different blockchains without introducing vulnerabilities. 

\subsection{Regulatory Concerns}  
Blockchain interoperability encounters considerable regulatory and legal obstacles, especially in the context of cross-border transactions and data sharing. Issues such as regulatory ambiguity, compliance demands, and jurisdictional challenges can impede the widespread adoption of $\mathcal{CCI}$ solutions. Collaboration between industry participants, policymakers, and regulatory authorities is essential for creating well-defined frameworks and standards.
For instance, data privacy regulations vary by region, such as the EU’s General Data Protection Regulation (GDPR) \cite{voigt2017eu} and the California Consumer Privacy Act (CCPA) \cite{goldman2020introduction} in the U.S., making compliance in cross-chain data sharing highly complex. 
Non-compliance with data privacy laws can lead to severe legal and financial repercussions. Another challenge involves jurisdiction: as data and transactions flow across blockchain networks in different countries, determining the applicable legal framework is not straightforward. Blockchain’s decentralized nature further complicates this issue, as pinpointing the exact location of data and transactions is difficult. Many regulators are still exploring blockchain regulations, and the lack of clarity makes it challenging for organizations to develop compliant interoperability solutions. Additionally, intellectual property and patent issues may pose regulatory obstacles, as proprietary technologies and protocols on various blockchain platforms are often protected by intellectual property rights. Developing interoperability solutions that respect these rights requires careful consideration and substantial legal expertise.

\subsection{Fragmentation of Interoperability Knowledge Framework} The current study presents a preliminary knowledge framework for blockchain interoperability, highlighting that this field remains relatively fragmented, even within specific applications or technologies (e.g., sidechains). Several studies have noted a lack of systematic information regarding types of interoperability, and the definition of interoperability itself is still debated \cite{ren2023interoperability,wang2023exploring,belchior2021survey}. No consensus has yet been reached on models and frameworks for interoperability—both conceptual models and cross-chain asset management models—particularly regarding their specific content and practical applicability. As discussed in Sect. \ref{totalsolution}, there is currently no optimal categorization that unifies all blockchain interoperability technologies.
Future research could aim to systematize knowledge on blockchain interoperability by adopting approaches inspired by frameworks like SEBOK (Systems Engineering Body of Knowledge) \cite{sebok}. Such frameworks could standardize general definitions of interoperability types, delineate their interrelationships, and specify the essential components of blockchain interoperability models and frameworks.

\section{Conclusion and Future Outlook}

In this paper, we conducted a comprehensive literature review on blockchain interoperability. By analyzing over one hundred relevant documents, we identified and categorized more than ten types of technologies, aiming to address the gaps in existing research concerning technical classification and interdisciplinary studies. We anticipate that this research will alleviate the burden for newcomers in the field and provide valuable insights for interdisciplinary researchers.

Looking ahead, as more scholars and organizations acknowledge the significance of interoperability, there is a growing trend in increasing research and development investment. 
Joint efforts from industry alliances and academic institutions will be essential in driving forward the development of $\mathcal{CCI}$ standards and solutions.
Furthermore, the integration of emerging technologies such as IoT, neural networks \cite{aldaej2024deep}, 6G \cite{dohler2024blockchains}, the metaverse \cite{li2023metaopera}, and AI \cite{zuo2023survey} with blockchain interoperability holds promising potential for new application scenarios. For instance, AI algorithms can optimize cross-chain transactions, while IoT devices can securely exchange data and automate processes through interoperable blockchains. 
Overall, the future of blockchain interoperability is highly promising, with the potential to transform interactions and collaborations within blockchain networks. 
In spite of all these difficulties, continued innovation and collaboration in the blockchain community are anticipated to propel the development of robust $\mathcal{CCI}$ solutions. By addressing concerns related to standardization, security, privacy, scalability, and compliance, the blockchain ecosystem can pave the way for new growth and innovation opportunities, ultimately realizing a decentralized future of interconnected devices.

\footnotesize

\bibliography{cas-refs}{}
\bibliographystyle{IEEEtran}

\section*{Author Biographies}

\textbf{Zhihong Deng} received the M.S. degree in the School of Mathematics and Computational Science, Hunan University of Science and Technology, Hunan, China, in 2021. He is currently working the Ph.D. degree at the School of Mathematics and Information Science, Guangzhou University, Guangzhou, China. His main research interests include blockchain, applied cryptography, algorithmic game theory and Web3.

\medskip

\textbf{Chunming Tang} received the Ph.D. degree in applied mathematics from Academy of Mathematics and Systems Science, Chinese Academy of Sciences, Beijing, China, in 2004. He is currently a professor of the School of Mathematics and Information Science, Guangzhou University, Guangzhou, China. He has authored or co-authored more than 100 research papers in refereed international journals and conferences, such as AsiaCrypt, IEEE \textsc{Transactions on Information Theory}, and IEEE \textsc{Transactions on Information Forensics and Security}. His main research interests include code theory, cryptography, and information security.

\medskip

\textbf{Taotao Li} received the Ph.D. degree in cyber security from Institute of Information Engineering, Chinese Academy of Sciences and University of Chinese Academy of Sciences, China, in 2022. He is currently a postdoc with the School of Software Engineering, Sun Yat-Sen University, Zhuhai, China. His main research interests include blockchain, Web3, applied cryptography.

\medskip

\textbf{Parhat Abla} received the Ph.D. degree in cyber security from Institute of Information Engineering, Chinese Academy of Sciences and University of Chinese Academy of Sciences, China, in 2022. He is currently an assistant resercher with the School of Software, South China Normal University. His main research interests include blockchain, Data security, cryptography, lattice-based cryptography.

\medskip

\textbf{Qi Chen} received the B.S. degree from the University of Information Engineering, China, in 2001, the M.S. degree from the National University of Defense Technology, China, in 2006, and the Ph.D. degree from Guangzhou University, China, in 2011, all in mathematics. Since 2017, he has been with Guangzhou University. His research interests include secret sharing, blockchain, cryptography, and coding theory.

\medskip

\textbf{Wei Liang} received a Ph.D. degree in computer science and technology from Hunan University in 2013, and a Postdoctoral Scholar with Lehigh University during 2014-2016. Dr. Liang is currently a Professor at the School of Computer Science and Engineering, Hunan University of Science and Technology. He has authored or co-authored more than 150 journal/conference papers. His research interests include intelligent transportation, security of IoV, blockchain, embedded systems and hardware IP protection, and security management in wireless sensor networks.

\medskip

\textbf{Debiao He} received the Ph.D. degree in applied mathematics from the School of Mathematics and Statistics, Wuhan University, Wuhan, China, in 2009. He is currently a Professor    with the School of Cyber Science and Engineering, Wuhan University. His research interests include cryptography and information security, in particular,   cryptographic protocols. He has authored or co-authored more than 100 research papers in refereed international journals and conferences, such as IEEE \textsc{Transactions on Dependable and Secure Computing}, IEEE \textsc{Transactions on Information Forensics and Security}, and USENIX Security Symposium. His work has been cited more than 15000 times at Google Scholar. He was the recipient of the IEEE \textsc{Systems Journal} 2018, 2019 Best Paper Award and IET Information Security 2019 Best Paper Award. He is with the Editorial Board of several international journals, such as IEEE \textsc{Transactions on Computers}, Journal of Information Security and Applications, Frontiers of Computer Science, and Human-centric Computing and Information Sciences.










\end{document}